\documentclass[12pt,preprint]{aastex}
\usepackage{apjfonts}
\usepackage{epsfig}
\usepackage{amsmath}
\begin{document}

\title{The three-parameter correlations about optical plateaus of gamma-ray bursts}
\author{Shu-Kun Si$^{1}$, Yan-Qing Qi$^{1}$, Feng-Xia Xue$^{1}$, Ya-Jie Liu$^{1}$, Xiao Wu$^{1}$, Shuang-Xi Yi$^{1}$, Qing-Wen Tang$^{2}$, Yuan-Chuan Zou$^{3}$, Fei-Fei Wang$^{3}$ And Xiang-Gao Wang$^{4}$}
\affil{$^{1}$School of Physics and Physical Engineering, Shandong Provincial Key Laboratory of Laser Polarization and Information Technology, Qufu Normal University, Qufu 273165, China; yisx2015@qfnu.edu.cn\\
       $^{2}$Department of Physics, Nanchang University, Nanchang 330031, China; qwtang@ncu.edu.cn\\
       $^{3}$School of Physics, Huazhong University of Science and Technology, Wuhan 430074, China; zouyc@hust.edu.cn \\
       $^{4}$GXU-NAOC Center for Astrophysics and Space Sciences, Department of Physics, Guangxi University, Nanning 530004, China; wangxg@gxu.edu.cn \\
       }

\begin{abstract}
Well-sampled optical light curves of 50 gamma-ray bursts (GRBs) with plateau features
are compiled from the literature. By empirical fitting, we obtained the parameters of
the optical plateaus, such as the decay slopes ($\alpha_{\rm 1}$ and $\alpha_{\rm 2}$), the break times ($T_{\rm b}$), and the
corresponding optical fluxes ($F_{\rm b}$) at the break times. The break time of optical plateaus ranges from tens
of seconds to $10^6$ seconds, with a typical value about $10^4$ seconds. We have calculated the break luminosity, and it mainly ranges from $10^{44}$ erg $s^{-1}$ to $10^{47}$ erg $s^{-1}$, which is generally two or three orders of magnitude less than the corresponding break luminosity of the X-ray afterglow plateaus. We reanalyzed the optical plateaus and also found that a significantly tighter correlation exists when we added the isotropic equivalent energy of GRBs $E_{\rm \gamma,iso}$ into the $L_{\rm b,z}-T_{\rm b,z}$ relation. The best fit correlation is obtained to be $L_{\rm b,z}\propto T_{\rm b,z}^{-0.9}E_{\rm \gamma,iso}^{0.4}$. We next explored the possible correlations among $L_{\rm b,z}$, $T_{\rm b,z}$ and $E_{\rm p,i}$, and found there is also a tight correlation between them, which takes the form of $L_{\rm b,z}\propto T_{\rm b,z}^{-0.9}E_{\rm p,i}^{0.5}$. We argue that these two tight $L_{\rm b,z}-T_{\rm b,z}-E_{\rm \gamma,iso}$ and $L_{\rm b,z}-T_{\rm b,z}-E_{\rm p,i}$ correlations are more physical, and it may be directly related to radiation physics of GRBs. The tight correlations are possible to be used as standard candles.

\end{abstract}


\section{Introduction}\label{sec:intro}

Gamma-ray bursts (GRBs) are the most luminous electromagnetic explosive events in the universe.
The widely accepted model of this phenomenon is the fireball model, which depicts the erratic,
transient events in gamma-rays (Piran 2004; M{\'e}sz{\'a}ros 2006; Zhang 2007; Kumar \& Zhang 2015),
and followed by long-lived, decaying afterglows in longer wavelengths (Rees \& M\'esz\'aros 1992;
M\'esz\'aros \& Rees  1997; Sari et al. 1998; Panaitescu et al. 1998; Zou et al. 2005; Yi et al. 2013;
Gao et al. 2013).  Lots of afterglow emissions were detected after decades of observations, which
have significantly improved our understanding of the physical origin of GRBs. In particular, after
the successful launch of the {\em Swift} satellite in 2004 (Gehrels et al. 2004), the canonical
X-ray light curves are proposed, such as several power-law segments followed by erratic flares
(Zhang et al. 2006; Nousek et al. 2006). Actually, the observed optical afterglow is also a mix of
various emission components, including the optical flares, the shallow decay segment, the afterglow
onset bump, and the late rebrightening component. The physical implications of the flares
and the plateau phase in X-ray and optical bands are discussed a lot, both particular phenomena
are related to the central engine of the GRB itself ( Dai \& Lu 1998a, b; Zhang \& M\'esz\'aros 2001;
Burrows et al. 2005; Dai et al. 2006; Falcone et al. 2006; Dall'Osso et al. 2011; Wang \& Dai 2013;
Rownlinson et al. 2013, 2014, 2017; Rea et al. 2015; Yi et al. 2016, 2017a).

The observed GRBs, whose redshifts up to eight (Salvaterra et al. 2009; Cucchiara et al. 2011),
make GRBs to be among the farthest known astrophysical sources, indicating GRBs may be good
candidates that can be used to probe our Universe. Several interesting empirical correlations
have been proposed by the observed GRB data. These physical correlations not only could help
the interpretation of the physical mechanisms responsible for the GRBs, but
also can infer important information about the nature of the emitting source
(e.g., Wang et al. 2015 for a recent review). There were several tight relations have been proposed
years ago, such as, $E_{\rm \gamma,iso}-E_{\rm p,i}$
(also called Amati Relation, Amati et al. 2002), $E_{\rm \gamma}-E_{\rm p,i}$ (Ghirlanda Relation, Ghirlanda et al. 2004) and $L_{\rm iso}-E_{\rm p,i}$(Yonekotu Relation, Yonekotu 2004). And also some tight correlations about the initial Lorentz factor $\Gamma_{0}$
among $E_{\rm \gamma,iso}$, $L_{\rm \gamma,iso}$ and $E_{\rm p,i}$ are obtained (Liang et al. 2010, 2015;
Ghirlanda et al. 2012; L{\"u} et al. 2012; Yi et al. 2015; Zou et al. 2015; Tang et al. 2015).
Some multi-variable correlations have been found for GRBs (e.g., Liang \& Zhang 2005; Rossi et al.
2008), which are useful to understand GRB physics. A shallow-decay (plateau) segment
is commonly seen in the X-ray afterglow light curves. Interestingly, a tight correlation
has been reported to exist between the break time of the plateau phase ($T_{\rm b,z}$, measured in the
rest frame) and the corresponding X-ray luminosity ($L_{\rm b,z}$, measured in the rest frame)
in the X-ray afterglows (the two dimensional Dainotti relation, Dainotti et al. 2008, 2010, 2011, 2013, 2015, 2017a). Li et al. (2012) also selected a group optical afterglows
with shallow-decay feature, overplot $L_{\rm b,z} - T_{\rm b,z}$ correlation in the burst frame, and
found optical data share a similar relation to the X-ray data. Later, Xu \& Huang (2012) compiled
a group of X-ray plateau sample, tried to add a third parameter, i.e. the isotropic energy release
$E_{\rm \gamma,iso}$, into the $L_{\rm b,z} - T_{\rm b,z}$ correlation, and found that the new three-parameter
correlation is much tighter than the previous correlation. Another three parameter relation, $L_{\rm peak}-L_{\rm b,z}-T_{\rm b,z}$ relation, is proposed in Dainotti et al. (2016) and Dainotti et al. (2017b), where the $L_{peak}$ is the peak luminosity in the prompt emission. The tighter new three-parameter correlations
may hopefully give a better measure for our universe. For a complete review on GRB correlations
also seen Dainotti \& del Vecchio (2017c) and Dainotti et al. (2018).

Therefore, it is interesting to continue to search for possible multi-variable correlation using the
optical plateaus, which are useful to understand GRB physics, and discuss their physical implications
for both X-ray and optical plateaus. In this paper, we try to compile a group of the optical afterglows
with plateau features, and obtain the results of the optical plateaus by the empirical fitting (Section 2).
In Section 3, we present the distributions of GRB optical plateaus, and study the correlations between
parameters of the optical plateaus, including two tight three - parameter correlations about the optical plateaus.
Conclusions and discussion are given in Section 4. A concordance cosmology with
parameters $H_0 = 71$ km s $^{-1}$  Mpc$^{-1}$, $\Omega_{\rm M}=0.30$, and $\Omega_{\Lambda}=0.70$ is adopted
in all part of this work.

\section{Data and Lightcurve Fitting}
According to Swift observations, lots of GRBs appear a plateau phase in the early X-ray afterglow,
followed by the normal decay phase (or a sharp decay) (Zhang et al. 2006; Nousek et al. 2006). The
similar shallow decay phase are also appeared in the optical light curves, but only a small fraction
of optical afterglows have the plateau phase, compared with X-ray light curves (Li et al. 2012). These particular shallow
decay phases may have similar physical origin, both of them are related to the central engine of the
GRB itself. The plateau phase is currently understood as being due to ongoing energy injection from the
central engine. One reasonable scenario is a fast rotating pulsar/magnetar as the the central engine,
which spins down through magnetic dipole radiation (Dai \& Lu 1998a, b; Zhang \& M\'esz\'aros 2001;
Fan \& Xu 2006; Liang et al. 2007; Rowlinson et al. 2013; L{\"u} \& Zhang 2014).

In this paper, we try to study the correlations about the optical plateaus, by extensively searching for
the remarkable feature of shallow decay phase from the published papers. Well-sampled light curves are available for 50 GRBs which
have such a shallow-decay segment. Most of the samples are taken from Li et al. (2012)
(see their Figure 7), and some GRBs are taken from Wang et al. (2015). According to Dai \& Liu (2012),
who have investigated that the sufficient angular momentum of the accreted matter is
transferred to the newborn millisecond magnetar and spins it up. It is this spin-up that leads to a dramatic increase of the
magnetic-dipole-radiation luminosity with time and thus significant brightening of an early afterglow.
We also selected some optical lightcurve with slight rising plateaus at early time. Therefore, the well-sampled afterglows which with the obvious plateaus in this paper are the transition in the optical afterglow light curves from a shallow decay (or a slight rising phase) to the normal decay (or an even steeper decay). Some optical light curves are usually composed of one or more power-law segments along with some flares, or rebrightening features, such as GRBs 030723, 081029, 100219A and so on. Here we make our fits only around the shallow decay feature, and exclude the mixed components when fitting the light curves. We fit the shallow decay with an empirical smooth broken power-law function (SBPL, Li et al. 2012)
\begin{equation}\label{SBPL1}
F_{\rm model}(t)=F_{b}\left[\left(\frac{t}{T_{b}}\right)^{\alpha_{1}\omega}+\left(\frac{t}{T_{b}}\right)^
{\alpha_{2}\omega}\right]^{-\frac{1}{\omega}},
\end{equation}
where $\alpha_{\rm 1}$ and $\alpha_{\rm 2}$ are the temporal slopes of the plateau and the followed decay,
$T_{\rm b}$ is the break time, $F_{\rm b}$ is the optical flux of the break time and $\omega$ represents the
sharpness of the peak of the light curve component. Actually, $\omega =3$ is applied when fitting lightcurve.
This method is very similar to the fitting method of GRB X-ray plateaus (Dainotti et al. 2016, 2017b). We provide the goodness-of-fit test and the residuals in each figures.
For the goodness-of-fit test, we take the $\chi^2$ test:
\begin{equation}
\chi^2 =  \sum_{1}^{\rm N_{bin}} \frac{\left[F_{\rm obs}(t_i)- F_{\rm model}(t_i)\right]^2}
{\left[\delta F_{\rm obs}(t_i)\right]^2}
\end{equation}
where $F_{\rm obs}(t_i)$ is the observational flux at time of $t_i$ and $\delta F_{\rm obs}(t_i)$ is the corresponding error at $68\%$ confidence level. The degrees of freedom (${\rm dof}$) is derived of $N_{bin}-4$, here 4 is number of the free parameters in the SBPL function. We assess a good fit when the value of $\chi^2/{\rm dof}$ is close to 1. Next, we calculate the residual as following:
\begin{equation}
{\rm Res} =  \frac{F_{\rm obs}(t_i)- F_{\rm model}(t_i)}{F_{\rm model}(t_i)}
\end{equation}
which can show the variation of residual flux. Those test are also plotted in Figure 1.

The optical plateaus are shown in Figure 1, and the fitting results for the shallow decay segments are
summarized in Table 1. We obtain the decay slopes ($\alpha_{\rm 1}$ and $\alpha_{\rm 2}$), the break times ($T_{\rm b}$),
and the corresponding optical flux ($F_{\rm b}$) at that moment. The luminosity at the break time ($L_{\rm b,z}$) of
our sample is derived from the equation:
\begin{equation}\label{SBPL2}
L_{b,z}=4\pi D_{L}^{2} F_{b}/(1+z),
\end{equation}
where $z$ is the redshift, and $D_L$ is the luminosity distance (also seen Oates et al. 2009). To produce the luminosities light curves of all optical afterglows, they converted the light curves (in count rate) into flux density and then into luminosity using Equ. 2 in Oates et al. (2009). However, we also calculated the luminosity at the optical break time using Equ. 4. in our paper, and the two motheds are very similar.

\section{The results}
Figure 2 shows the distributions of the break times ($T_{\rm b}$), the luminosity at the break time ($L_{\rm b,z}$) and
the decay slopes ($\alpha_{\rm 1}$ and $\alpha_{\rm 2}$). The break time ranges from tens of seconds to $10^6$ seconds
after the GRB trigger, with a typical value about $10^4$ seconds, which is matching the break time distribution
of X-ray plateaus (Liang et al. 2007; Dainotti et al. 2010; L{\"u} \& Zhang 2014). The break luminosity of the
optical plateaus mainly range from $10^{44}$ erg $s^{-1}$ to $10^{47}$ erg $s^{-1}$, generally two or three
orders of magnitude less than the corresponding luminosity of the X-ray afterglow plateaus. The typical slope values
of the two segments are about $-0.4$ and $-1.3$, which is consistent with the features of the plateaus.

Figure 3 presents two correlations about $F_{\rm b}-T_{\rm b,z}$ and $L_{\rm b,z}-T_{\rm b,z}$ for optical plateaus.
$L_{\rm b,z}$ and $T_{\rm b,z}$ are all transferred to the rest frame, and the fitting results are shown in Table 2.
The break optical flux is anti-correlated with the break of optical plateaus, with the slope index 0.71.
The optical break luminosity $L_{\rm b,z}$ is anti-correlated with $T_{\rm b,z}$, as shown in Figure 3. The best fit shows
in Table 2, with a Spearman correlation coefficient of $R = 0.84 $, which is clearly stronger than the $F_{\rm b}-T_{\rm b,z}$ correlation.
According to Dainotti et al. (2010), who have considered the evolution with the redshift in different bins about the selected sample, they found the correlation coefficient of the correlation about the luminosity at the break time and break time (hereafter LT) is quite large in the different redshift bins, thus arguing in favor of the existence of LT correlation at any redshift.
To properly investigate the intrinsic nature of the correlation it is necessary to apply the Efron \& Petrosian (1992) method which is able to overcome the problem of redshift evolution in the variables such as time and luminosity. For an approach successfully tested of this method see Dainotti et al. (2013) and Dainotti et al. (2015). However, this investigation goes beyond the scope of the present paper. The slope of  $L_{\rm b,z}-T_{\rm b,z}$ is roughly $-1$, which indicates the corresponding R-band energy $E_{\rm R,iso} \equiv L_{\rm b,z} T_{\rm b,z}$ is roughly a standard energy reservoir. This tight correlation between $L_{\rm b,z}$ and
$T_{\rm b,z}$ for optical plateaus is almost the same as the corresponding correlation of X-ray plateaus
(Dainotti et al. 2010; Li et al. 2012), and suggests that the longer time of plateau associates with
the dimmer break luminosity.

Interestingly, the subsample of 19 long GRBs associated with supernovae (SNe) also presents a very high correlation
coefficient between the luminosity at the end of the plateau and the end time of the plateau of the X-ray afterglows (Dainotti et al. 2017a).
Although, some difference in slopes between the normal long GRBs with no SNe and long GRBs with SNe, and the debate about this difference
remains open and it may be resolved with more SNe data, the tighter LT correlation about long GRBs with SNe is a significant finding and
may hopefully give a better measure for our universe.

According to Xu \& Huang (2012), they studied a group of X-ray afterglows with plateau features, and added
the isotropic energy release into $L_{\rm b,z} - T_{\rm b,z}$. They finally obtained even more tighter the new
three-parameter correlation, called $L_{\rm b}-T_{\rm b}-E_{\rm \gamma,iso}$ correlation. In this paper, we try to
search for possible multi-variable correlation using the optical plateaus, which depending on the
correlation of $L_{\rm b,z}-T_{\rm b,z}$ for optical plateaus. Therefore, we selected the isotropic energy
for each GRB with the optical plateau (see Figure 1). We investigate whether an intrinsic correlation
exists between the three parameters of $L_{\rm b,z}$, $T_{\rm b,z}$ and $E_{\rm \gamma,iso}$ for optical plateaus as,
\begin{equation}\label{SBPL3}
L_{\rm b,z}=A+B \log T_{\rm b,z}+C \log E_{\rm \gamma,iso},
\end{equation}
where $A$, $B$, and $C$ are constants to be determined from the fit to
the observational data. In this equation, $A$ is the constant of the while $B$ and $C$ are actually the power-law indices of
break time and isotropic equivalent energy when we approximate $L_{\rm b,z}$ as power-law functions of $T_{\rm b,z}$ and $E_{\rm \gamma,iso}$.
More details seen Xu \& Huang (2012) and Liang et al. (2015). In order to find more significant correlation, we gave the Spearman coefficient and the related hypothesis test p-value. If the p-value is smaller than 0.1, it means the correlation has very high probability to be true. At the same time, if the absolute value of Spearman coefficient is closer to 1, the correlation is tighter. We use adjusted $R^{2}$ to stand for the goodness of the regression model. Adjusted $R^{2}$ means the variance percentage explained considering the parameter freedom. After that, we did hypothesis tests for all the regression coefficients and the whole linear regression model. Similarly, if the p-value is smaller than 0.1, it means the model has very high probability to be true. By using the method discussed above, an even more tighter correlation
about optical plateau is obtained between the three parameters with,
\begin{equation}\label{SBPL4}
\log L_{\rm b,z}=(29.22\pm 5.04)+(-0.92\pm0.08)\times \log T_{\rm b,z}+(0.37\pm0.09)\times \log E_{\rm \gamma,iso}.
\end{equation}
The adjusted $R^{2}$ is 0.77. The F-test p-value for the whole linear model is $3.4 \times 10^{-16}$. The regression coefficient for $T_{\rm b,z}$ has t-test, the p-value is $6.6 \times 10^{-15}$. The regression coefficient for $E_{\rm \gamma,iso}$ has t-test, the p-value is $3 \times 10^{-4}$. All the linear regression model and coefficients pass the hypothesis tests. However, for $L_{\rm b,z}-T_{\rm b,z}$ correlation, the adjusted $R^{2}$ is 0.7. It implies appending $E_{\rm \gamma,iso}$ is meaningful. The fitting result is shown in Figure 4, and it clearly indicates that this three-parameter correlation is more
tighter than for $L_{\rm b,z}-T_{\rm b,z}$ with the 50 optical plateaus. According to the tight $E_{\rm \gamma,iso}-E_{\rm p,i}$
correlation, called `Amati Relation' (Amati et al. 2002), we also selected the peak energy ($E_{\rm p,i}$) in the $\nu f_{\rm \nu}$ spectrum from
the literature, and the data are shown in Table 1. Next, we explored the possible correlations among $L_{\rm b,z}$, $T_{\rm b,z}$
and $E_{\rm p,i}$, and found there is also a tight correlation between them, and
\begin{equation}\label{SBPL5}
\log L_{\rm b,z}=(47.48\pm 0.56)+(-0.91\pm0.09)\times \log T_{\rm b,z}+(0.48\pm0.16)\times \log E_{\rm p,i}.
\end{equation}
The adjusted $R^{2}$ is 0.75. The F-test p-value for the whole linear model is $4.3 \times 10^{-15}$. The regression coefficient for $T_{\rm b,z}$ has t-test, the p-value is $1.1 \times 10^{-13}$. The regression coefficient for $E_{\rm p,i}$ has t-test, the p-value is $4.6 \times 10^{-3}$. All the linear regression model and coefficients also pass the hypothesis tests. This correlation also becomes better compared with $L_{\rm b,z}-T_{\rm b,z}$ correlation. As shown in Table 2, the Spearman correlation coefficients are 0.89 and 0.87 respectively with chance probability smaller than $10^{-4}$, which suggest these two correlations are quite tight. With the luminosities and energies inside the correlations, they can be used as standard candles. Using these two  standard candle relations, one could perform the cosmological parameters independently comparing with other methods.

The tight three - parameter of $L_{\rm b,z}$, $T_{\rm b,z}$ and $E_{\rm \gamma,iso}$ is found using X-ray afterglows with plateau
phase (Dainotti et al. 2010; Xu \& Huang 2012). In our selected optical sample, there also exists the same tight three - parameter
correlation for the optical plateaus with the Spearman correlation coefficient $R=0.89$ (see Table 2). The
similar tight correlations between X-ray and optical plateaus indicate that both of them may have the same physical
origin. Another tight correlation is obtained among $L_{\rm b,z}$, $T_{\rm b,z}$ and $E_{\rm p,i}$ of optical plateaus, with
the Spearman correlation coefficient $R=0.87$ (see Figure 4 and Table 2). The redshift range covered by our GRB optical sample is not very large, therefore, we also consider the evolution with the redshift of the sample. We divided the optical sample into three redshift bins
(0-0.98, 1-1.92, 2-4,67) and five redshift bins (0-0.70, 0.72-1.24, 1.25-2.03, 2.1-2.7, 2.71-4,67), respectively. The results summarized in Table 3 and shown in Figures 5, 6. The correlation coefficient is quite large in all the redshift bins for both correlations, the correlations of sub-sample are even tighter than the whole sample. The slopes for the different bins of the two correlations are consistent with the whole sample, thus arguing in favor of the existence of the three correlations at any z. We also plotted the slope of the correlation vary with the redshift in different bins in Figure 7.

The tight correlations of $L_{\rm b}-T_{\rm b}$ and $L_{\rm b}-T_{\rm b}-E_{\rm \gamma,iso}$ are both found in X-ray and optical plateaus, respectively (Dainotti et al. 2008, 2010, 2015, 2017a, b; Xu \& Huang 2012; Li et al. 2012). Notice our correlation is about $L_{\rm b}-T_{\rm b}$ when considering the prompt isotropic energy $E_{\rm \gamma,iso}$ for the optical plateaus, while long GRBs with a plateau phase in their X-ray afterglows also obey $L_{\rm b}-T_{\rm b}-L_{peak}$ relation (3D relation), where $L_{peak}$ is the peak luminosity of prompt emission. According to Dainotti et al. (2016) and (2017a), the 3D relation planes are not statistically different for sub-categories when the sample are divided into X-ray flashes, GRBs associated with supernovae, ordinary long-duration GRBs, and short GRBs with extended emission. The similar tight correlations between X-ray and optical plateaus indicate that both of them may have the same physical origin. However, this is not the case when the sample of GRBs associated with SNe Ib/c is taken into consideration. In the case
of the X-ray correlation the slope is roughly -2 when there is a strong spectroscopic association between GRB and SNe Ib/c.

From Eqs. (\ref{SBPL4}) and (\ref{SBPL5}), they show the coefficients of $\log T_{\rm b,z}$ are $-0.92$ and $-0.91$ respectively, which are quite close to $-1$. We doubt the combination of $L_{\rm b,z}$ and $T_{\rm b,z}$ has any correlation with  $\log E_{\rm \gamma,iso}$ and $\log E_{\rm p,i}$. Then we performed the statistics on the $E_{\rm R,iso}$ related to $\log E_{\rm \gamma,iso}$ and $\log E_{\rm p,i}$ respectively. The results are shown in the last two rows of Table 2 and in Figure 8. From the figure, one can see some positive correlations. The correlations should come from the total kinetic energy. However, the Spearman correlation coefficients are 0.55 and 0.45 respectively, which are clearly looser than the 3-parameter correlations. This indicates the 3-parameter correlations are more meaningful. The energy of the R band and the energy of the $\gamma-$ray band may not have very tight correlation, and the same to the peak energy of prompt emission. It might show the prompt emission is not straightforward related to the optical emission.

\section{Conclusions and Discussion}

We have compiled the optical afterglow light curves for 50 GRBs with obvious plateau phases, by fitting
the light curves with a empirical smooth broken power-law function, we obtain the parameters of the optical
plateaus, such as the decay slopes ($\alpha_{\rm 1}$ and $\alpha_{\rm 2}$), the break times ($T_{\rm b}$), and the
corresponding optical flux ($F_{\rm b}$) at that moment. The break time of optical plateaus ranges from tens
of seconds to $10^6$ seconds, with a typical value about $10^4$ seconds, and the corresponding luminosity ($L_{\rm b,z}$)
mainly range from $10^{44}$ erg $s^{-1}$ to $10^{47}$ erg $s^{-1}$, generally two or three orders of magnitude
less than the corresponding luminosity of the X-ray afterglow plateaus. We added the isotropic energy into
the correlation of $L_{\rm b,z}- T_{\rm b,z}$ for optical plateaus, and found the new three-parameter correlation is also
existed in the GRBs with an obvious optical plateau phase. This $L_{\rm b,z}-T_{\rm b,z}-E_{\rm \gamma,iso}$ correlation is
tighter than the $L_{\rm b,z}-T_{\rm b,z}$ correlation for optical plateaus. We next explore the possible correlations
among $L_{\rm b,z}$, $T_{\rm b,z}$ and $E_{\rm p,i}$, and found there is also a tight correlation between them. The similar
tight correlations between X-ray and optical plateaus indicate that both of them may have the same physical origin.
We argue that this two tight $L_{\rm b,z}-T_{\rm b,z}-E_{\rm \gamma,iso}$ and $L_{\rm b,z}-T_{\rm b,z}-E_{\rm p,i}$ correlations are more physical, and it may be directly related to radiation physics of GRBs. We suggest these two relations are possible to be used as standard candles.

In order to identify shallow decays, one needs to systematically explore temporal breaks in the afterglow light curves.
Theoretically there are another two types of temporal breaks except the shallow decays.  The first type is a transition
from the normal decay phase (the slope $\sim-1$) to a steeper phase (the slope $\sim-2$), which is
best interpreted as a jet break (Rhoads 1999; Sari et al. 1999; Frail et al. 2001; Wu et al. 2004; Liang et al. 2008; Racusin et al.
2009; Yi et al. 2017b; Xi et al. 2017). The second type connects the onset afterglow with a smooth bump in the early time.
The afterglow onset is produced by GRB fireball as decelerated by the circumburst medium (Sari \& Piran 1999; Molinari et al. 2007;
Liang et al. 2010; Yi et al. 2013). Differentiation of these types of breaks is usually straightforward,
but sometimes can be more complicated  (see Wang et al. 2015), therefore some disguised breaks with plateau features may
also mixed together with the optical plateaus.

The optical light curve behavior is quite different from the X-rays. In the X-rays, they are mainly plateaus followed by steep decays, which indicates the origins before and after the steep decays are different. Including both the X-ray flares and plateaus, they are mainly believed being the long lasting activity of the central engine, either by the direct emission, by the late internal shocks, or by the electromagnetic energy injection. After the steep decays, they are categorized into the normal late afterglow emitted by the external shock into the ambient material. However, from the light curves of the optical band, there is no clear evidence for this gap, even there are also breaks for many afterglows. Especially, after the breaks, there are no steep decays followed by normal afterglow decays. These might indicate the different origins for the breaks for X-rays and the optical band. For the X-rays  the breaks, they mainly show the properties of the central engine, while for the optical breaks, they are mainly the combination of the central engine and the ambient materials. Their also exist 3-parameter correlations for the optical band, which indicates either the central engine and the ambient for GRBs are connected, or the inner radiation mechanism dominates the correlations. For the former case, the breaks might be caused by the deceleration of the GRB jets, while the deceleration time is contributed by both the total kinetic energy and medium density. For the later case, the breaks might come from  the typical frequencies crossing the observational band. These two hypothesizes are distinguishable. One can figure out by checking the spectral indices before and after the breaks. With accumulated data, especially multi-band spectra by more powerful telescopes, one may reveal the inner mechanism. The most promising situation might be that the optical breaks might be divided into several groups by taking care of the spectral index and the light curve shapes. The correlations of sub-groups might be even tighter. This may give yet another even tighter standard candle relation.

\section*{Acknowledgments}
We thank the anonymous referee for constructive suggestions.
This work is supported by the National Basic Research Program of China (973
Program, grant No. 2014CB845800), the National Natural Science Foundation
of China (Grant Nos. 11703015, 11547029, 11673006), the China Postdoctoral Science Foundation (Grant
No. 2017M612233), the Natural Science Foundation of Shandong Province (Grant No. ZR2017BA006),
and the Programs of Qufu Normal University (xkj201614, 201710446105), the Guangxi Science Foundation
(grant No. 2016GXNSFFA380006), the Innovation Team and Outstanding Scholar Program in Guangxi Colleges, and
the One-Hundred-Talents Program of Guangxi colleges.
.

\begin{figure*}
\centering
\includegraphics[angle=0,scale=0.30]{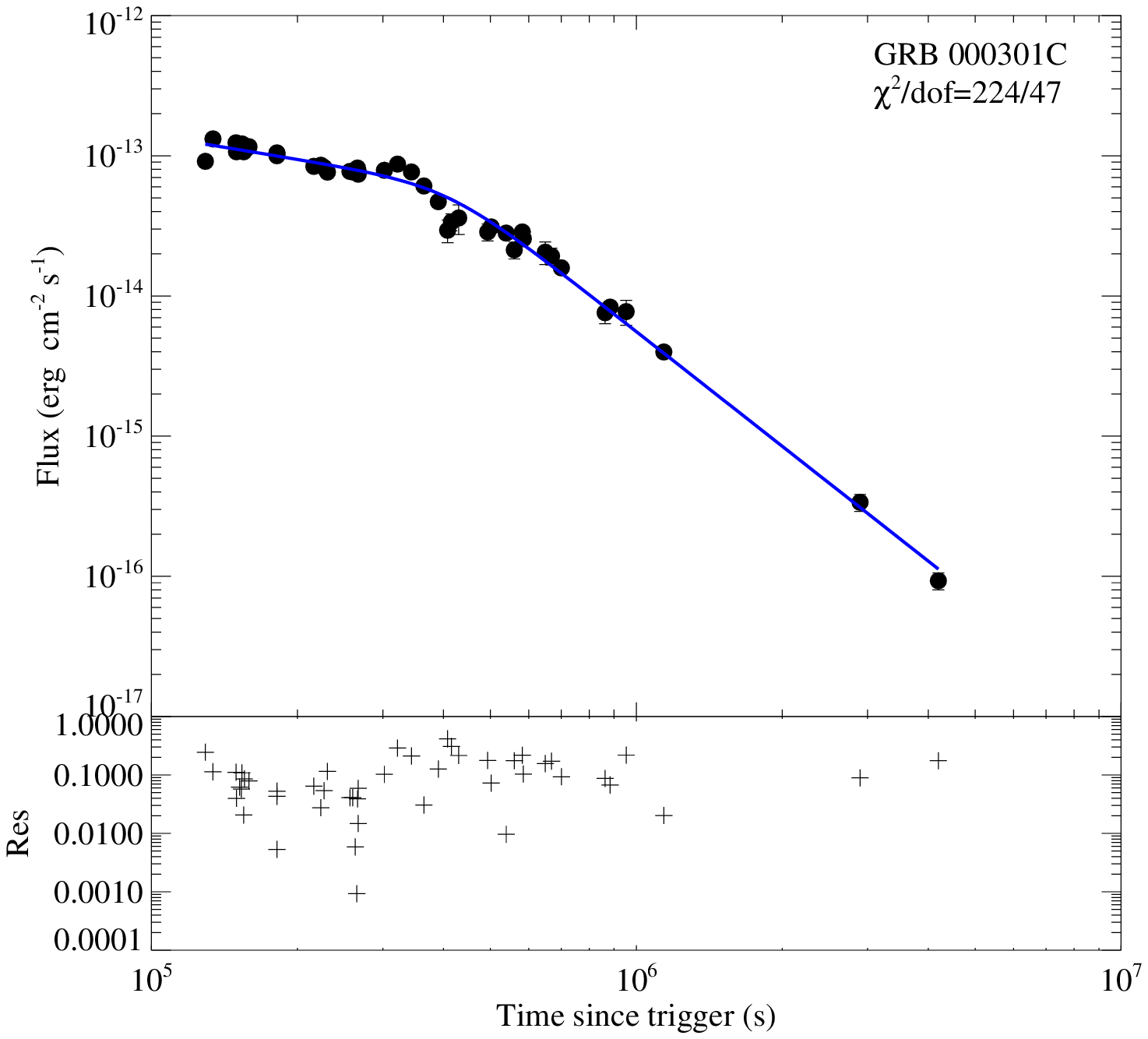}
\includegraphics[angle=0,scale=0.30]{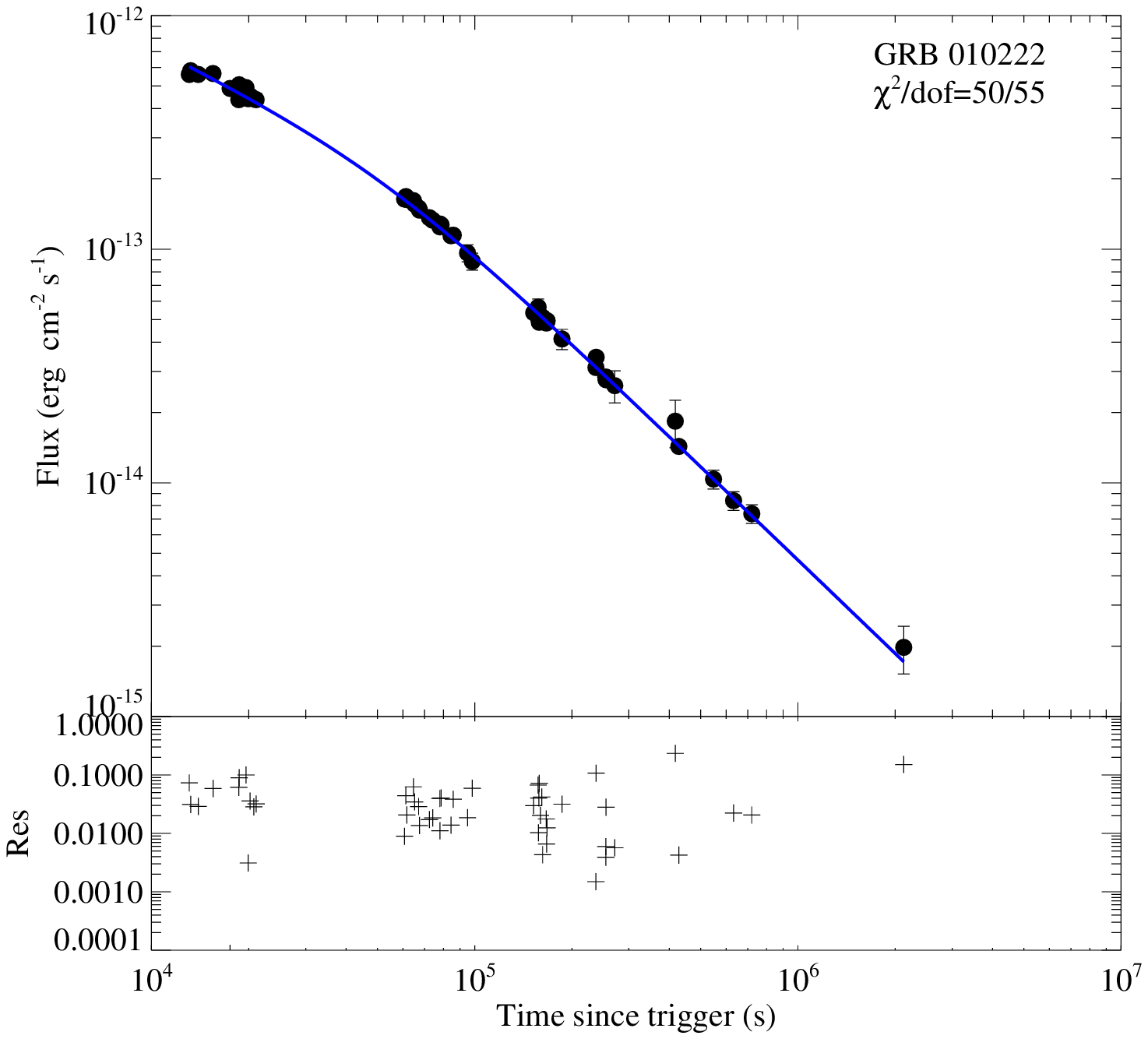}
\includegraphics[angle=0,scale=0.30]{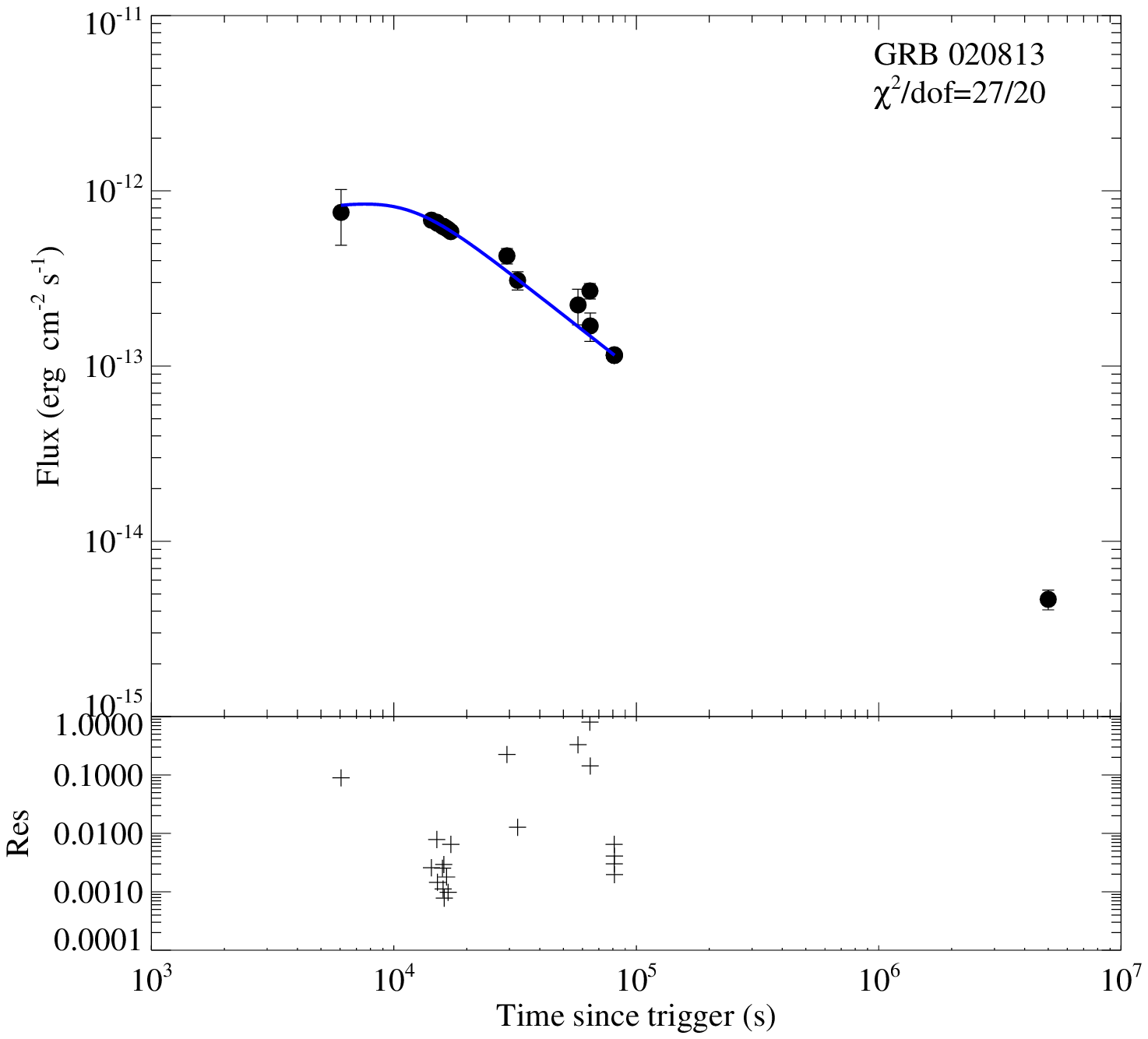}\\
\includegraphics[angle=0,scale=0.30]{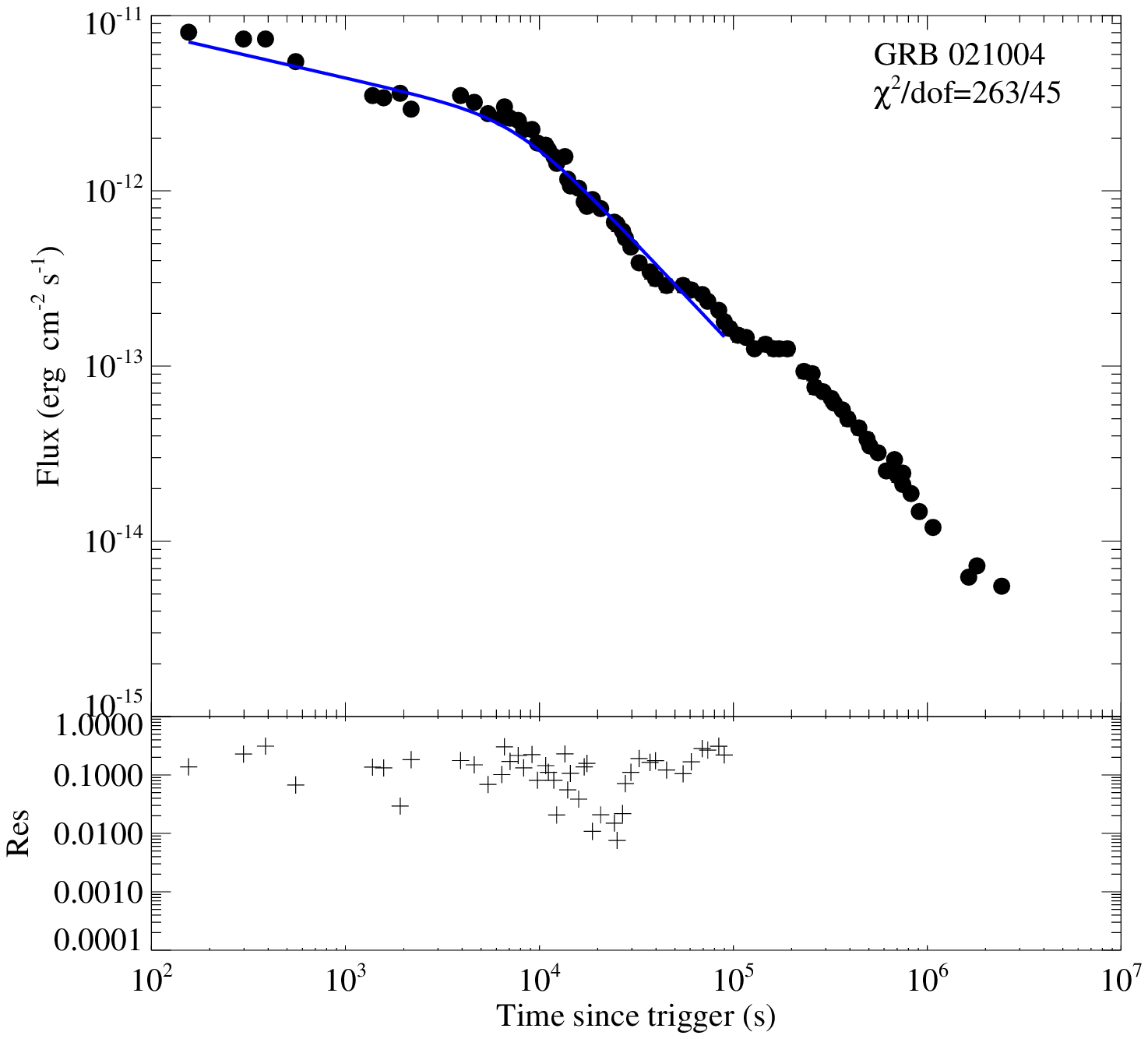}
\includegraphics[angle=0,scale=0.30]{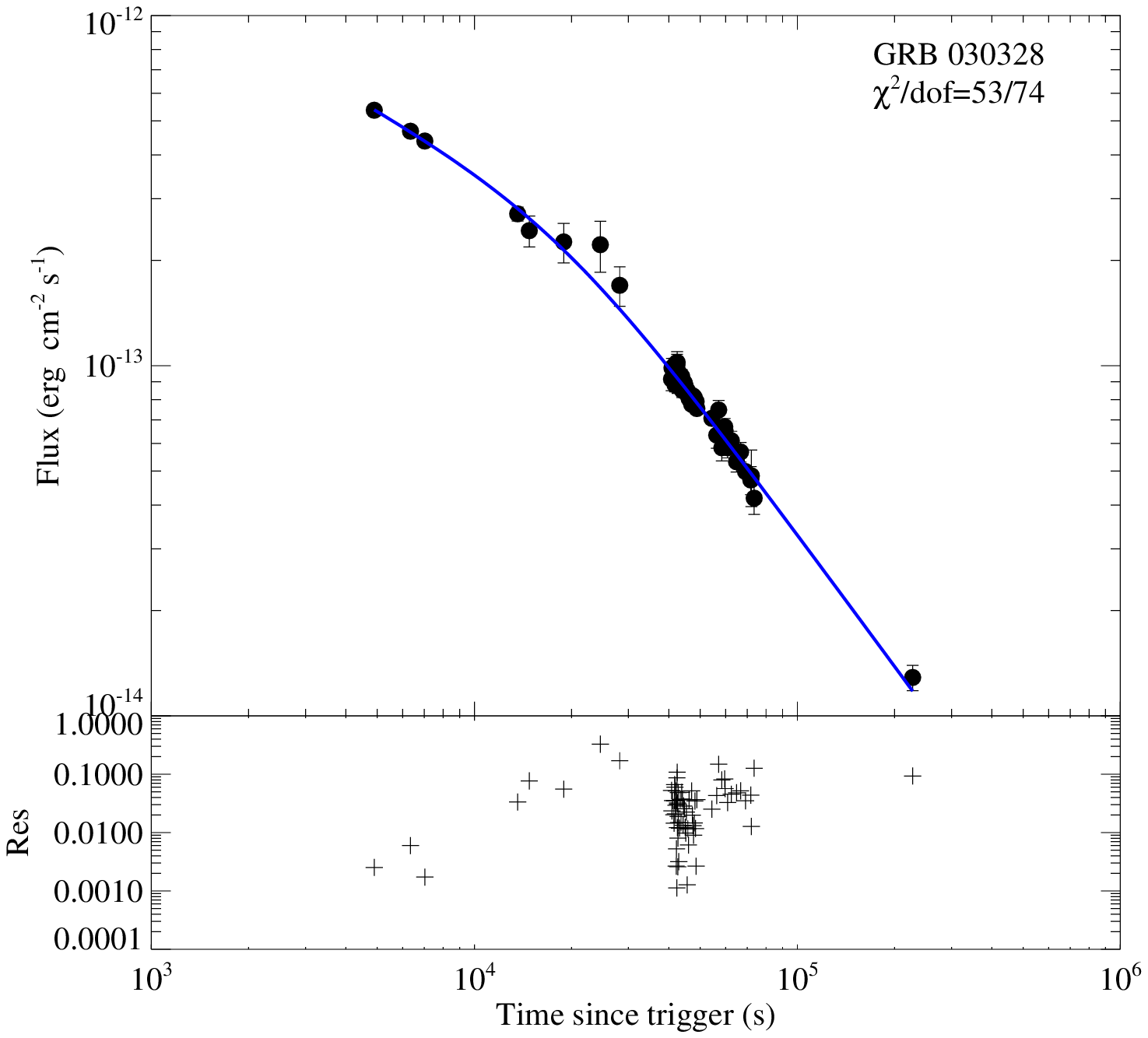}
\includegraphics[angle=0,scale=0.30]{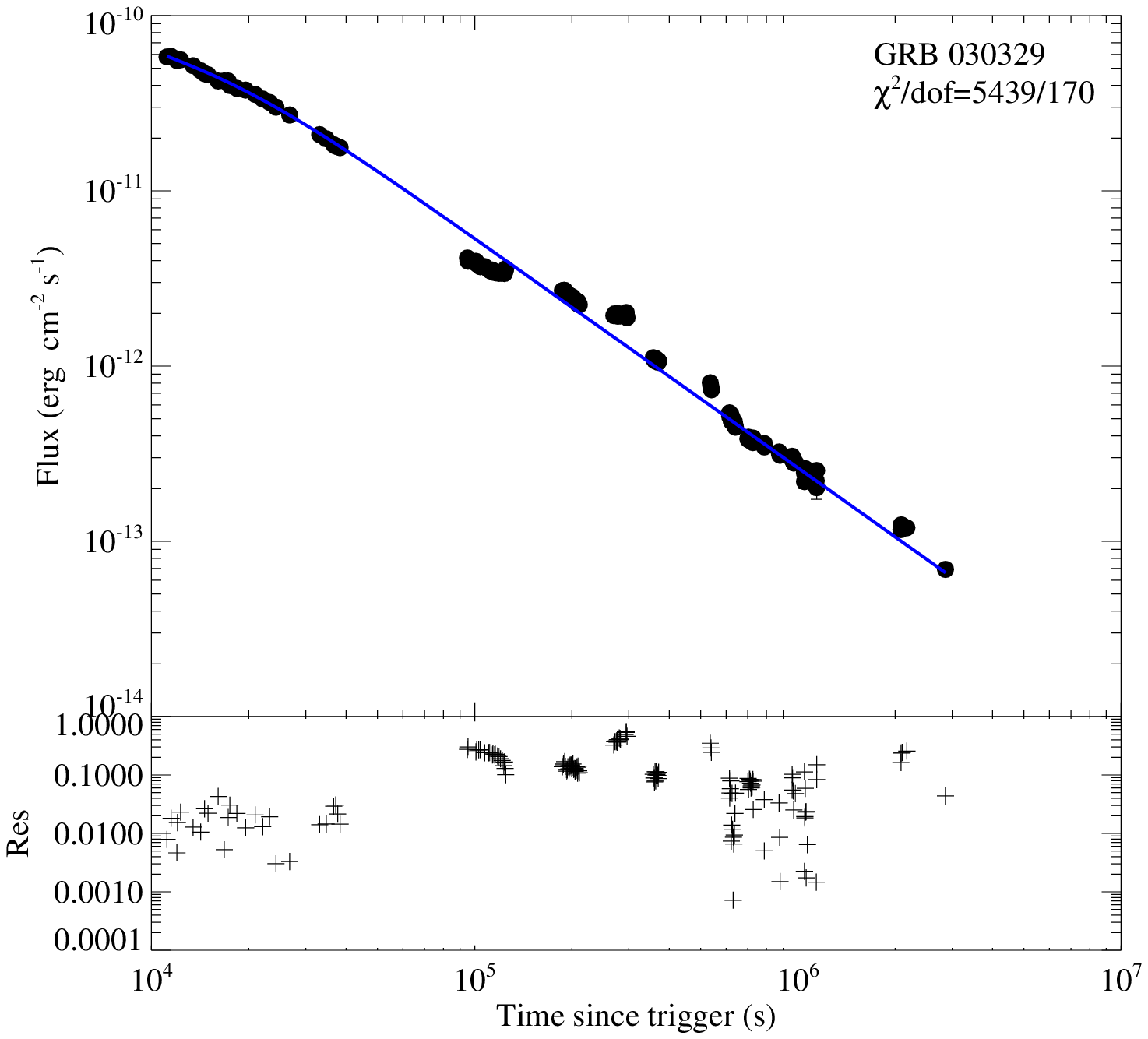}\\
\includegraphics[angle=0,scale=0.30]{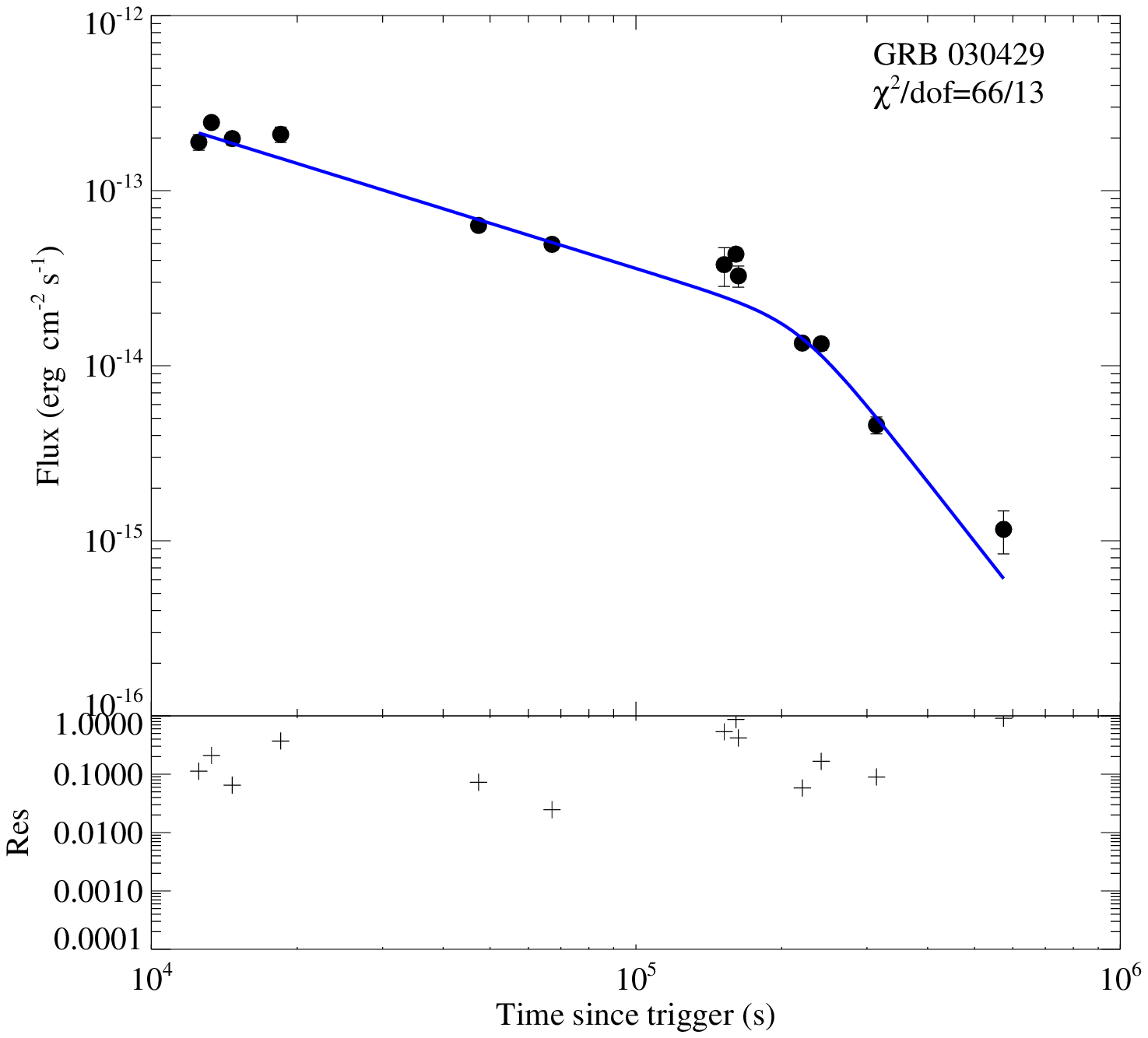}
\includegraphics[angle=0,scale=0.30]{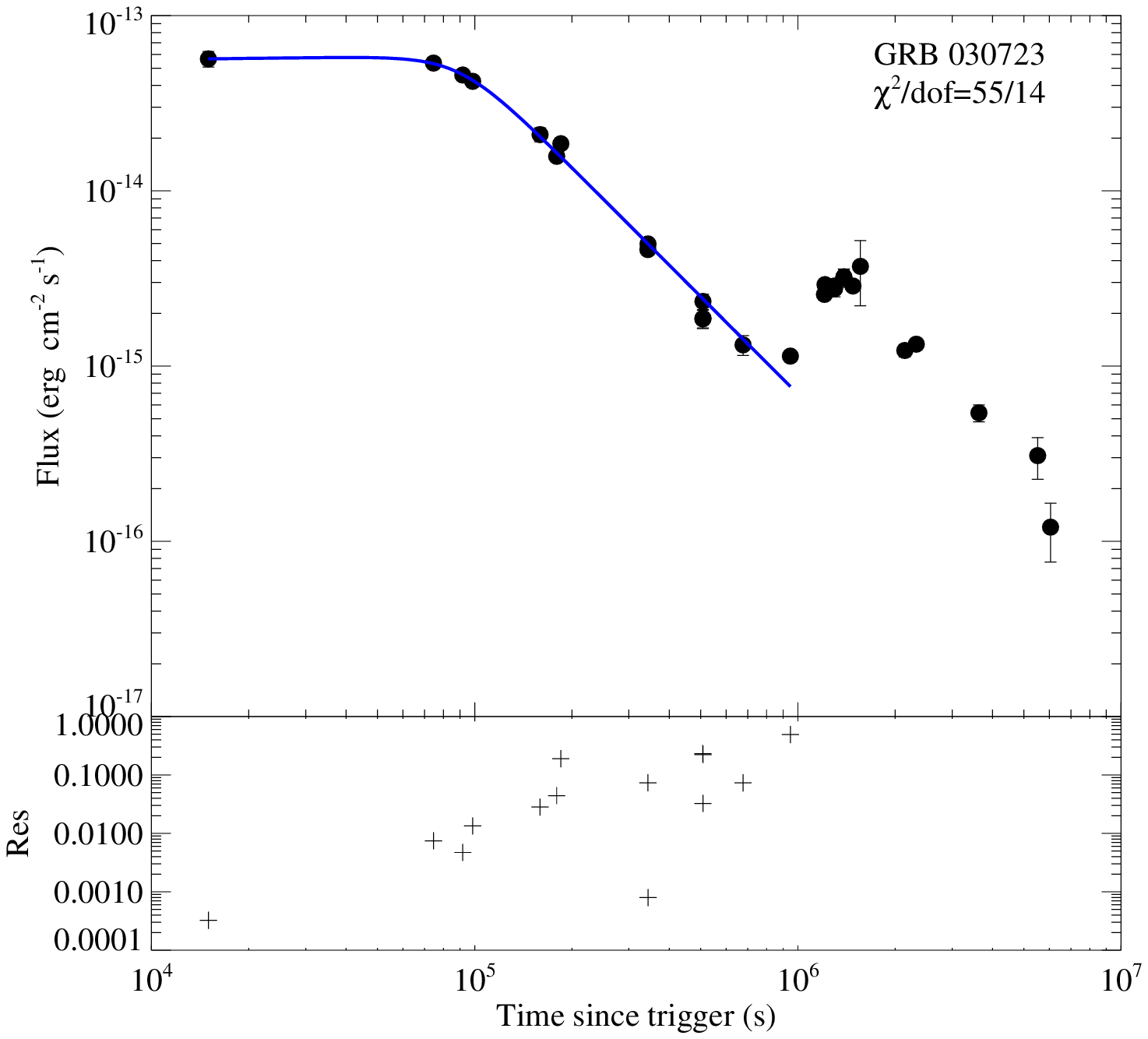}
\includegraphics[angle=0,scale=0.30]{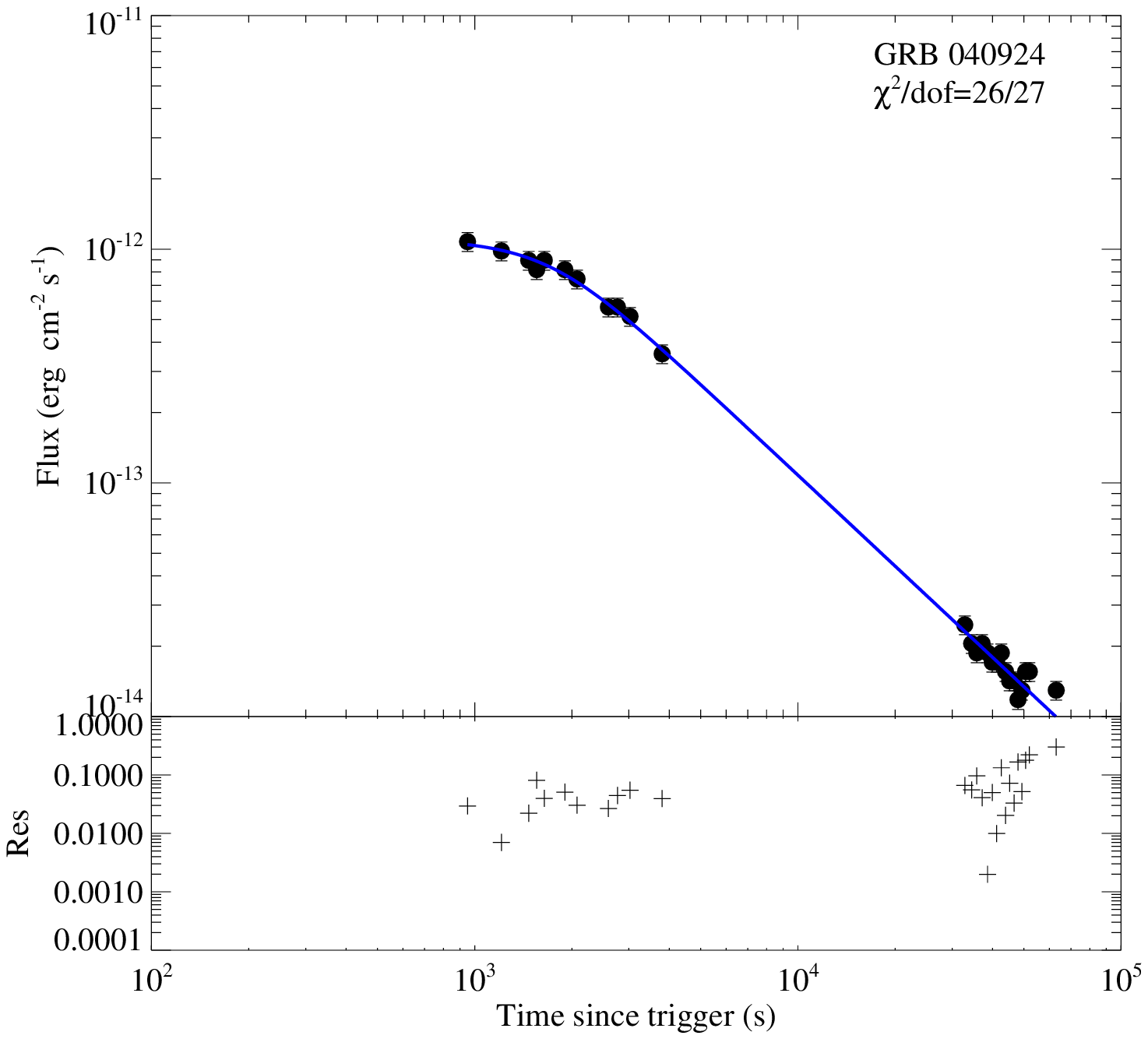}\\
\includegraphics[angle=0,scale=0.30]{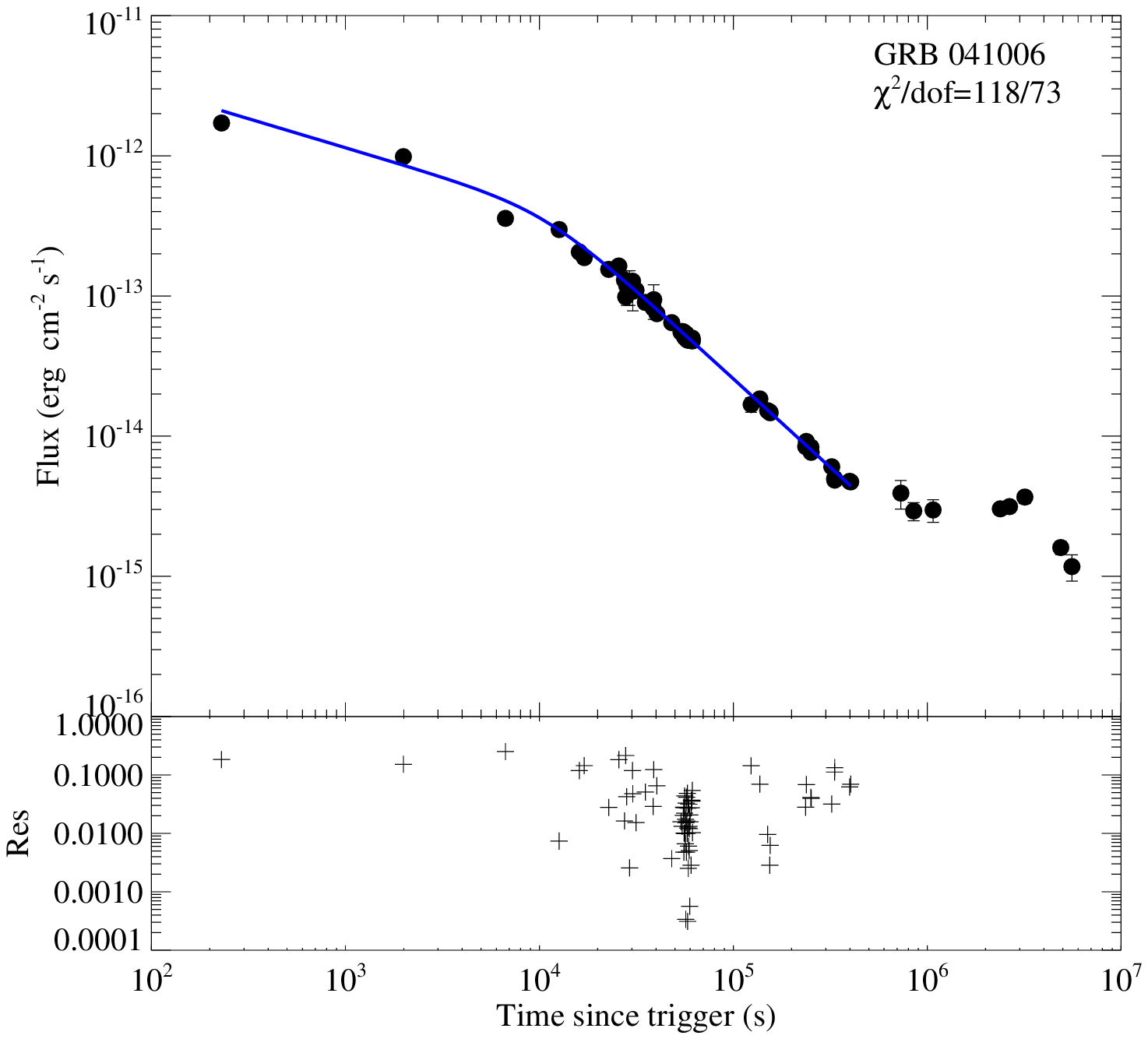}
\includegraphics[angle=0,scale=0.30]{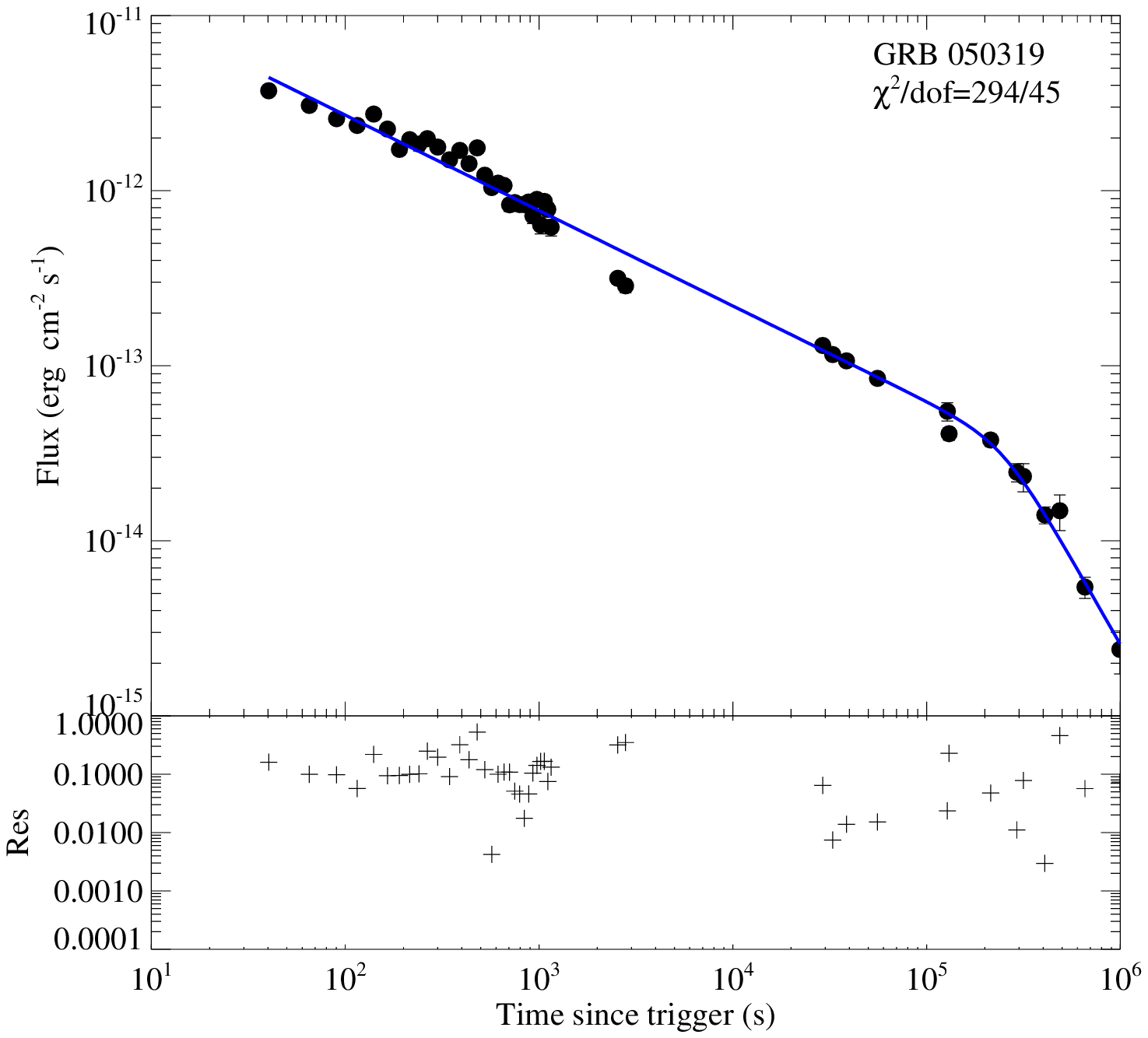}
\includegraphics[angle=0,scale=0.30]{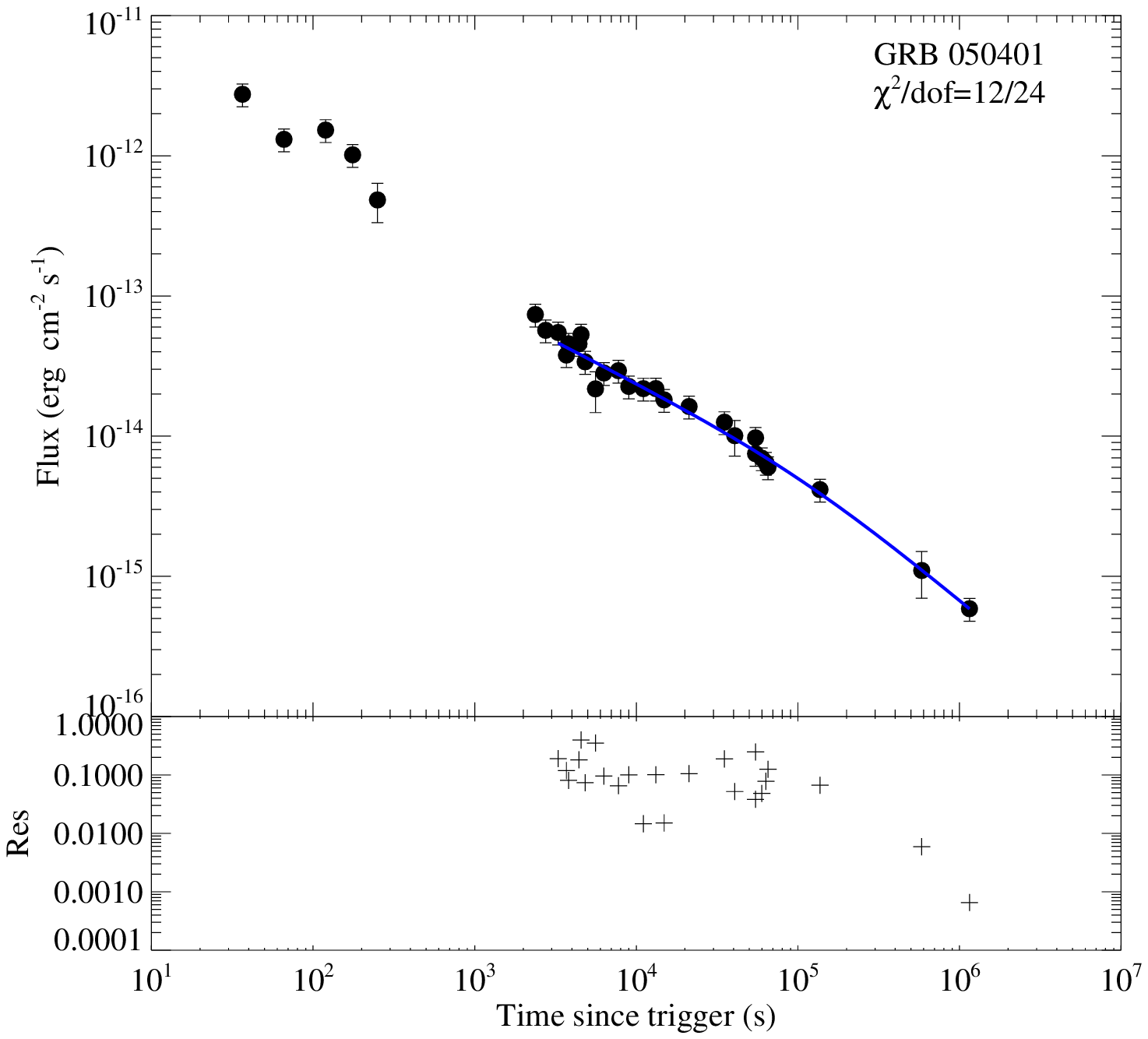}\hfill
\caption{The fitting results of optical light curves with plateau components. Most of the samples are taken from Li et al. (2012), and some are taken from Wang et al. (2015). We used an smooth broken power-law function to fit the light curves, and the solid lines represent the best fit to the optical data.}
\end{figure*}

\begin{figure*}
\centering
\includegraphics[angle=0,scale=0.30]{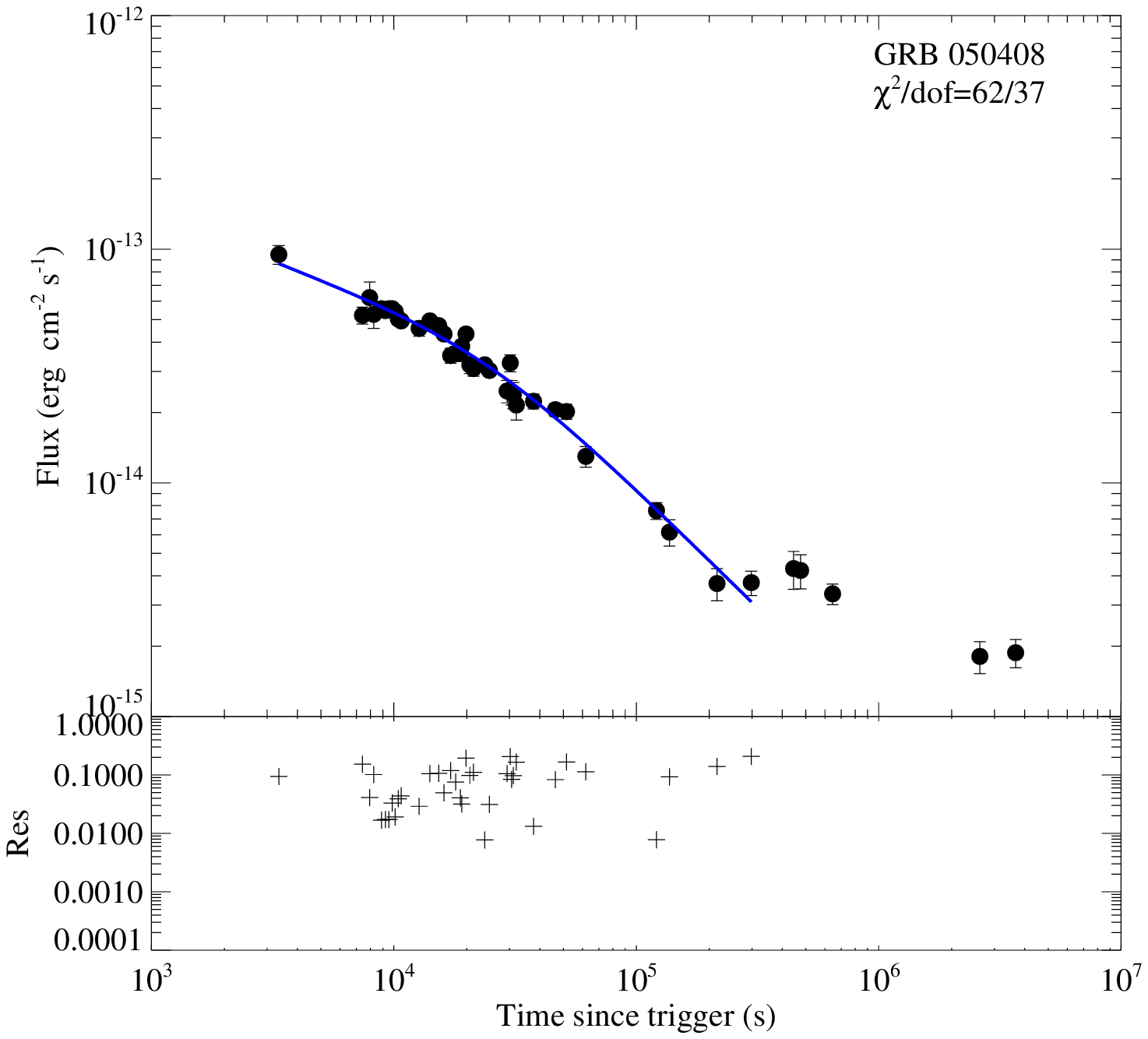}
\includegraphics[angle=0,scale=0.30]{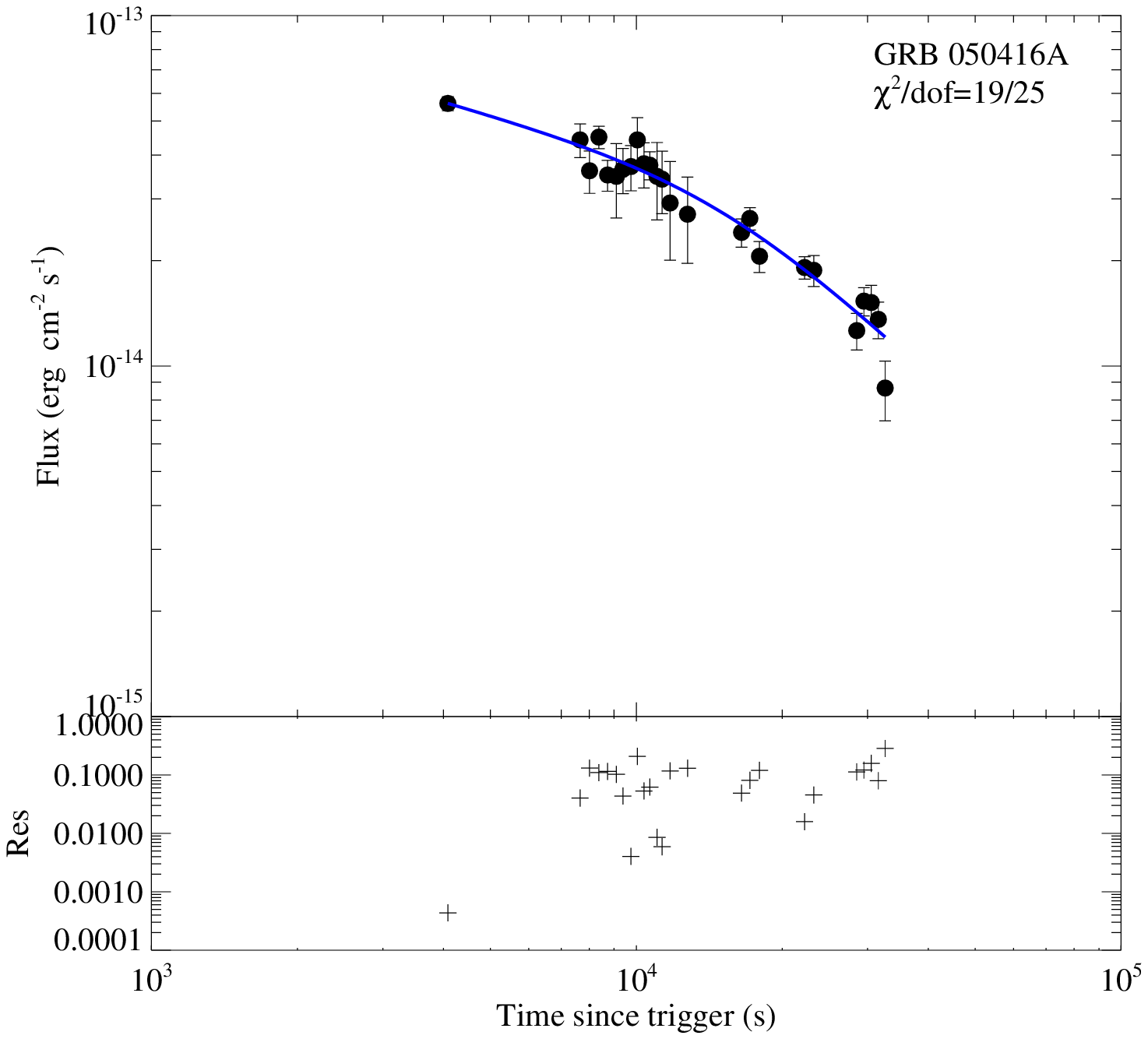}
\includegraphics[angle=0,scale=0.30]{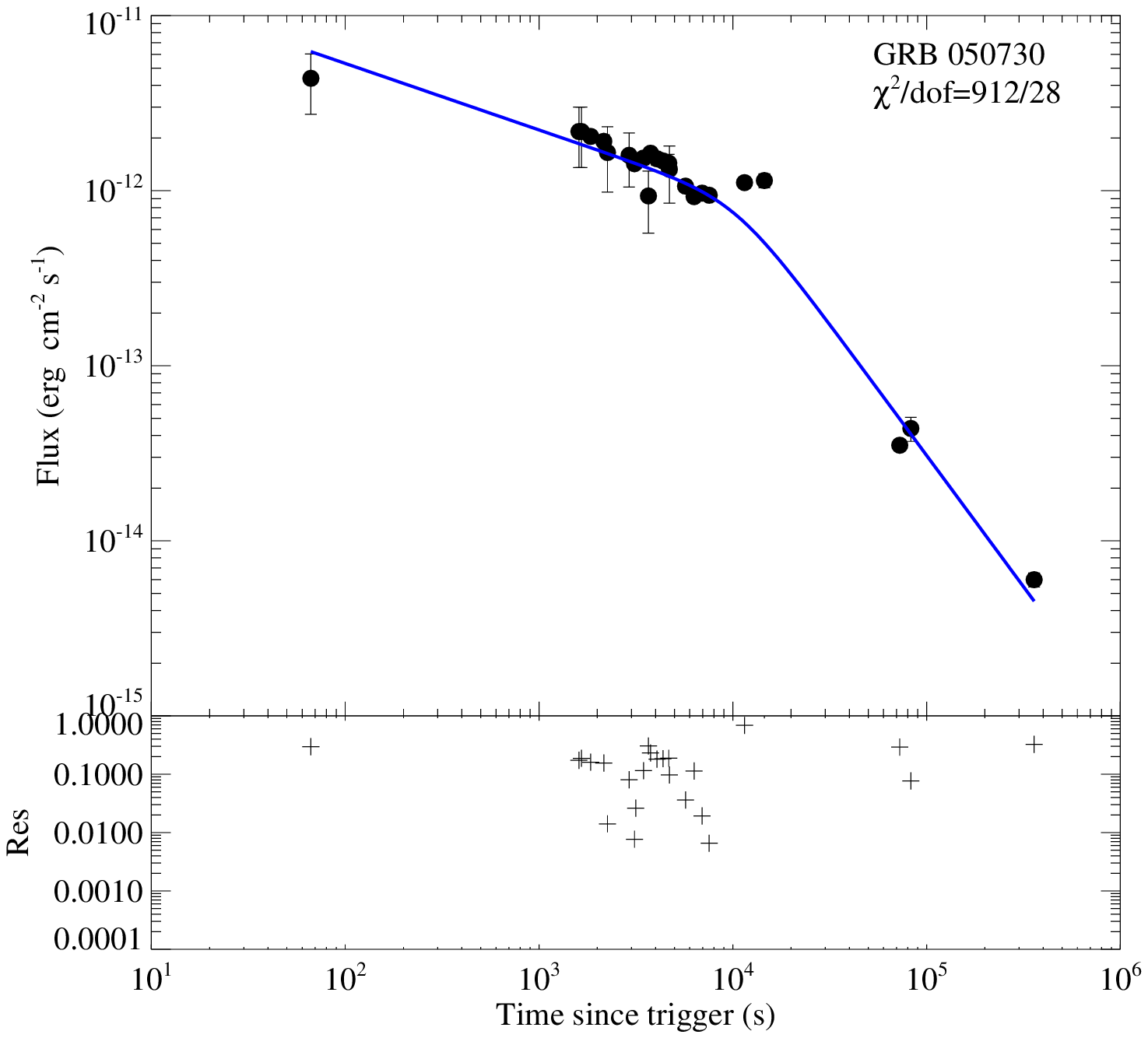}
\includegraphics[angle=0,scale=0.30]{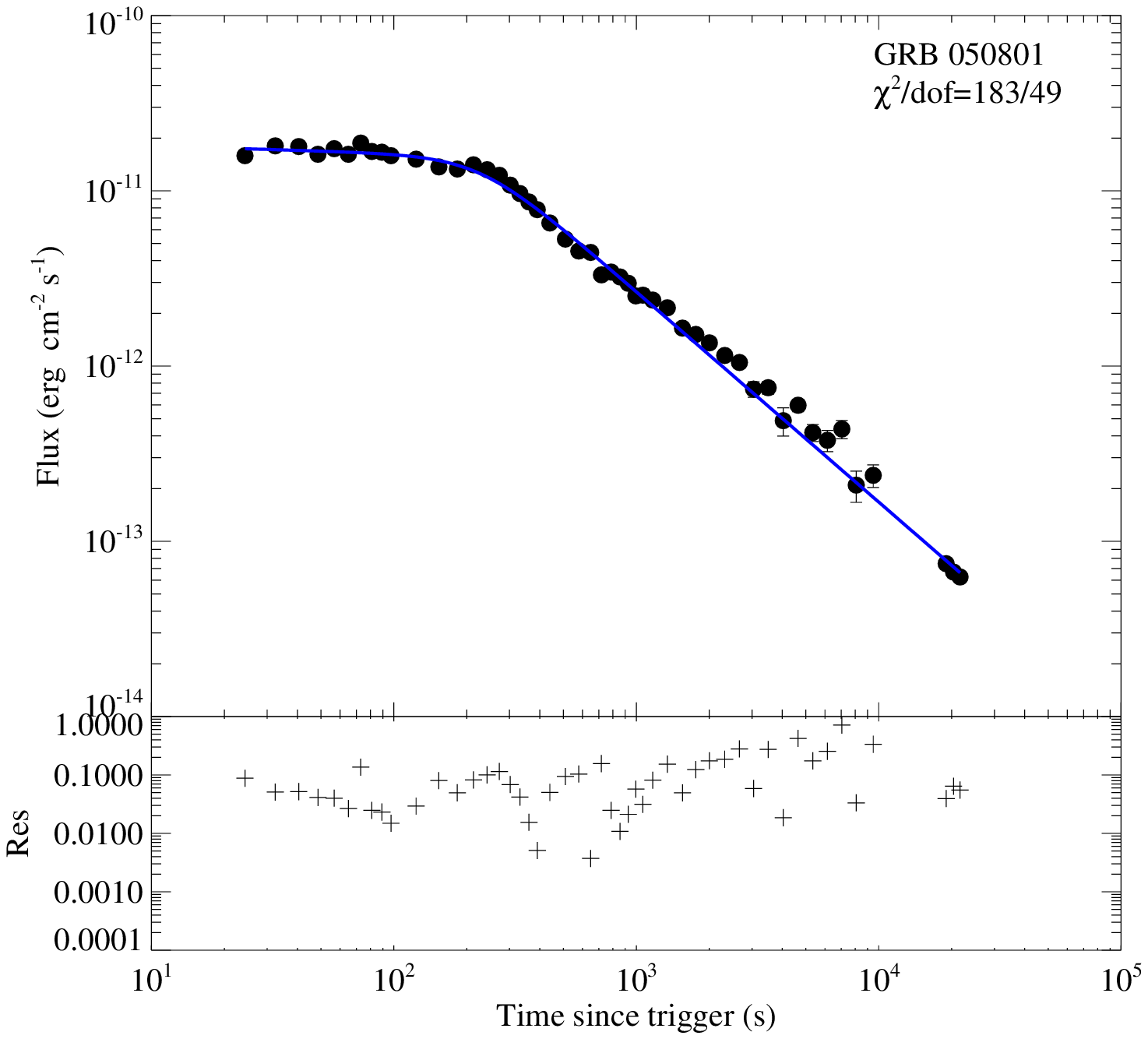}
\includegraphics[angle=0,scale=0.30]{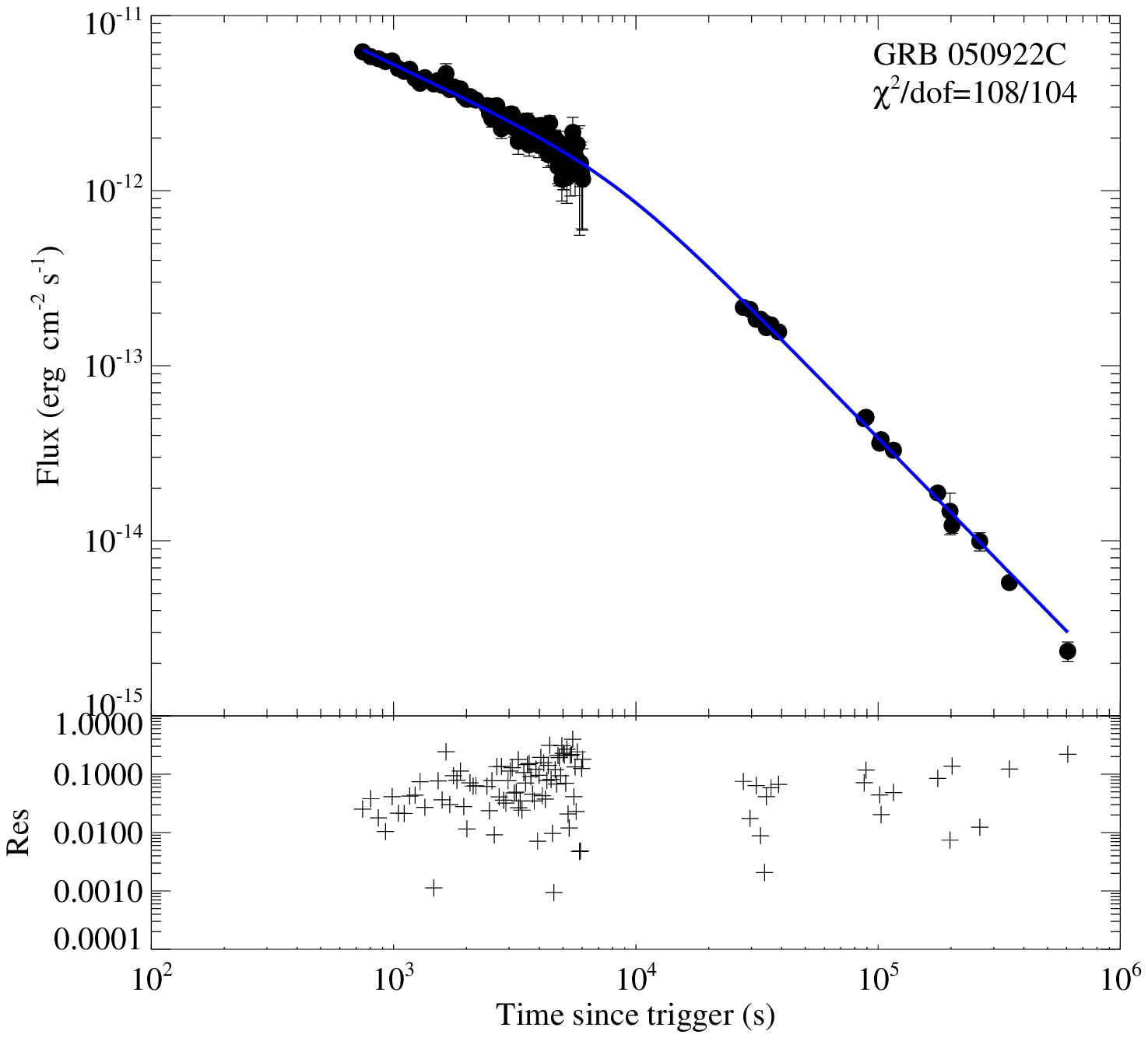}
\includegraphics[angle=0,scale=0.30]{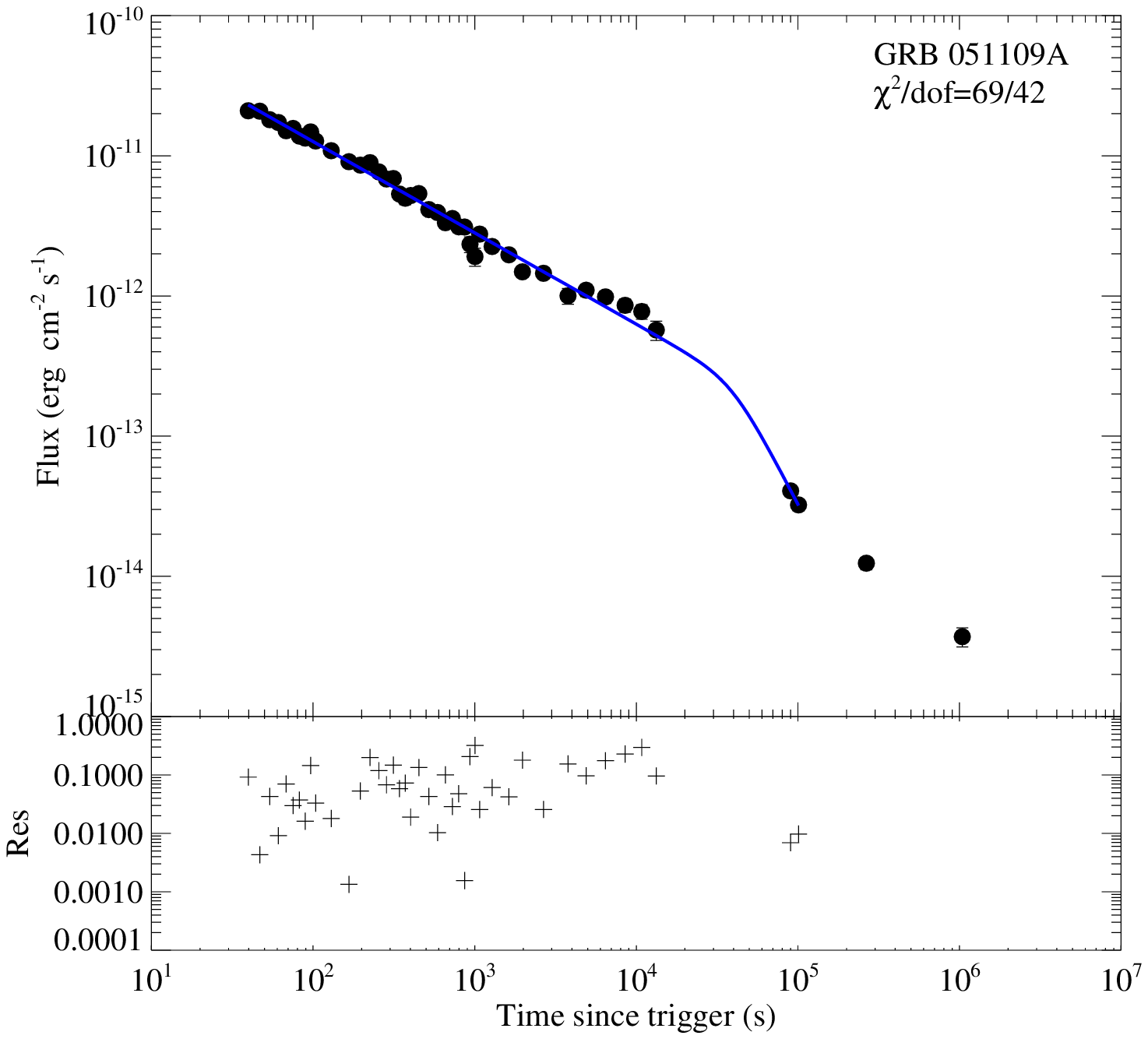}
\includegraphics[angle=0,scale=0.30]{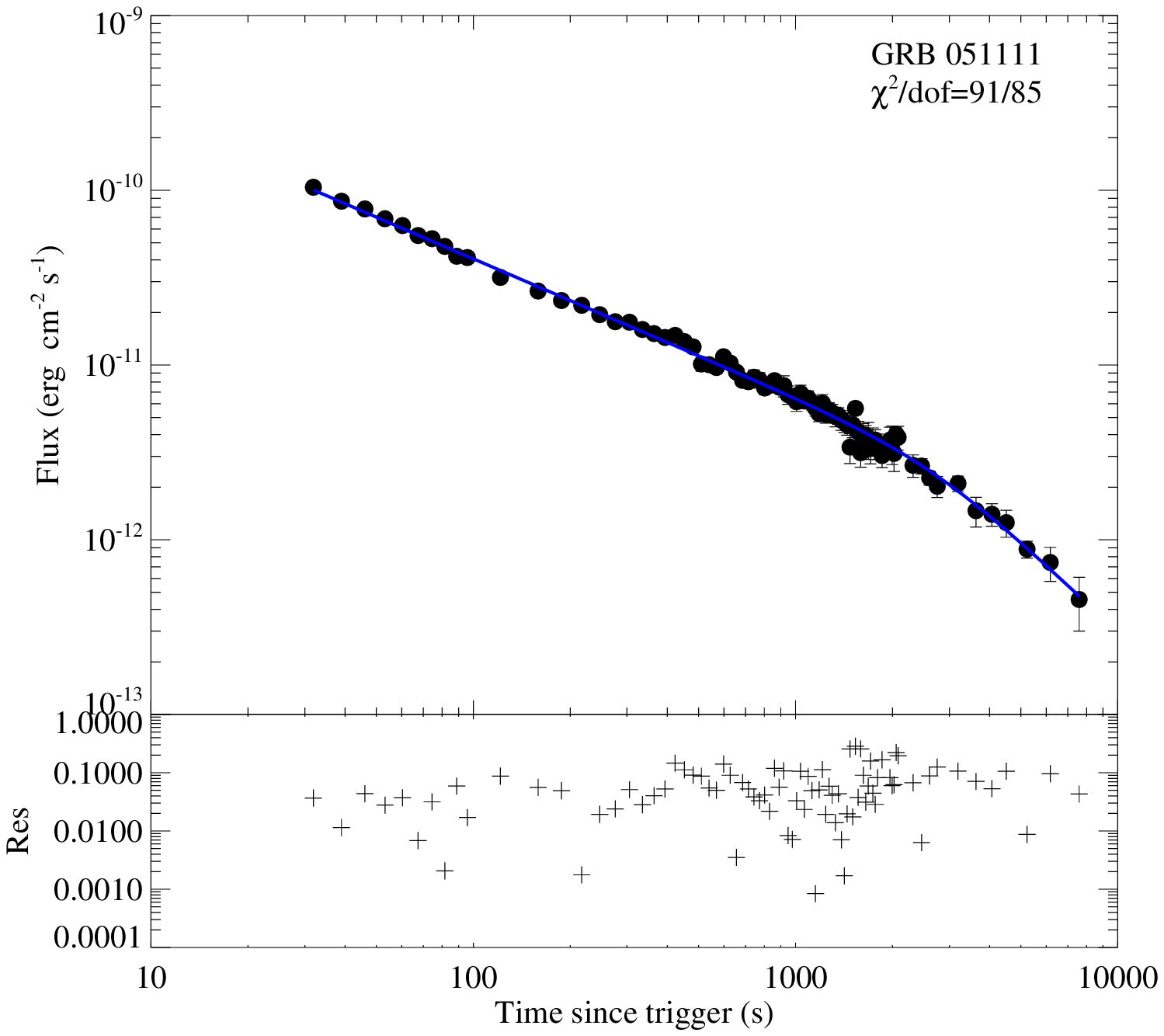}
\includegraphics[angle=0,scale=0.30]{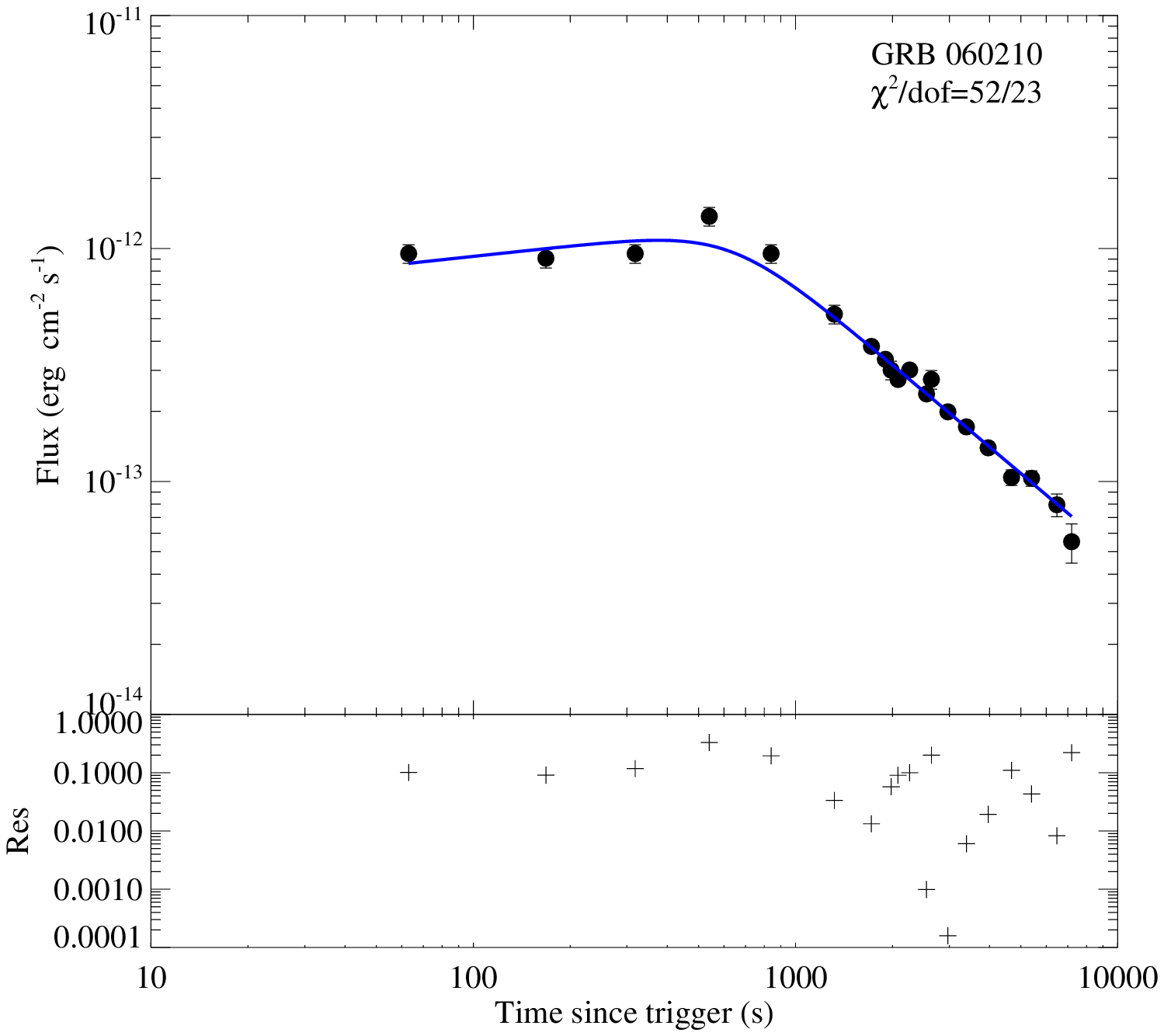}
\includegraphics[angle=0,scale=0.30]{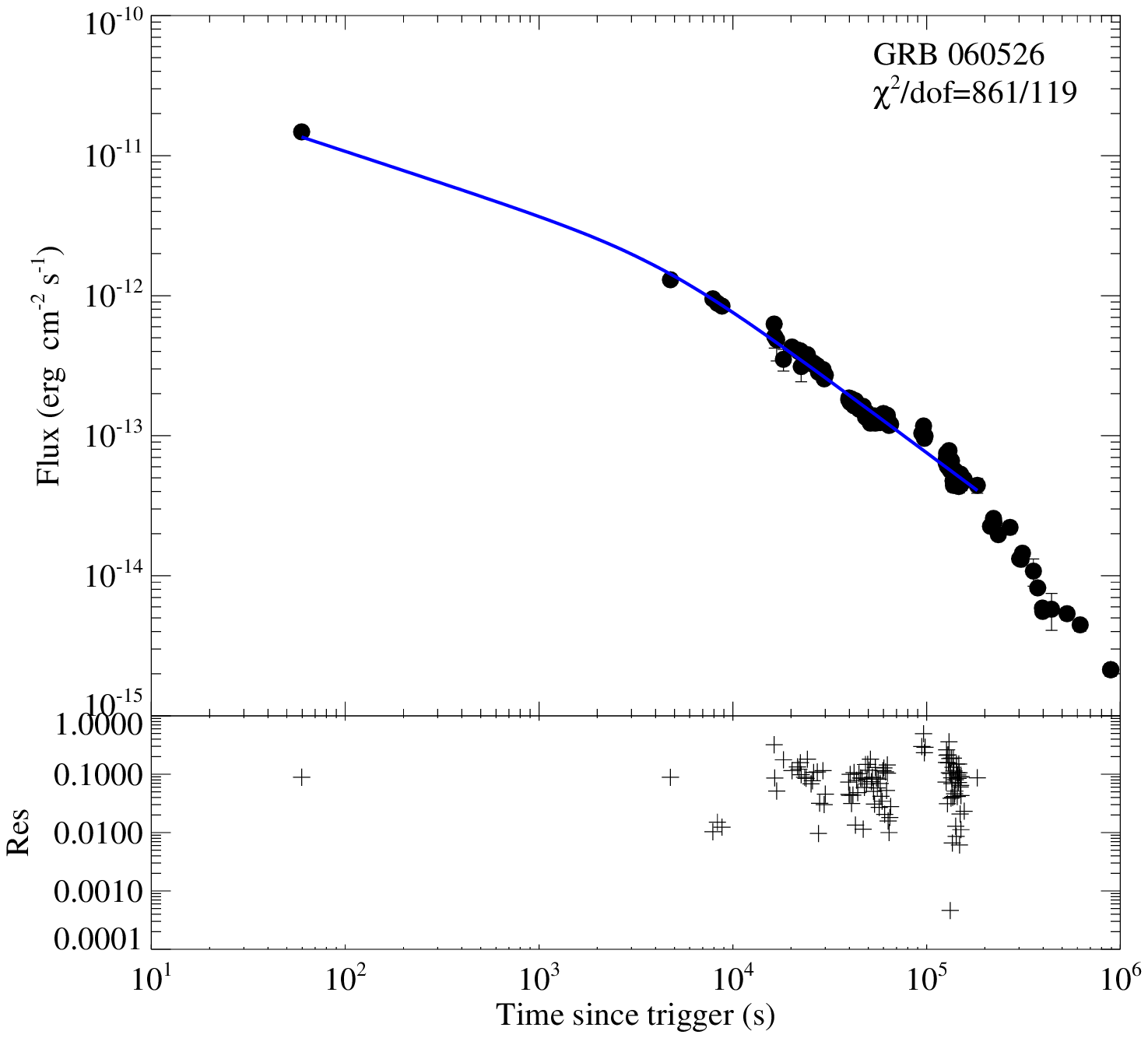}
\includegraphics[angle=0,scale=0.30]{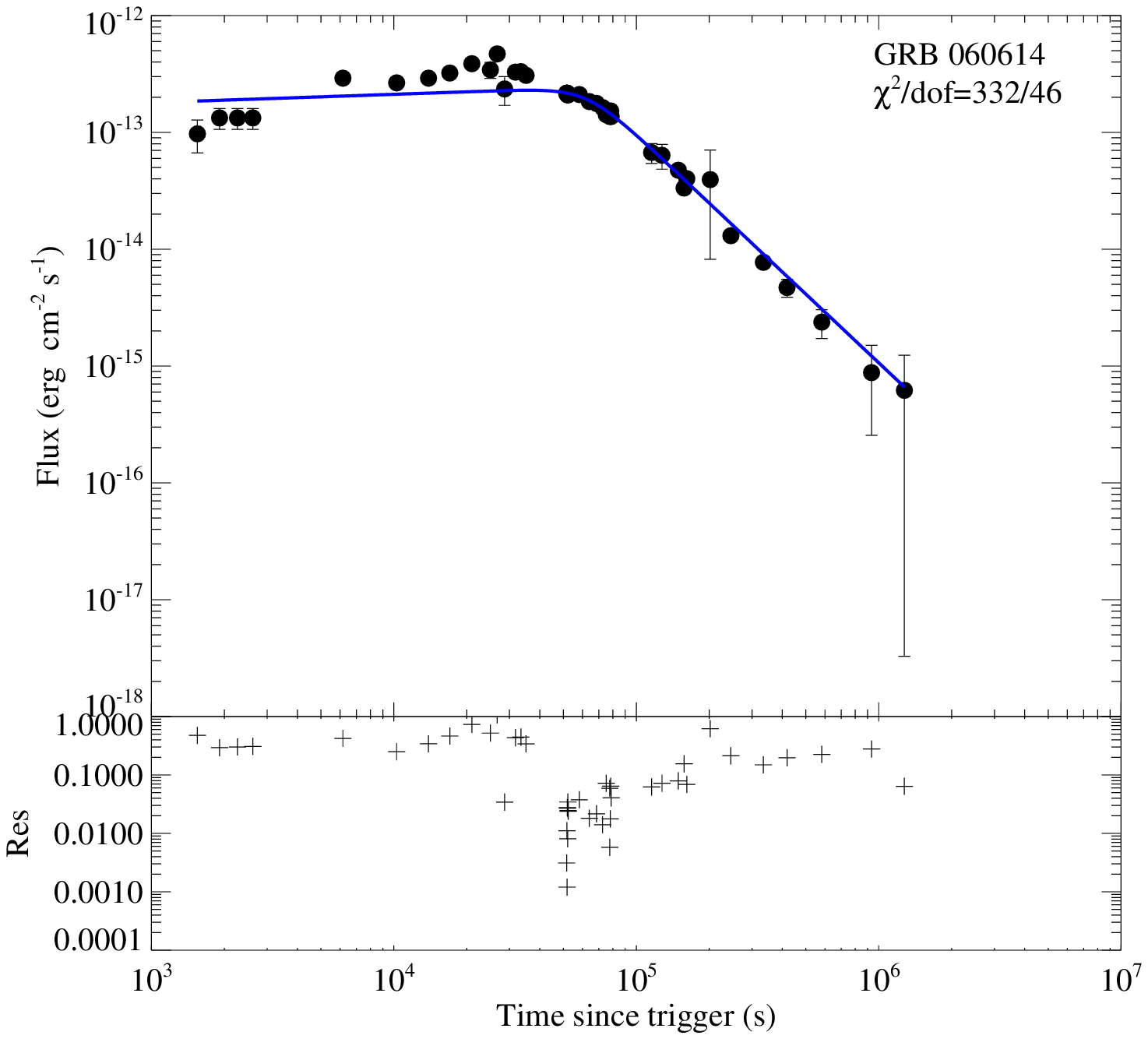}
\includegraphics[angle=0,scale=0.30]{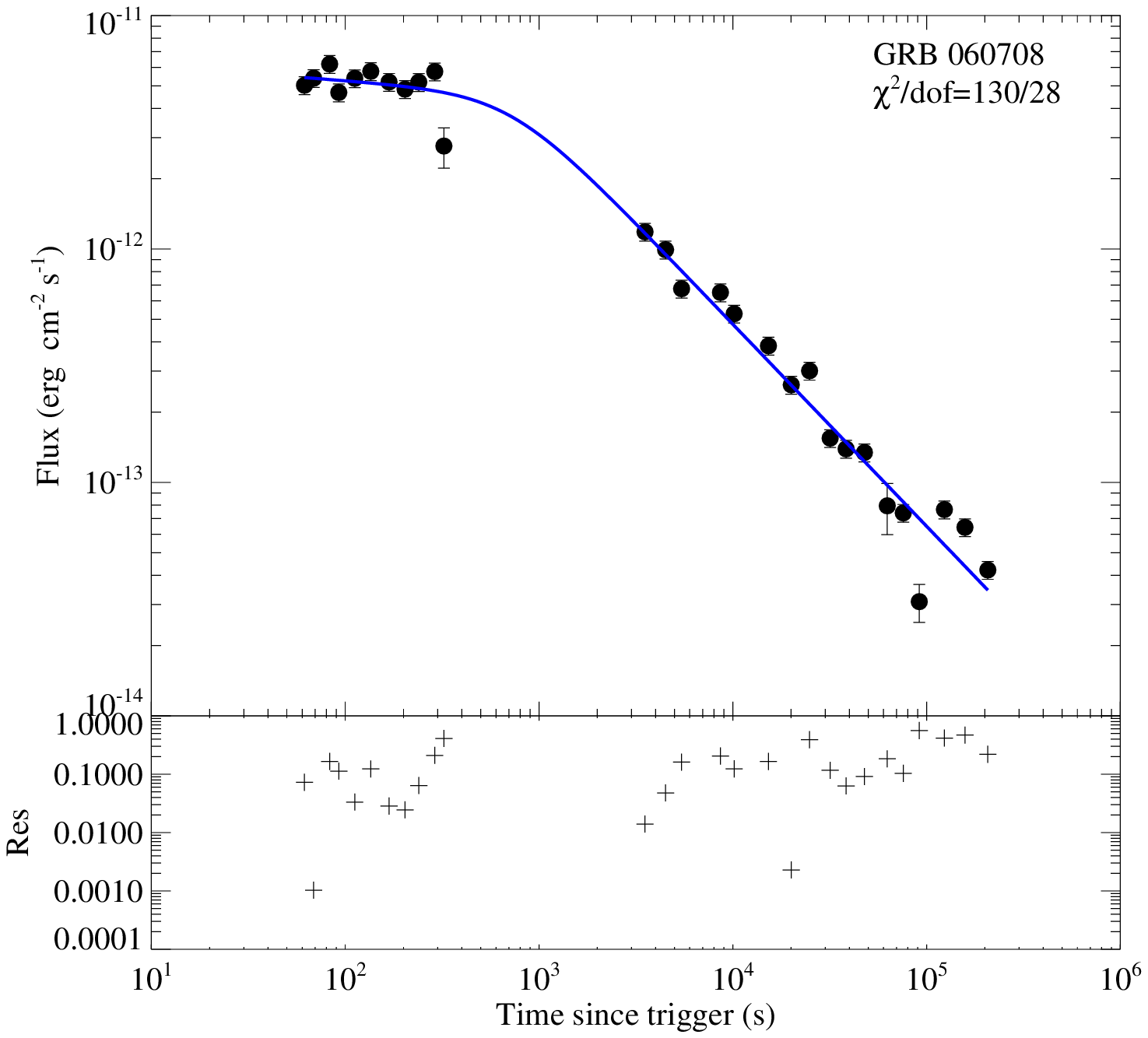}
\includegraphics[angle=0,scale=0.30]{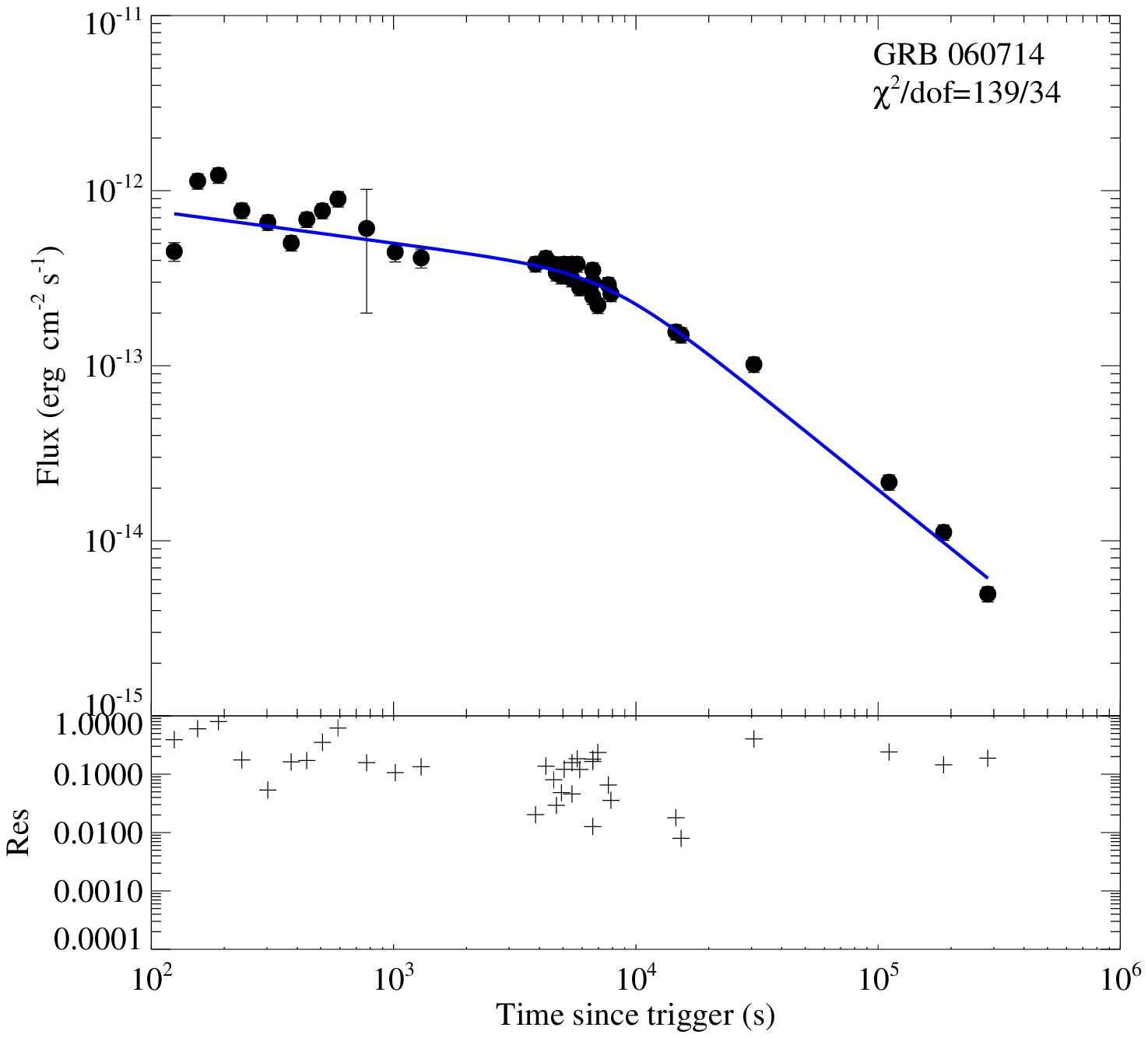}\hfill
\center{Fig. 1--- Continued}
\end{figure*}

\begin{figure*}
\centering
\includegraphics[angle=0,scale=0.30]{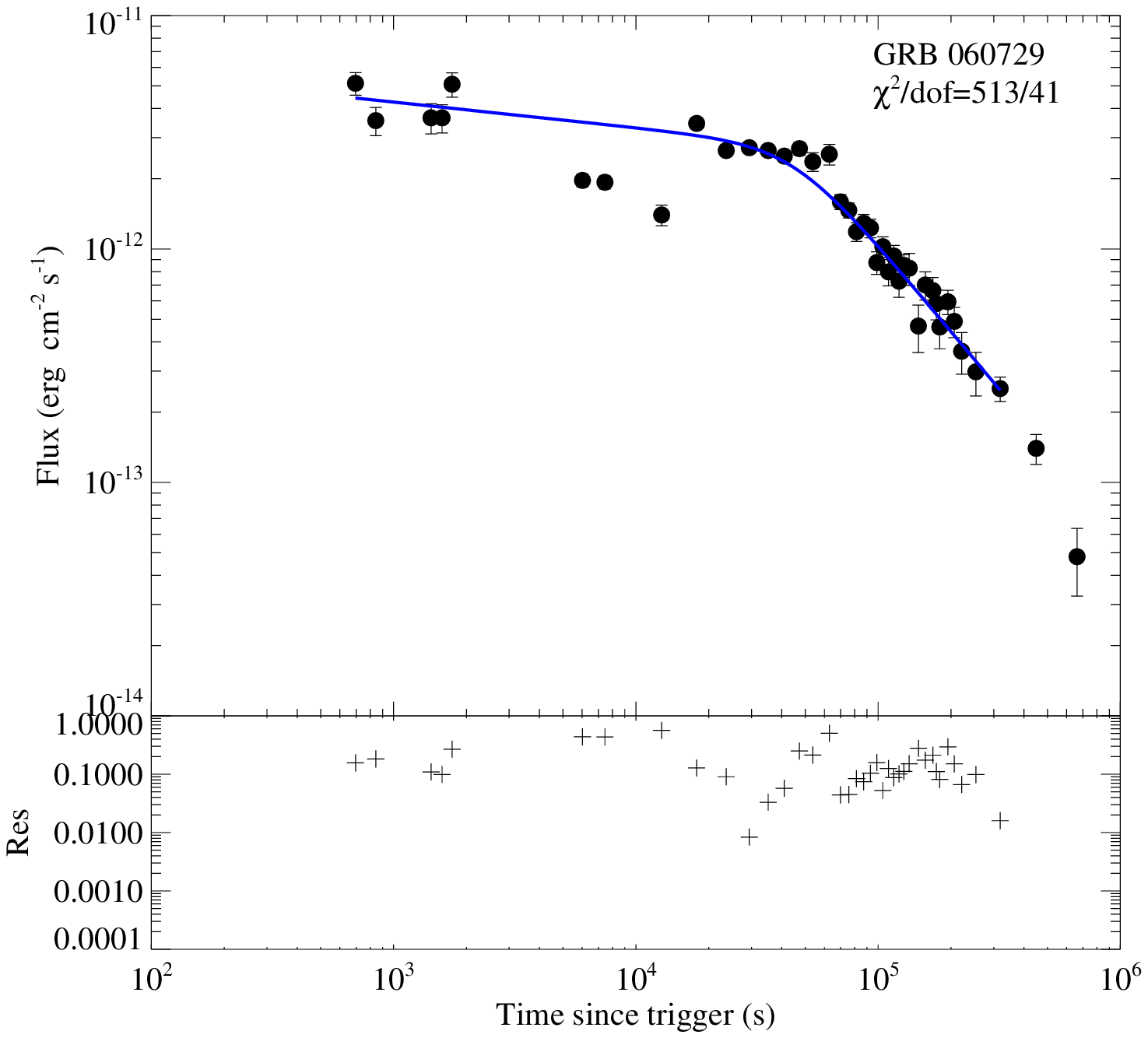}
\includegraphics[angle=0,scale=0.30]{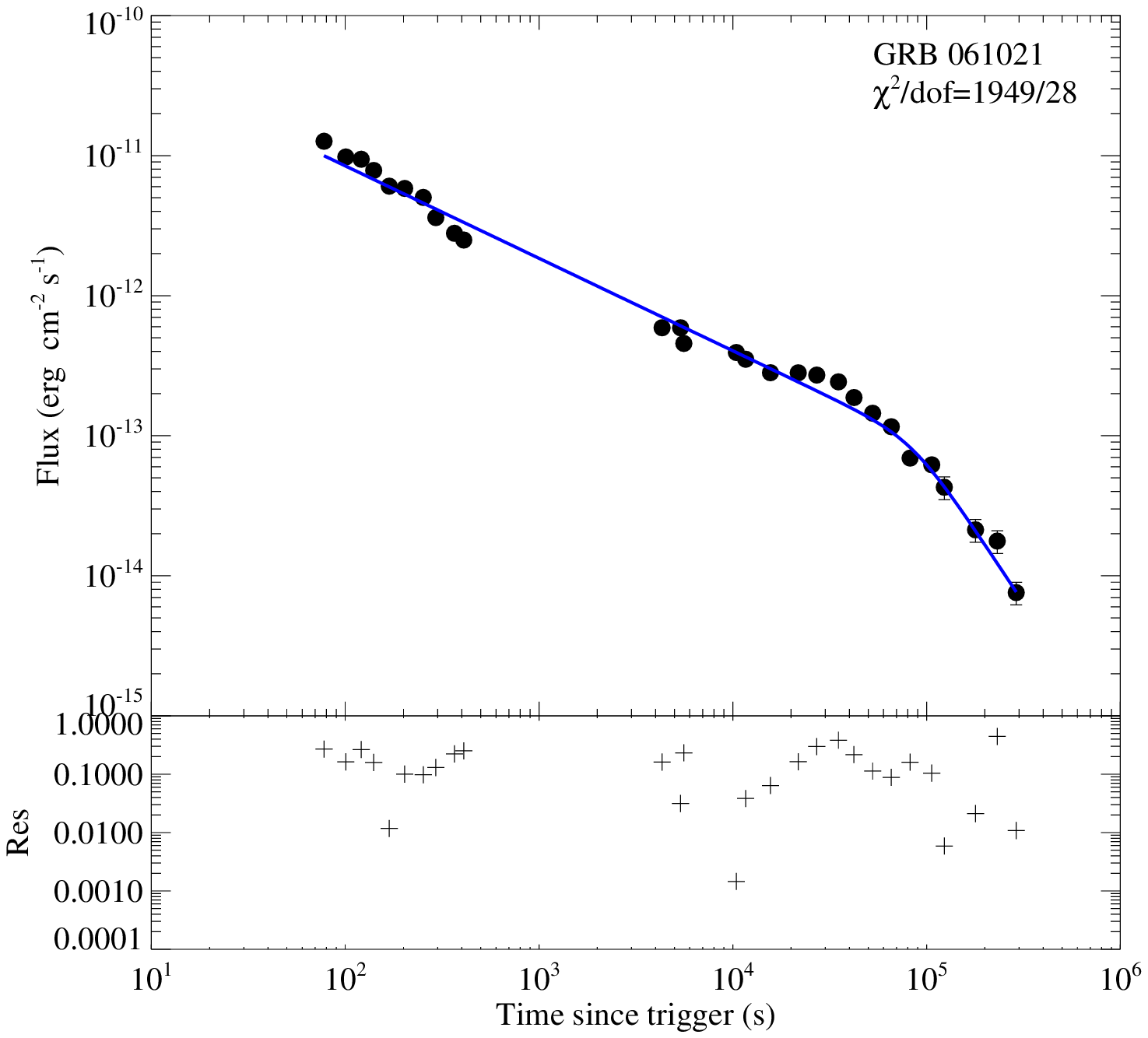}
\includegraphics[angle=0,scale=0.30]{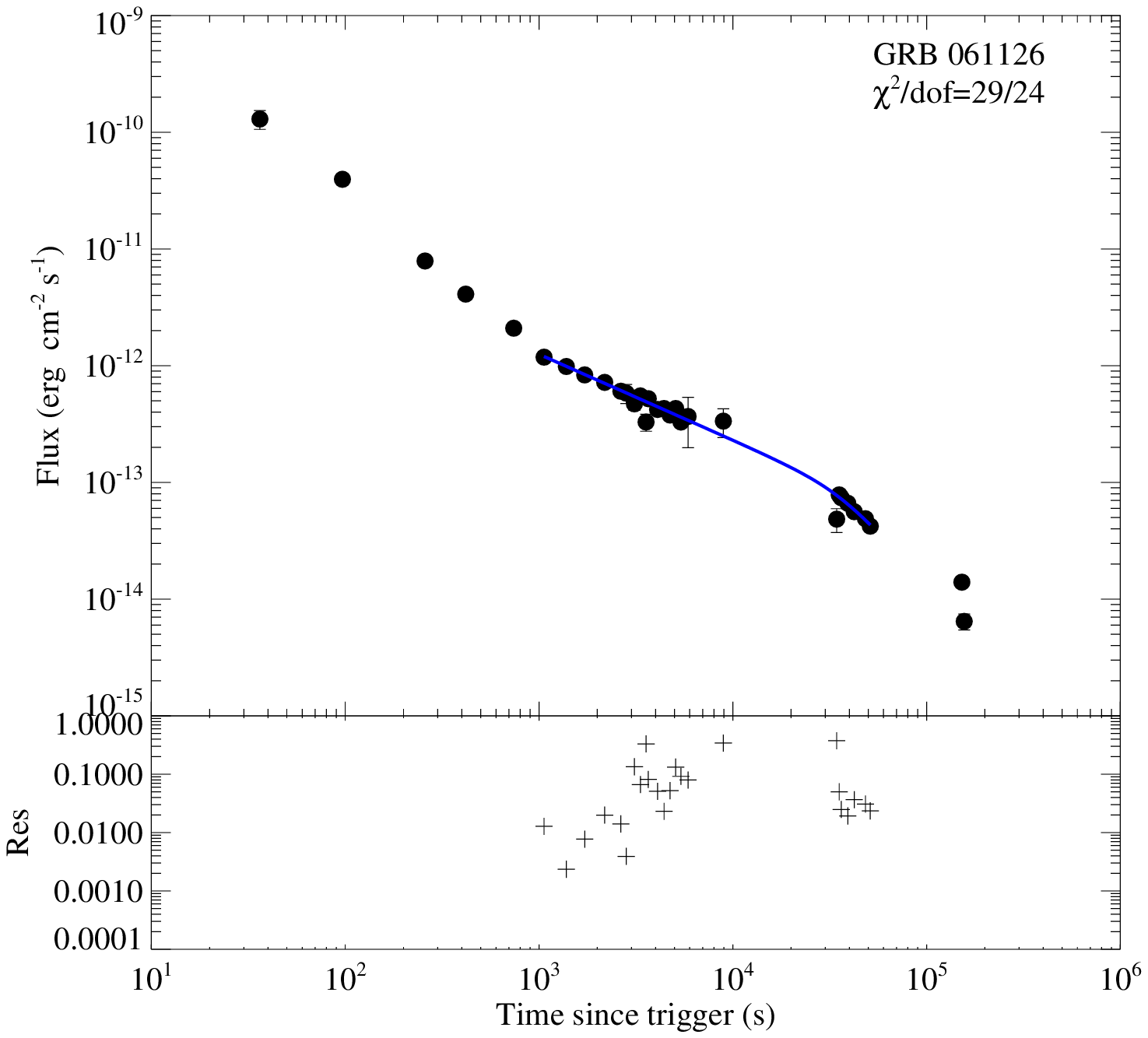}
\includegraphics[angle=0,scale=0.30]{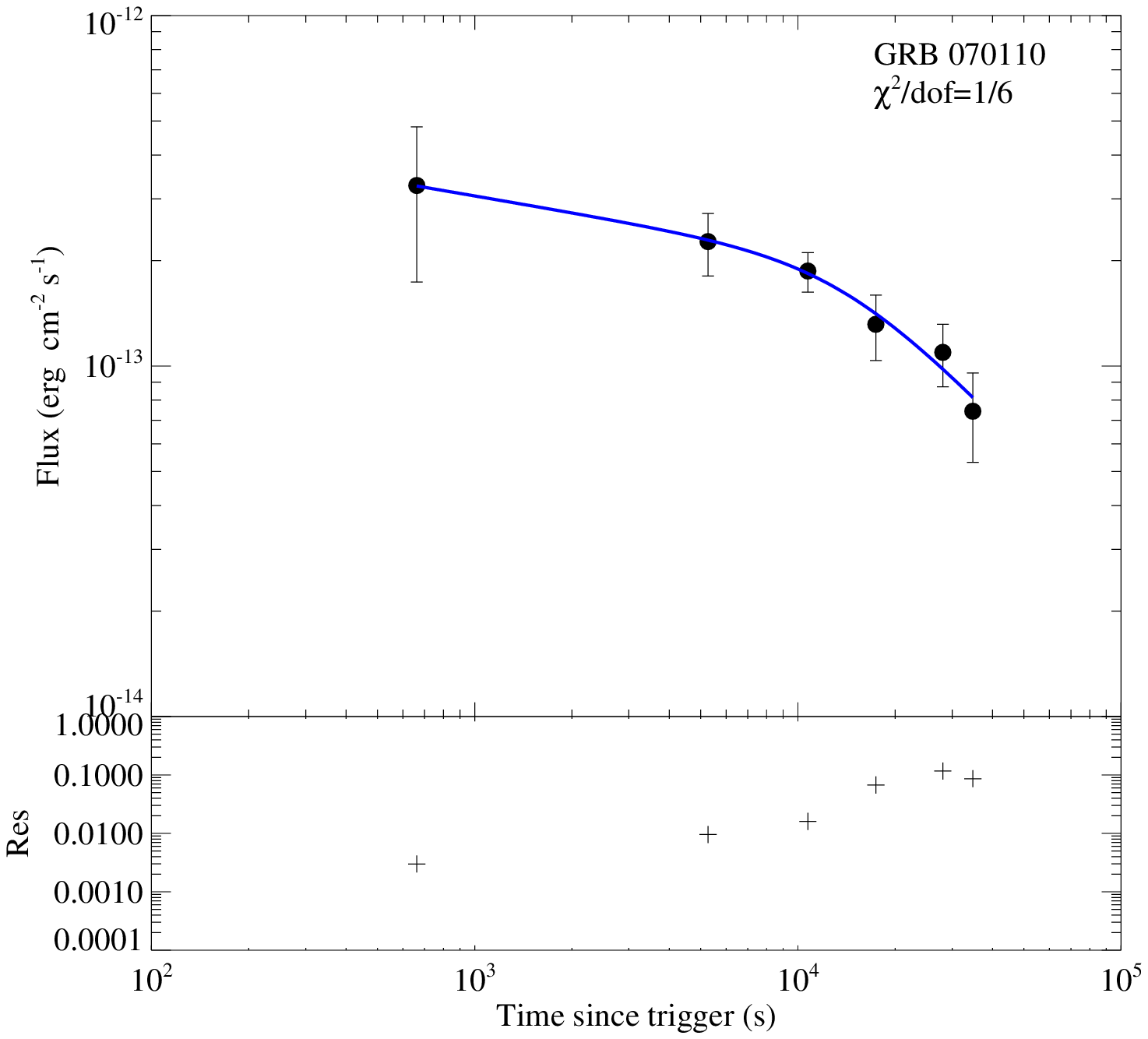}
\includegraphics[angle=0,scale=0.30]{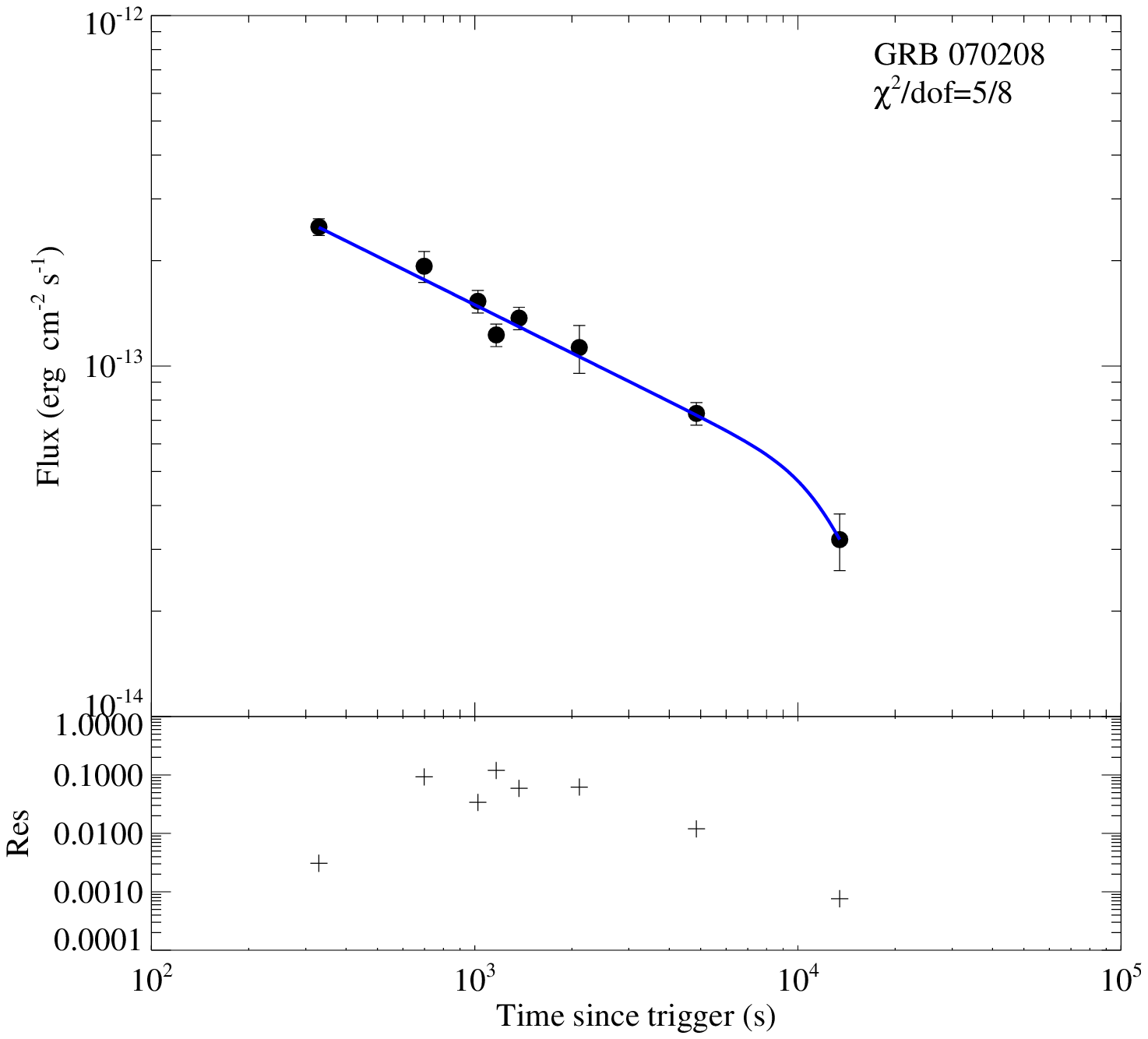}
\includegraphics[angle=0,scale=0.30]{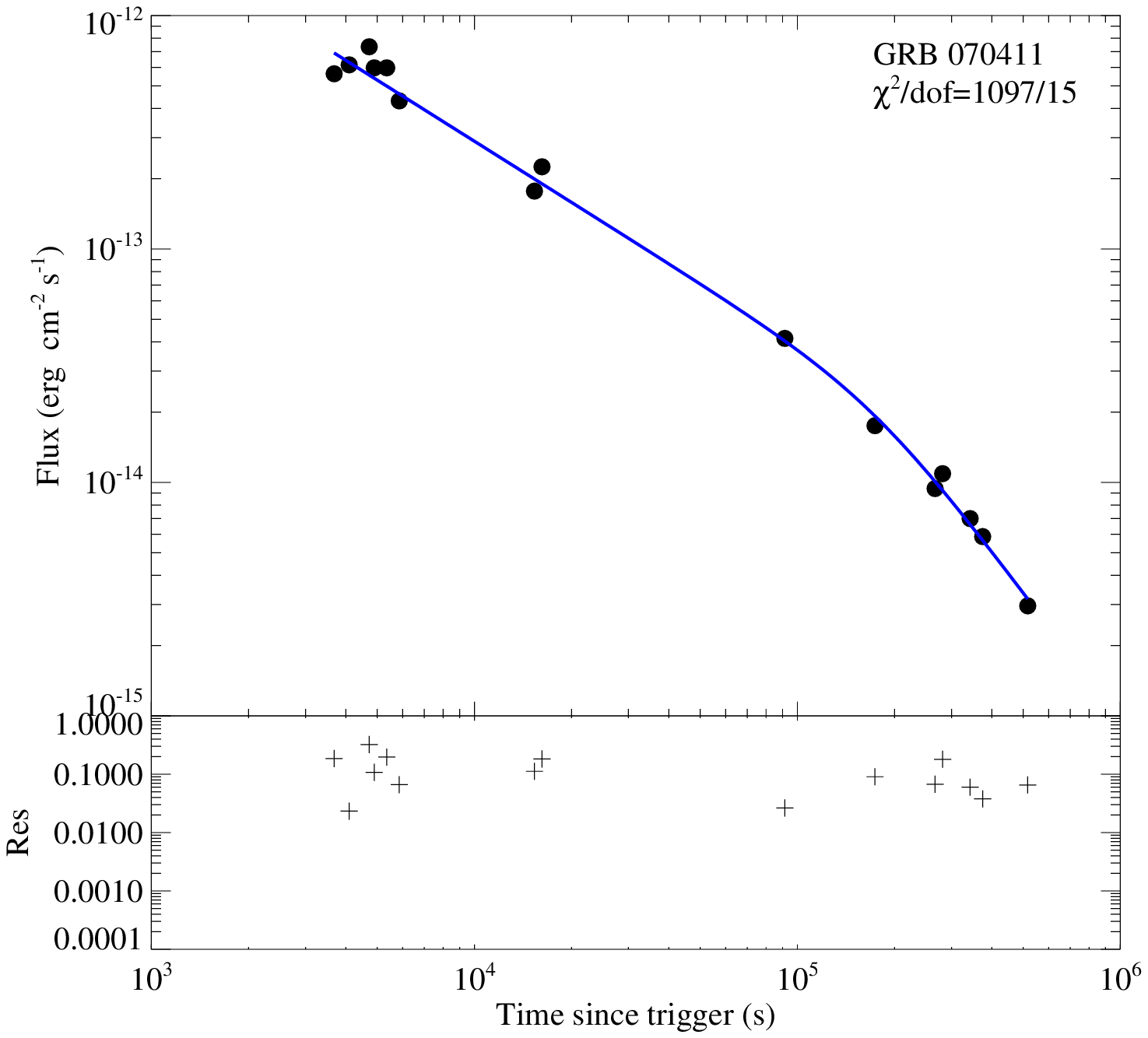}
\includegraphics[angle=0,scale=0.30]{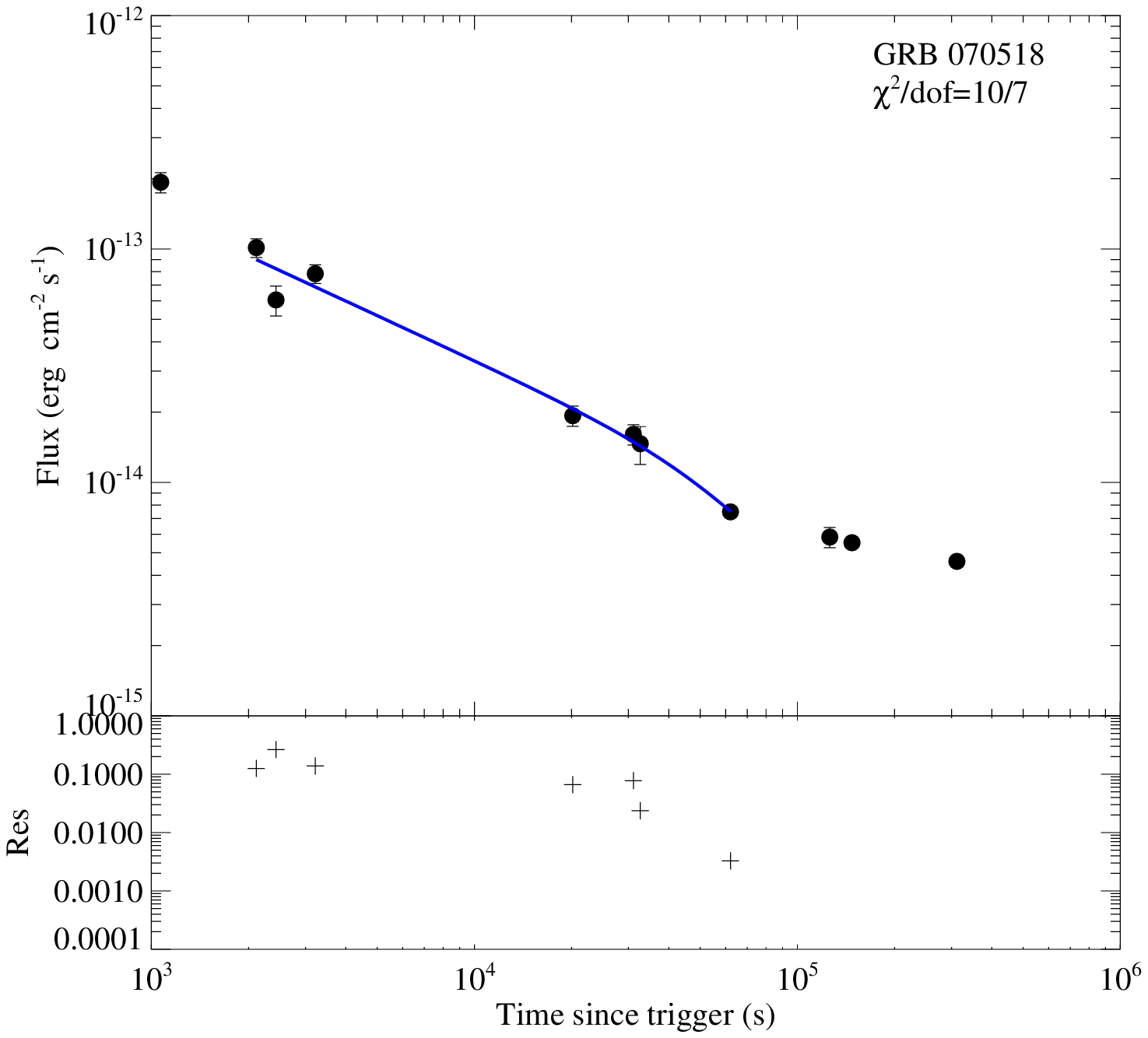}
\includegraphics[angle=0,scale=0.30]{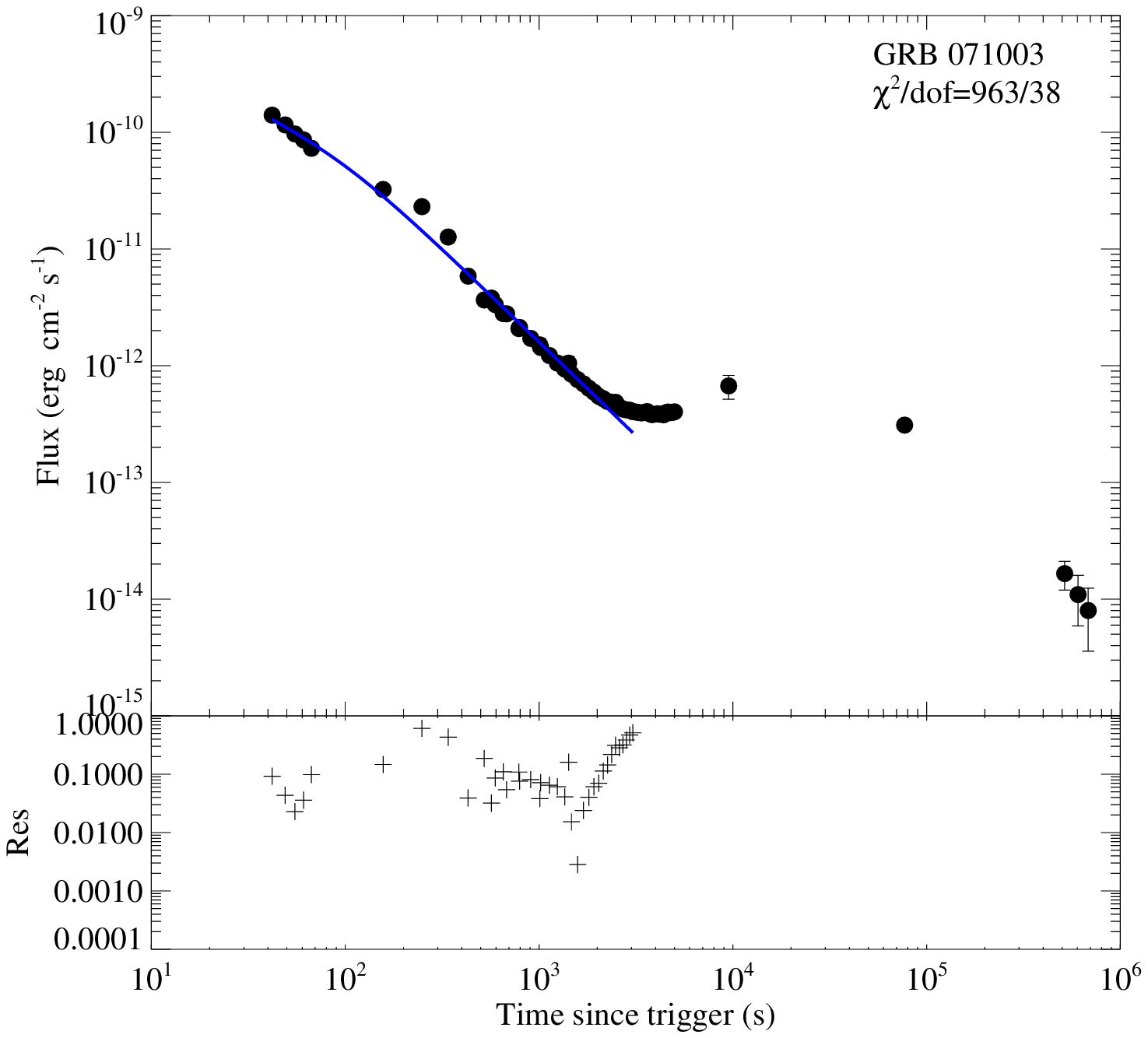}
\includegraphics[angle=0,scale=0.30]{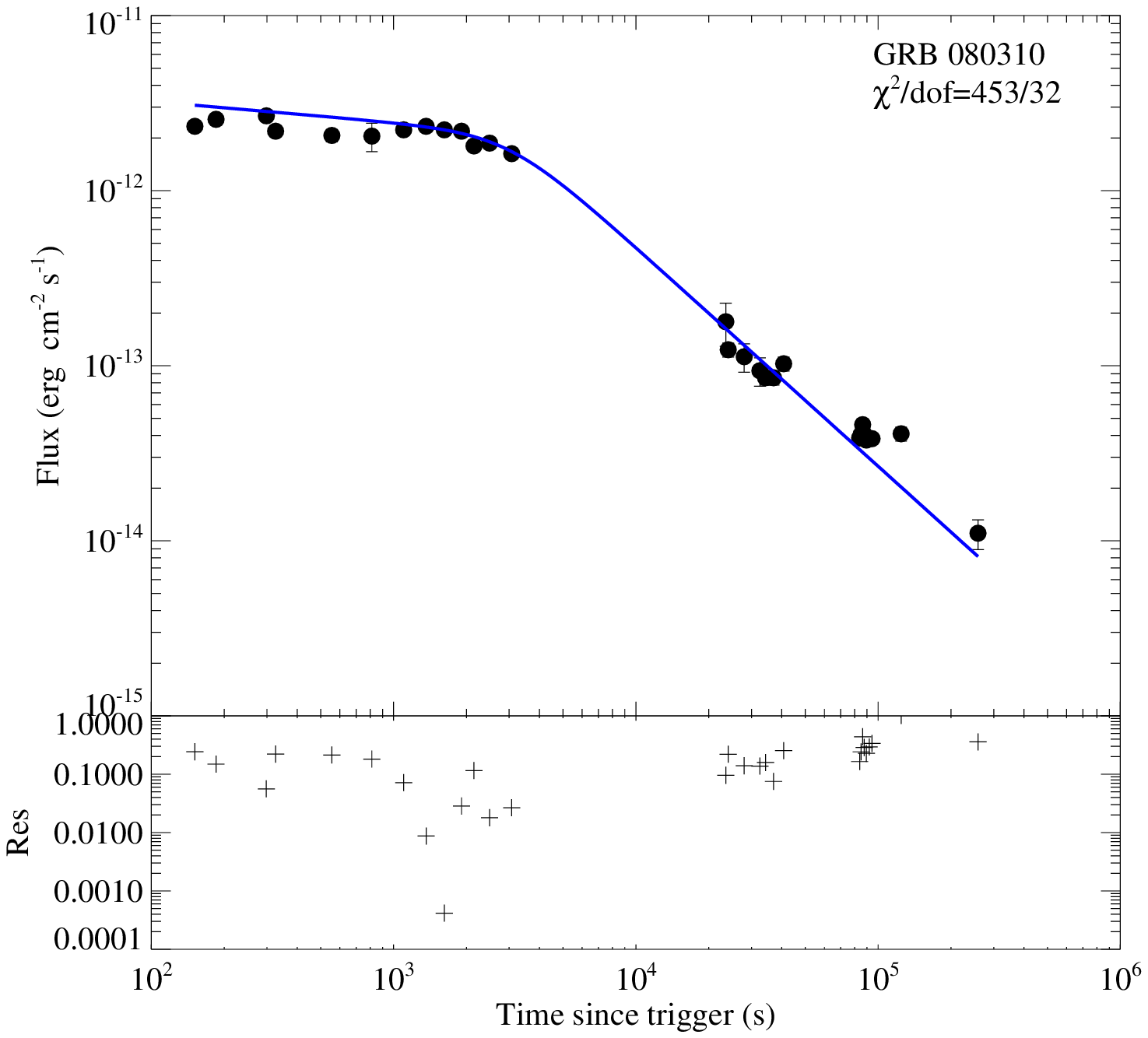}
\includegraphics[angle=0,scale=0.30]{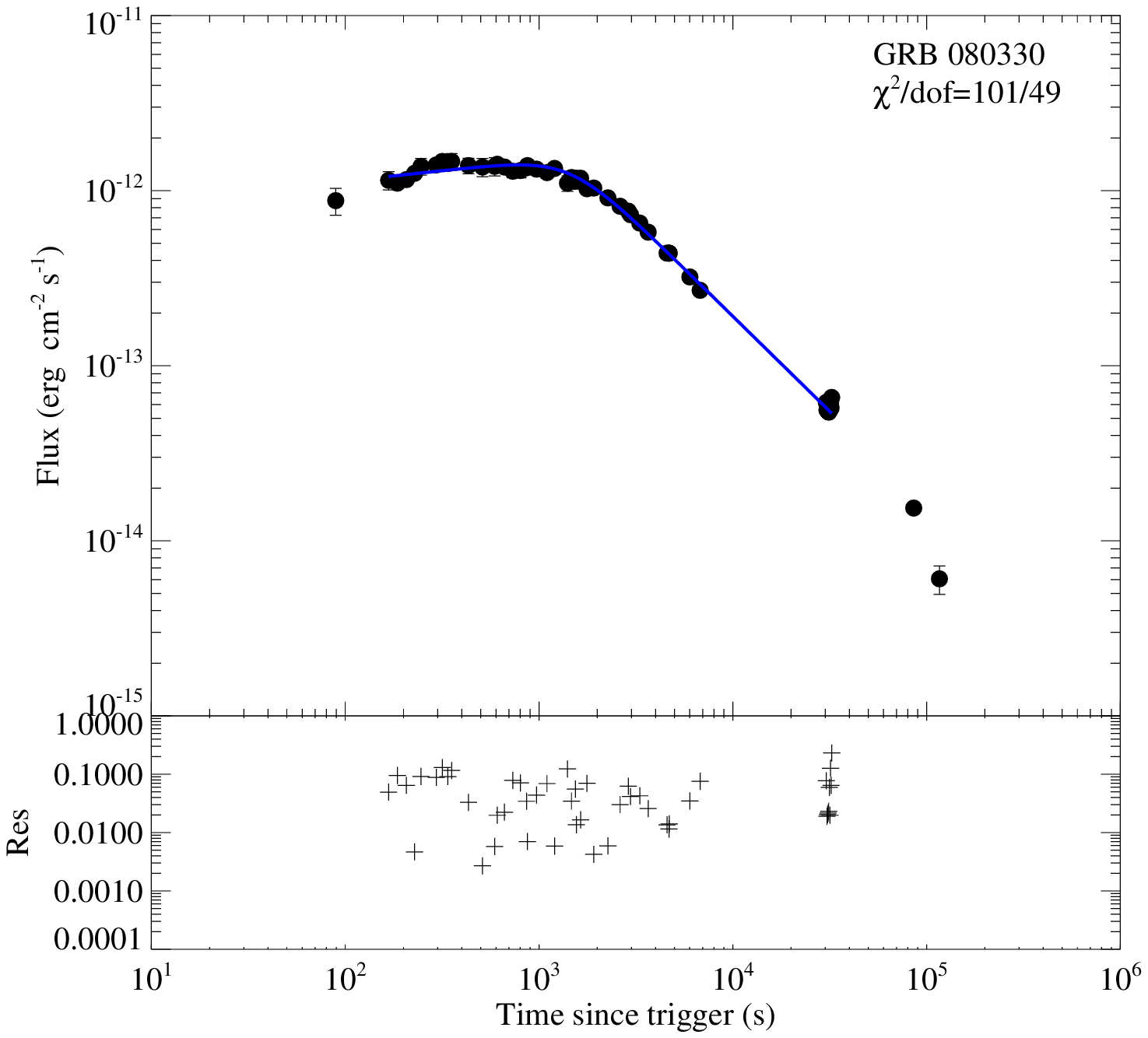}
\includegraphics[angle=0,scale=0.30]{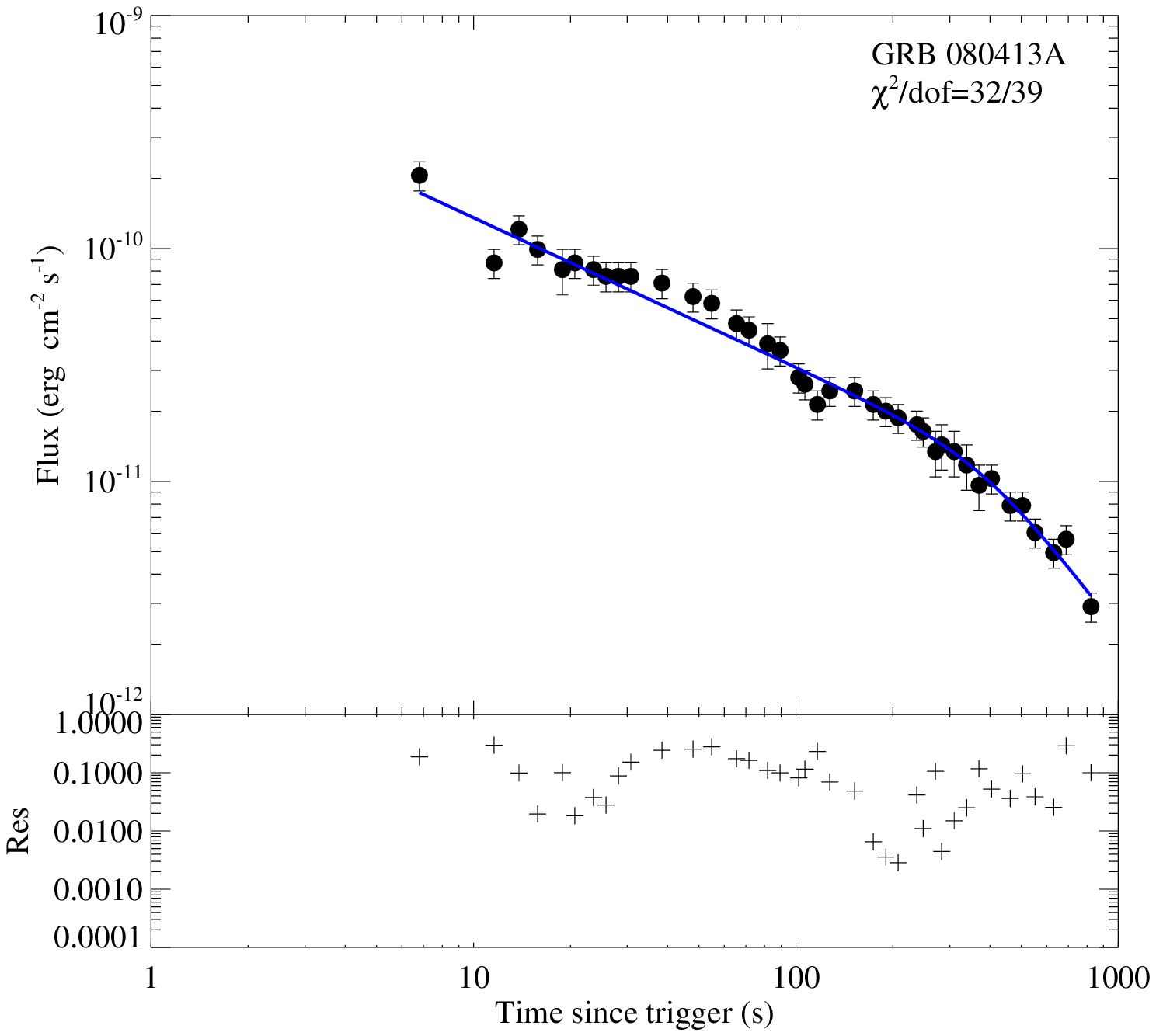}
\includegraphics[angle=0,scale=0.30]{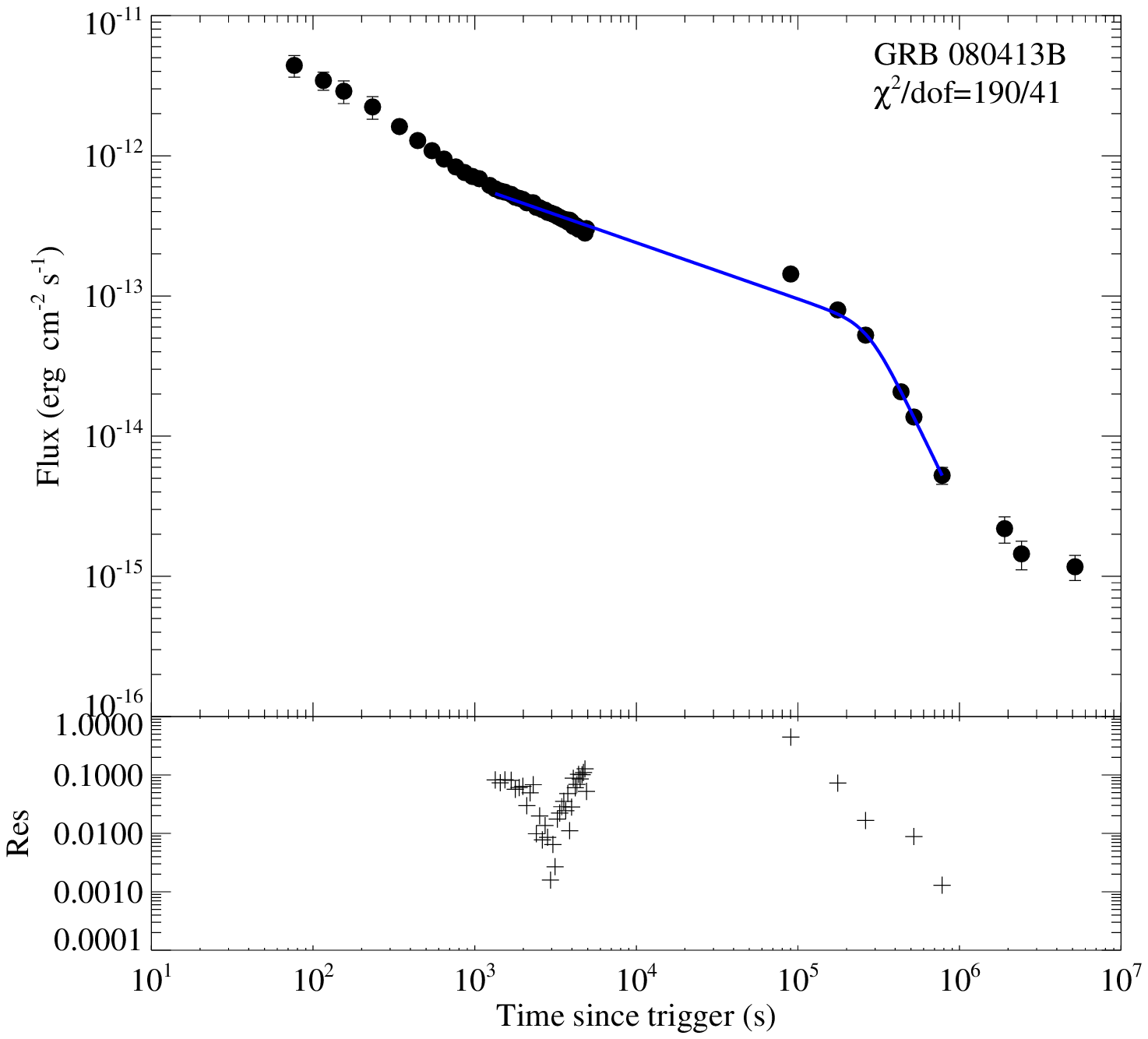}\hfill
\center{Fig. 1--- Continued}
\end{figure*}

\begin{figure*}
\centering
\includegraphics[angle=0,scale=0.30]{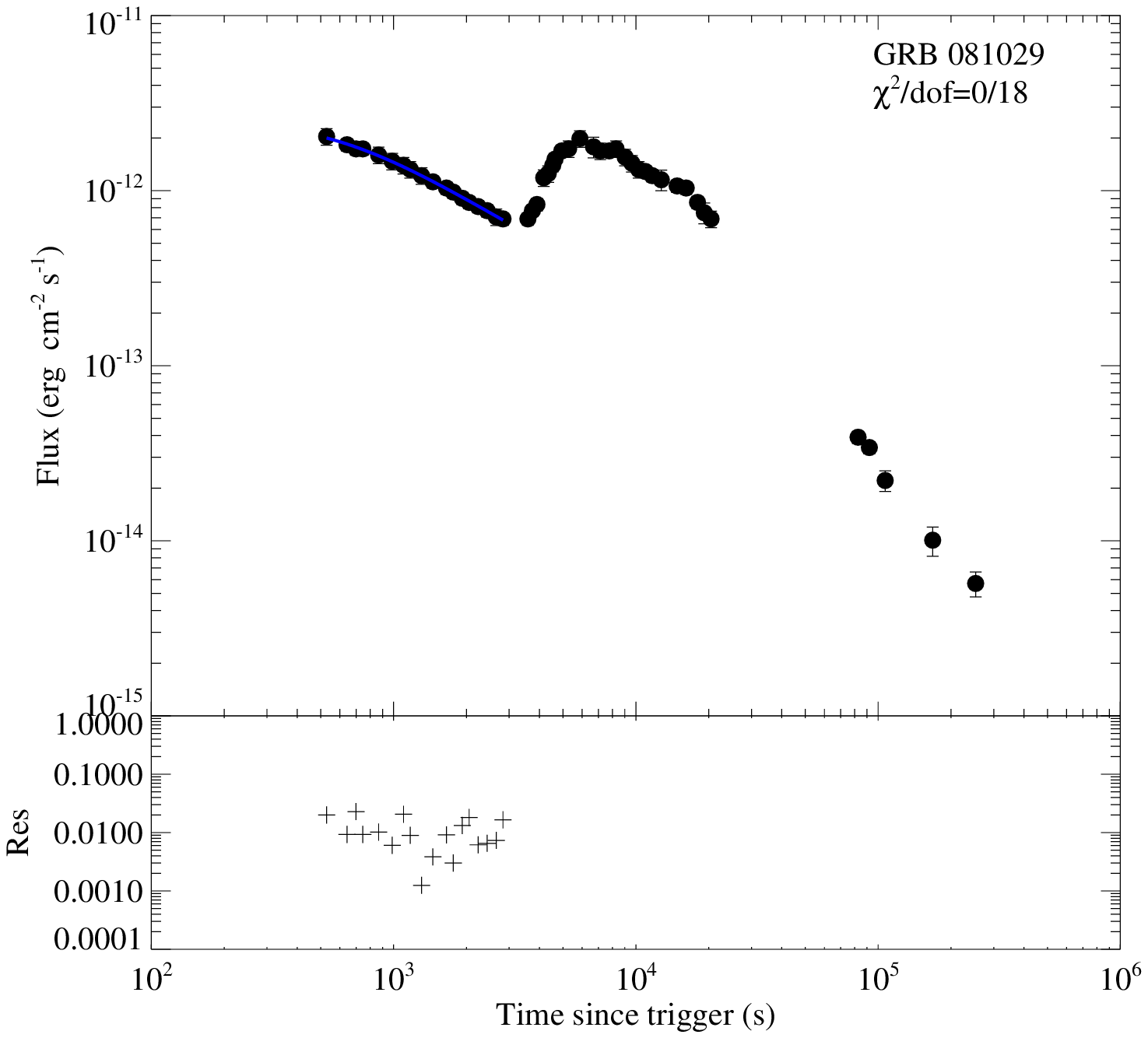}
\includegraphics[angle=0,scale=0.30]{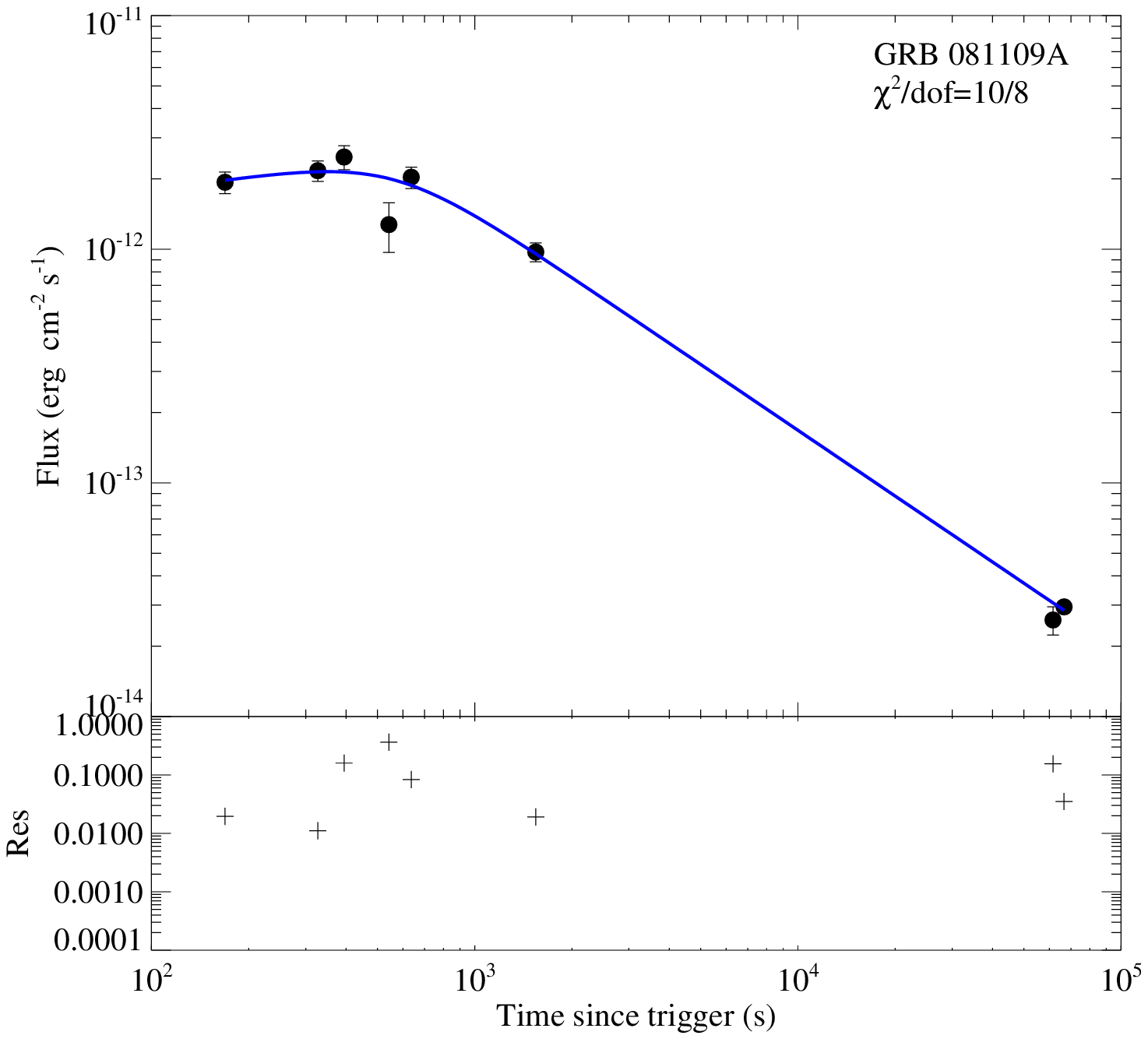}
\includegraphics[angle=0,scale=0.30]{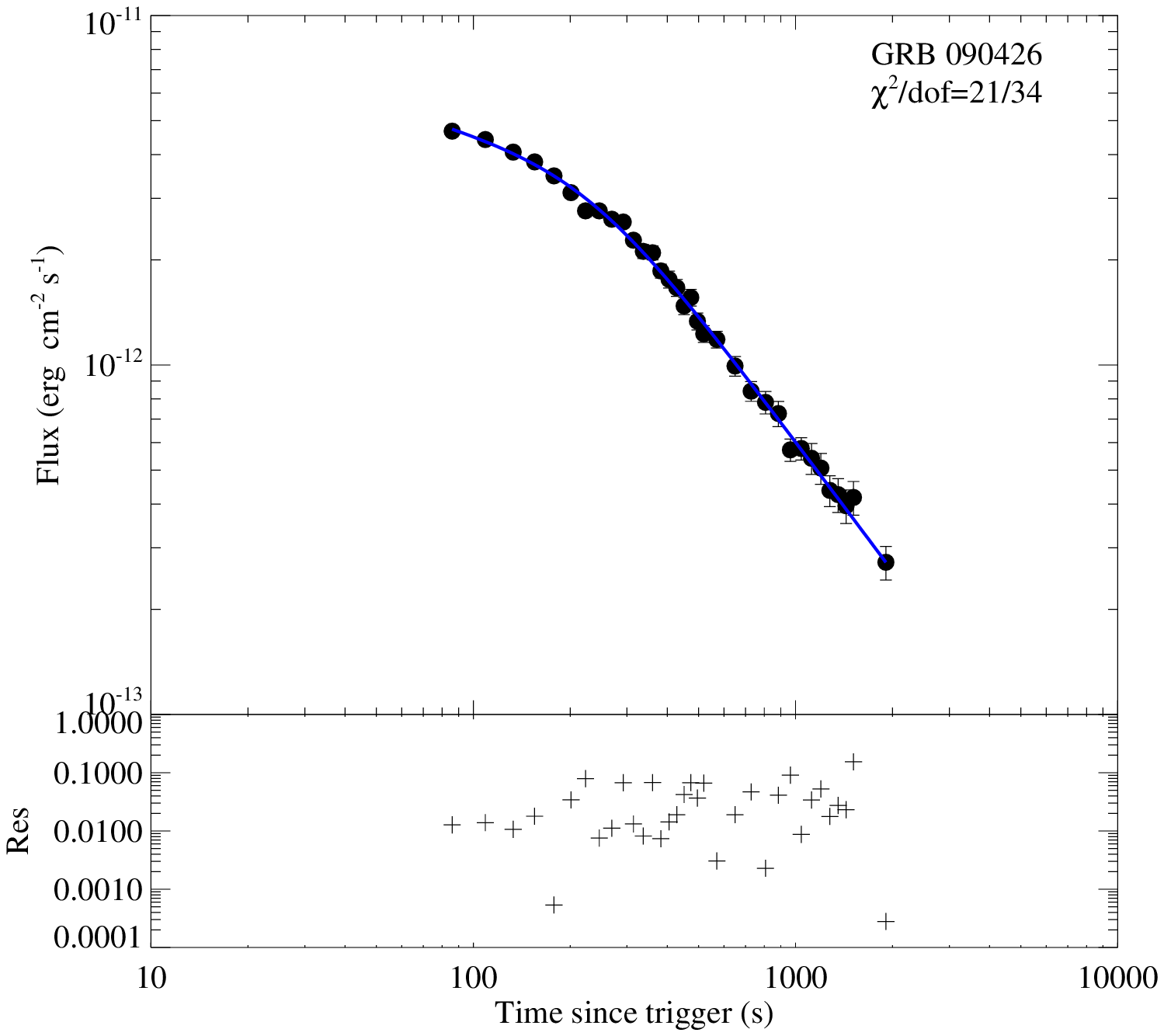}
\includegraphics[angle=0,scale=0.30]{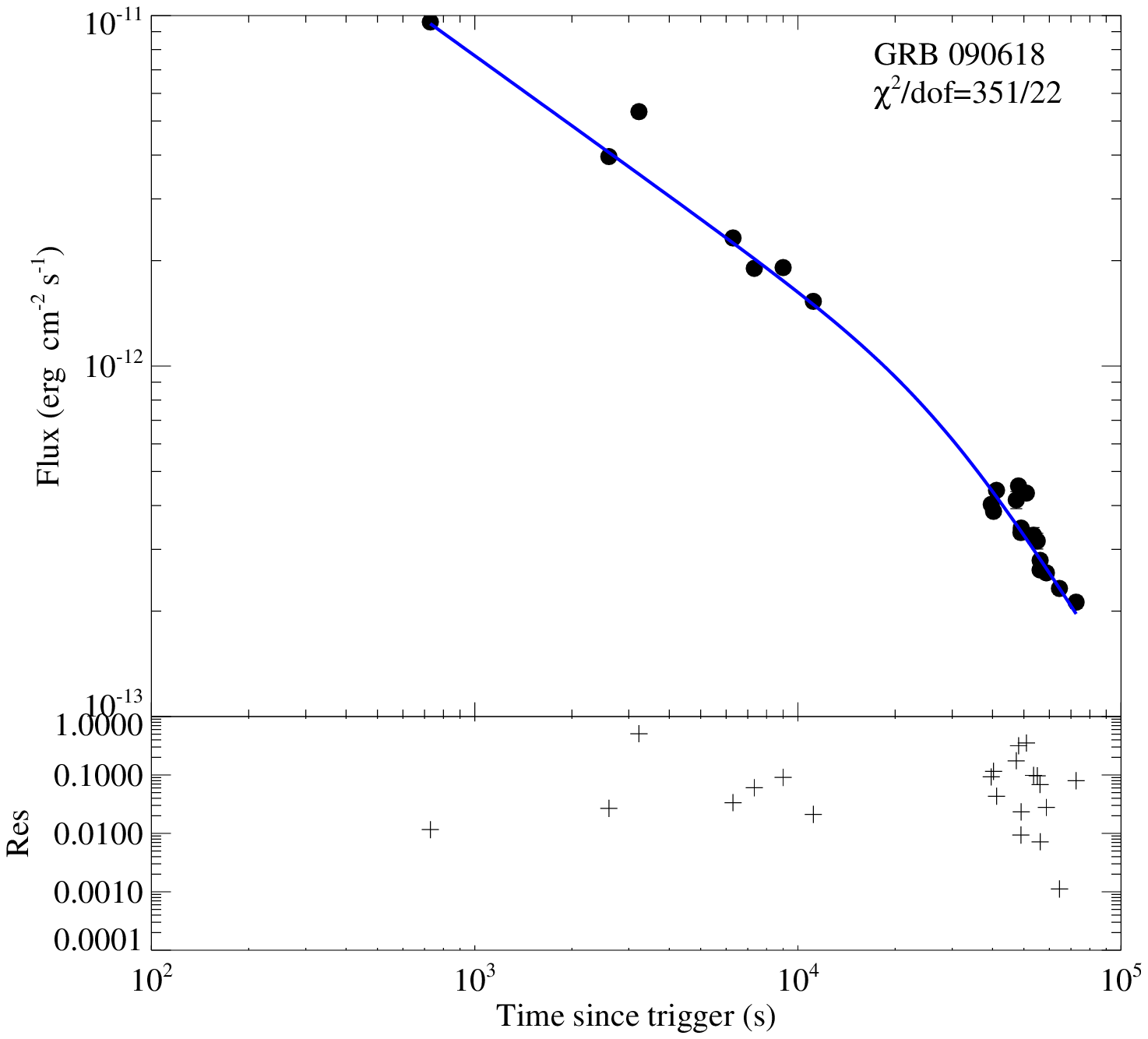}
\includegraphics[angle=0,scale=0.30]{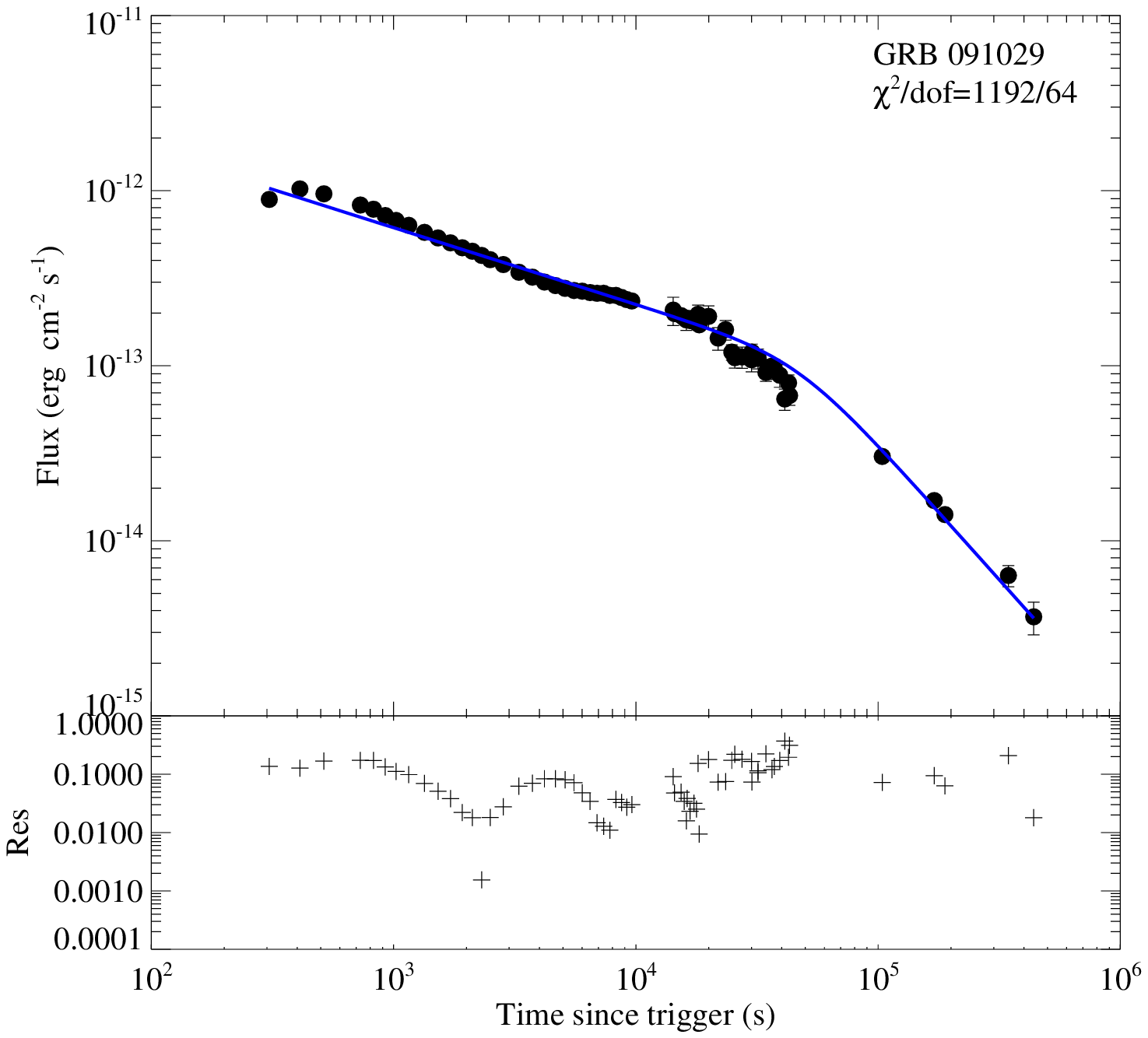}
\includegraphics[angle=0,scale=0.30]{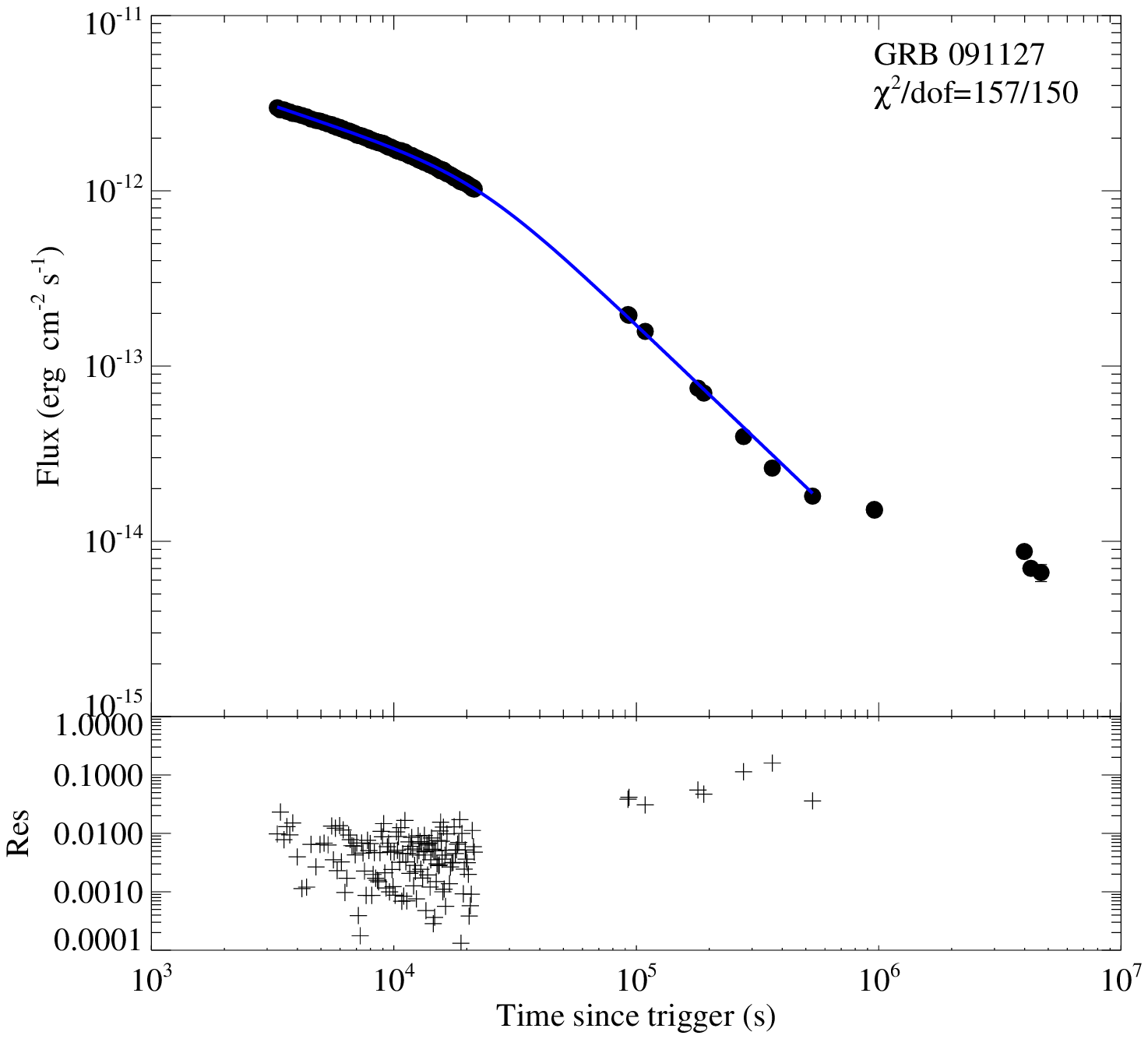}
\includegraphics[angle=0,scale=0.30]{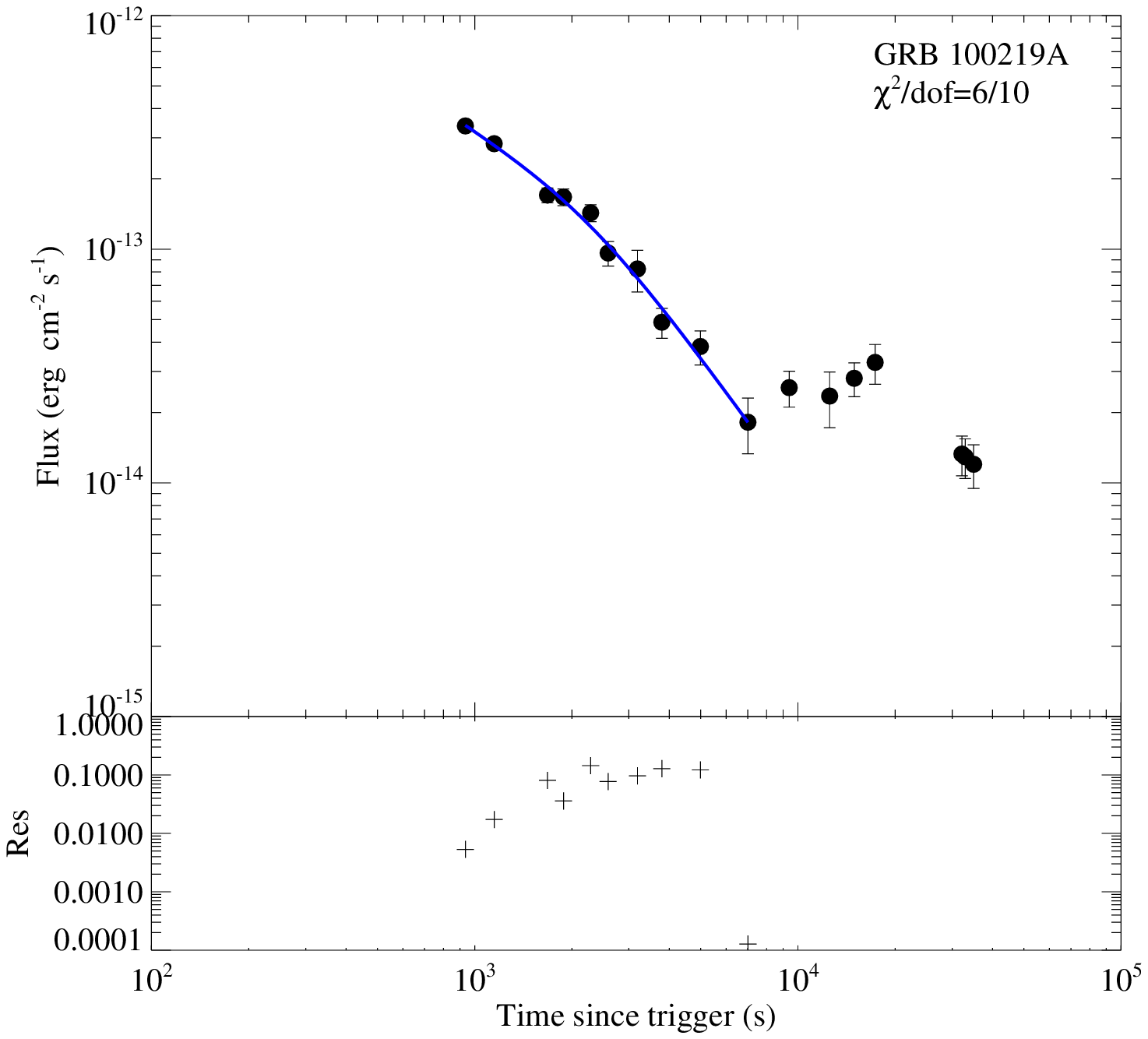}
\includegraphics[angle=0,scale=0.30]{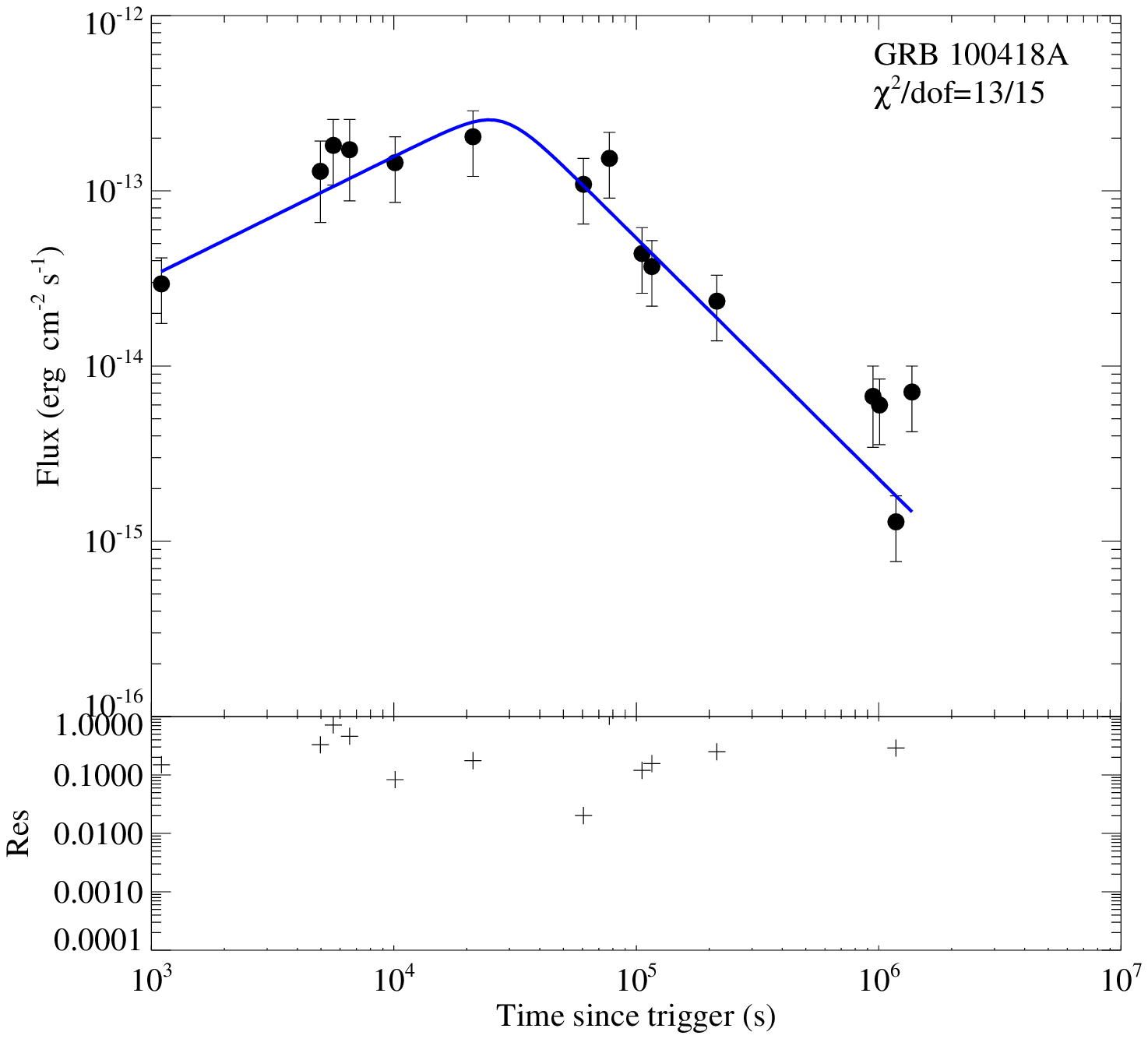}
\includegraphics[angle=0,scale=0.30]{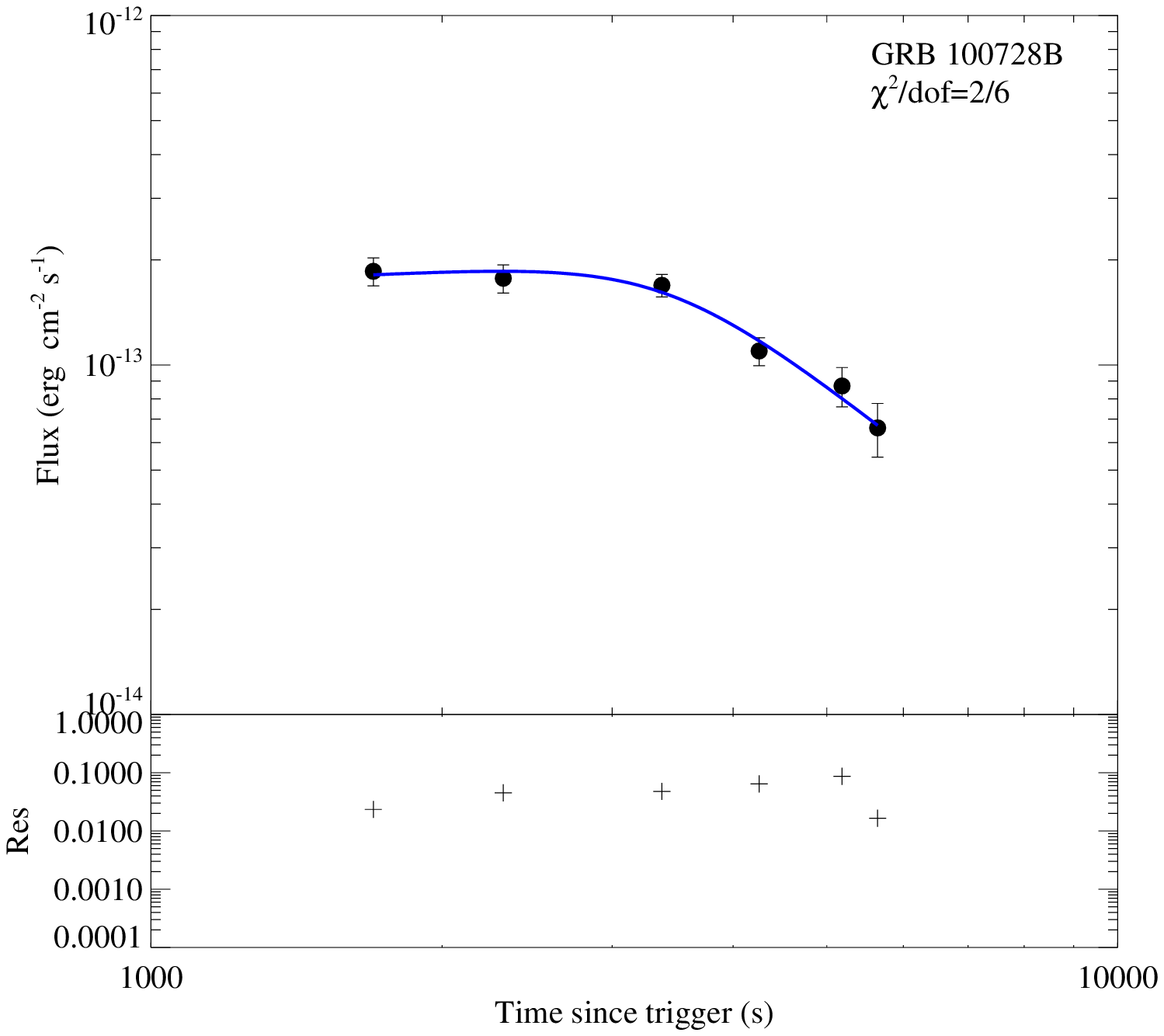}
\includegraphics[angle=0,scale=0.30]{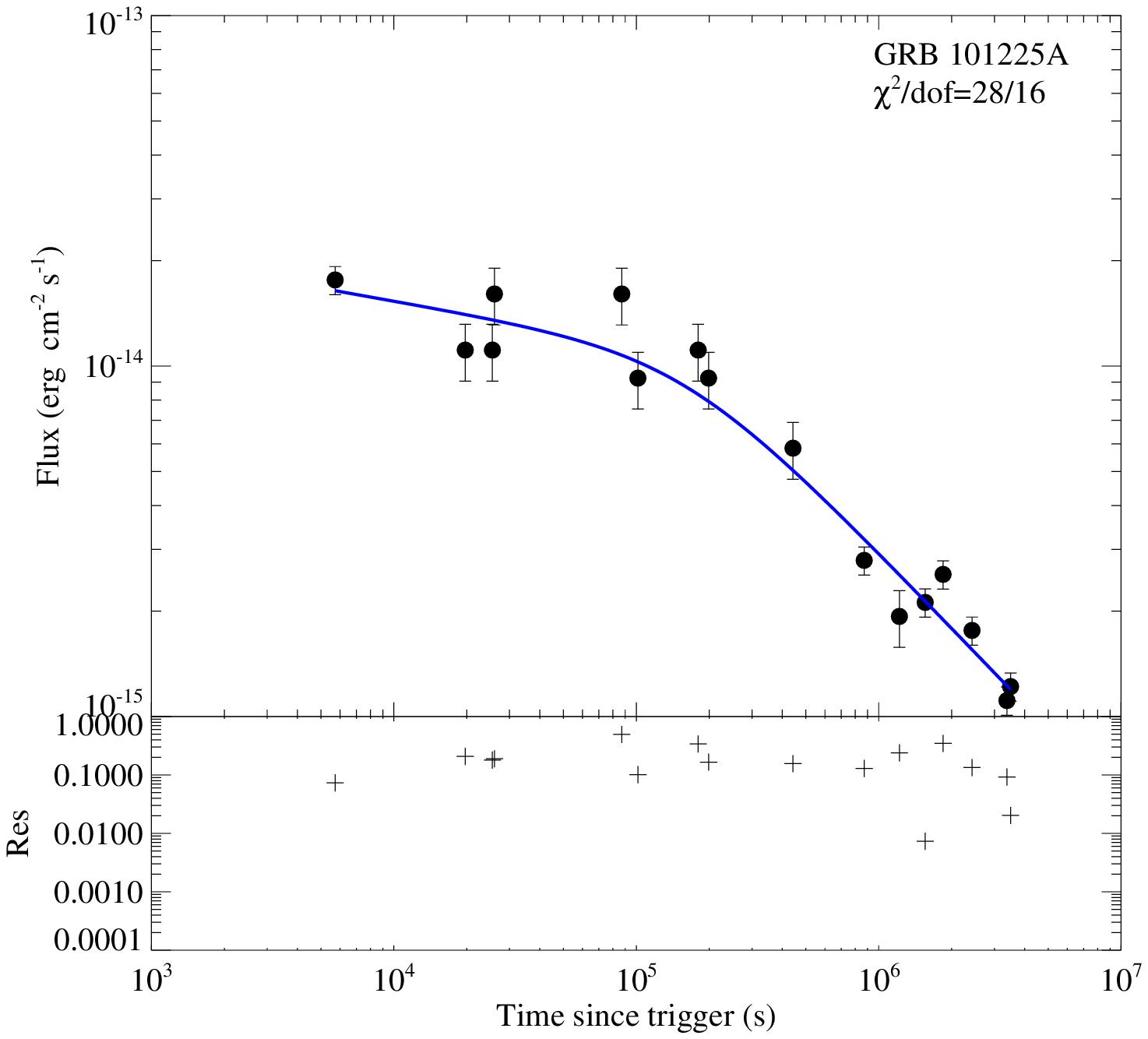}
\includegraphics[angle=0,scale=0.30]{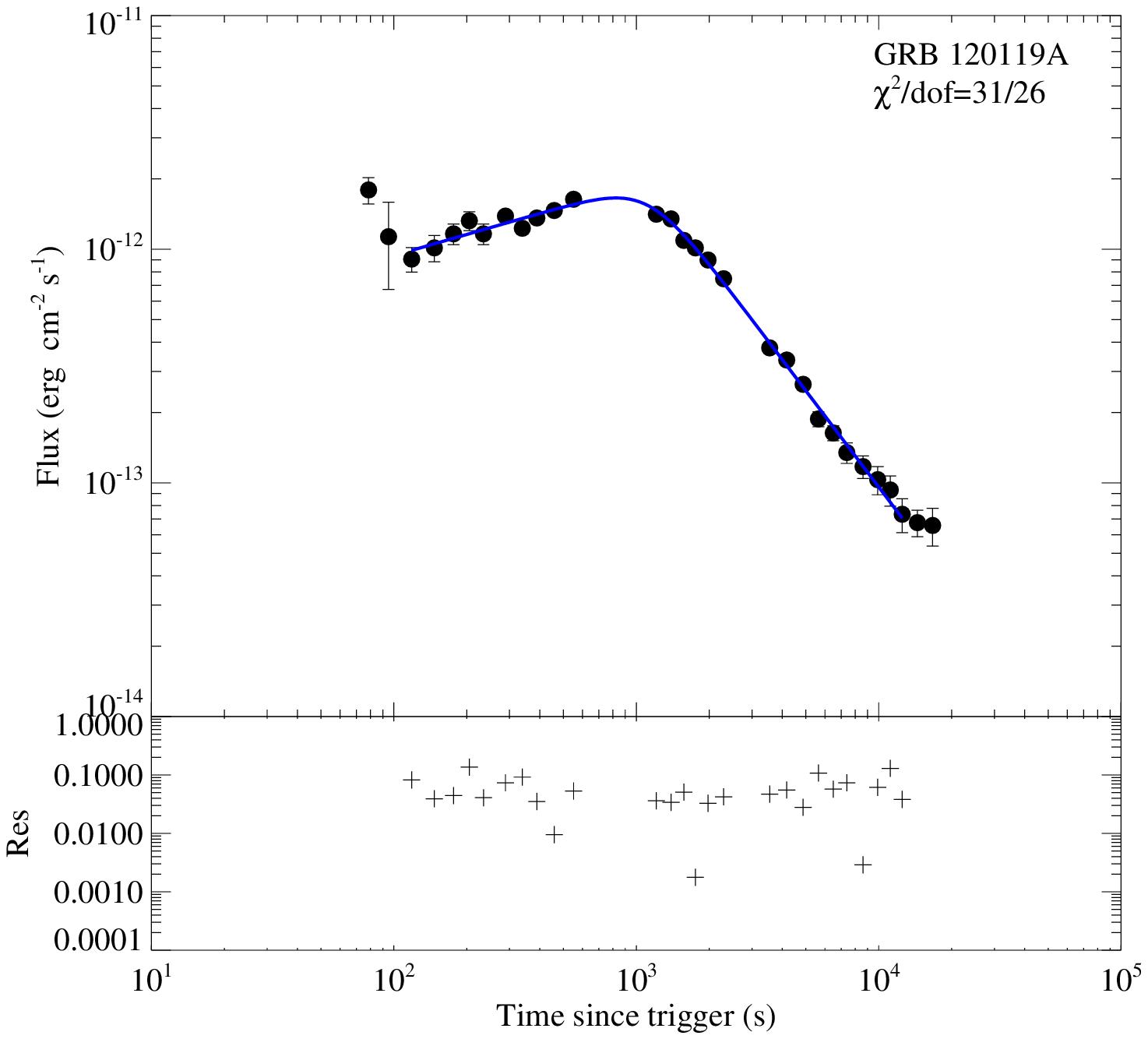}
\includegraphics[angle=0,scale=0.30]{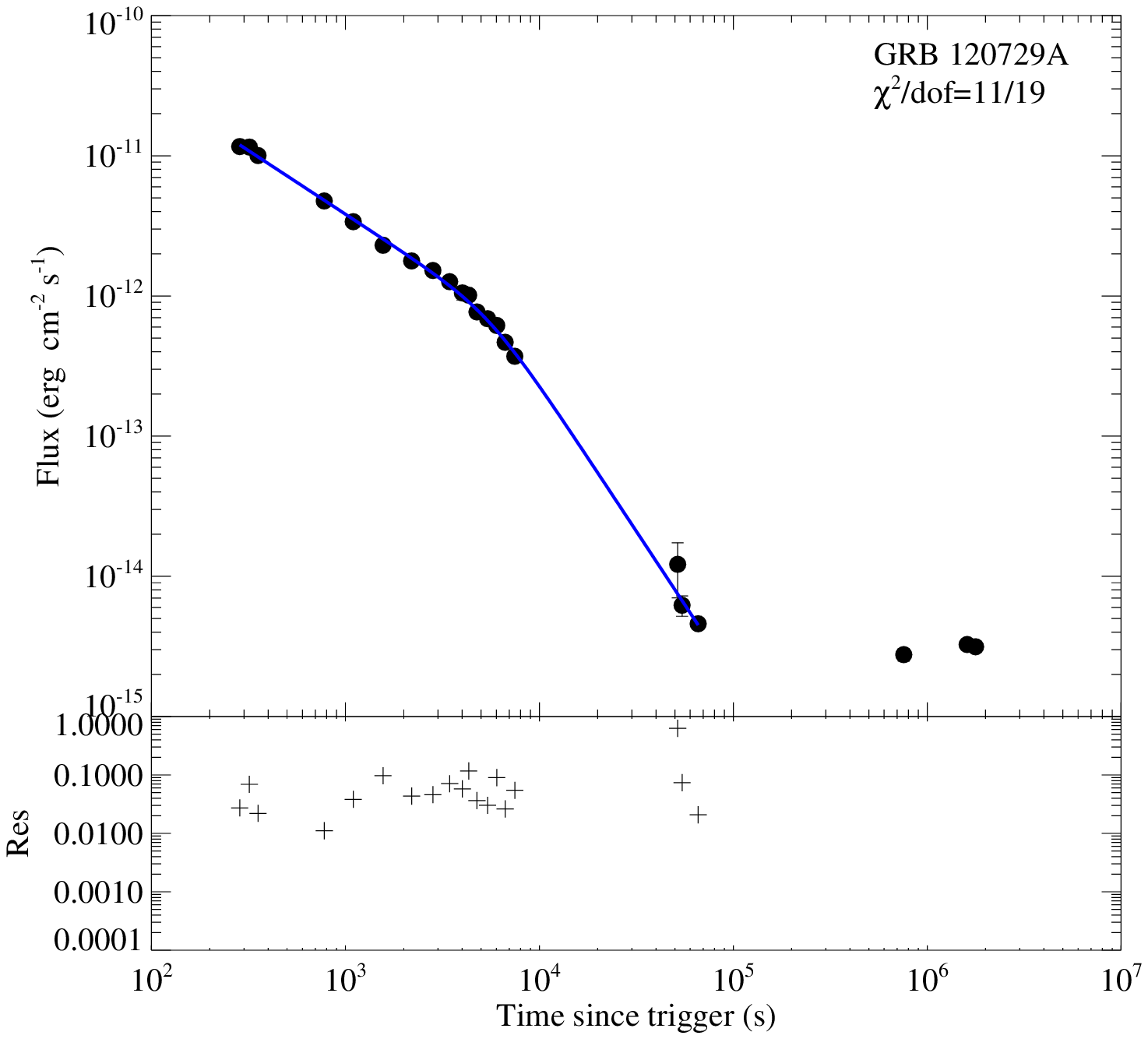}\hfill
\center{Fig. 1--- Continued}
\end{figure*}

\begin{figure*}
\includegraphics[angle=0,scale=0.30]{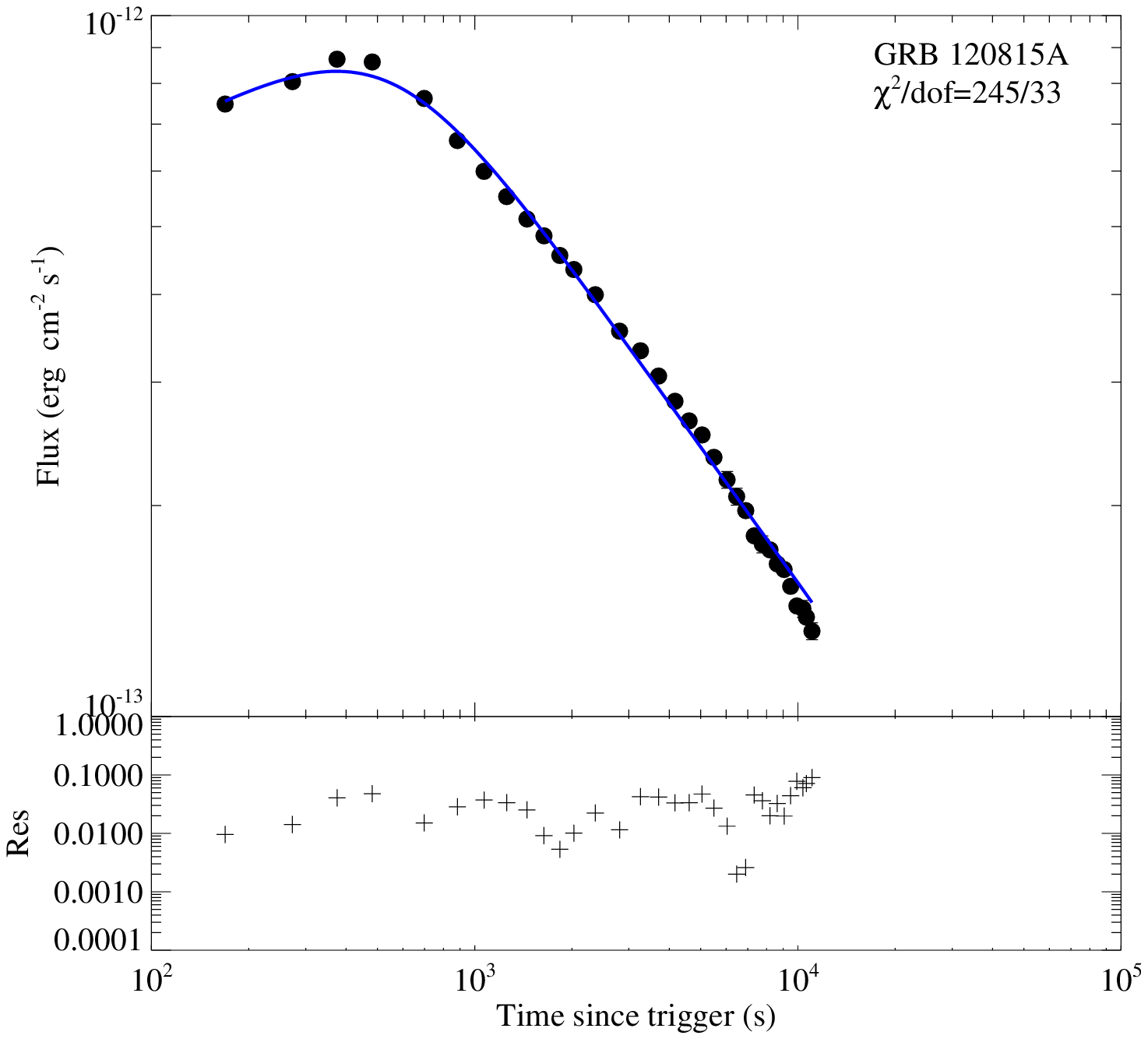}
\includegraphics[angle=0,scale=0.30]{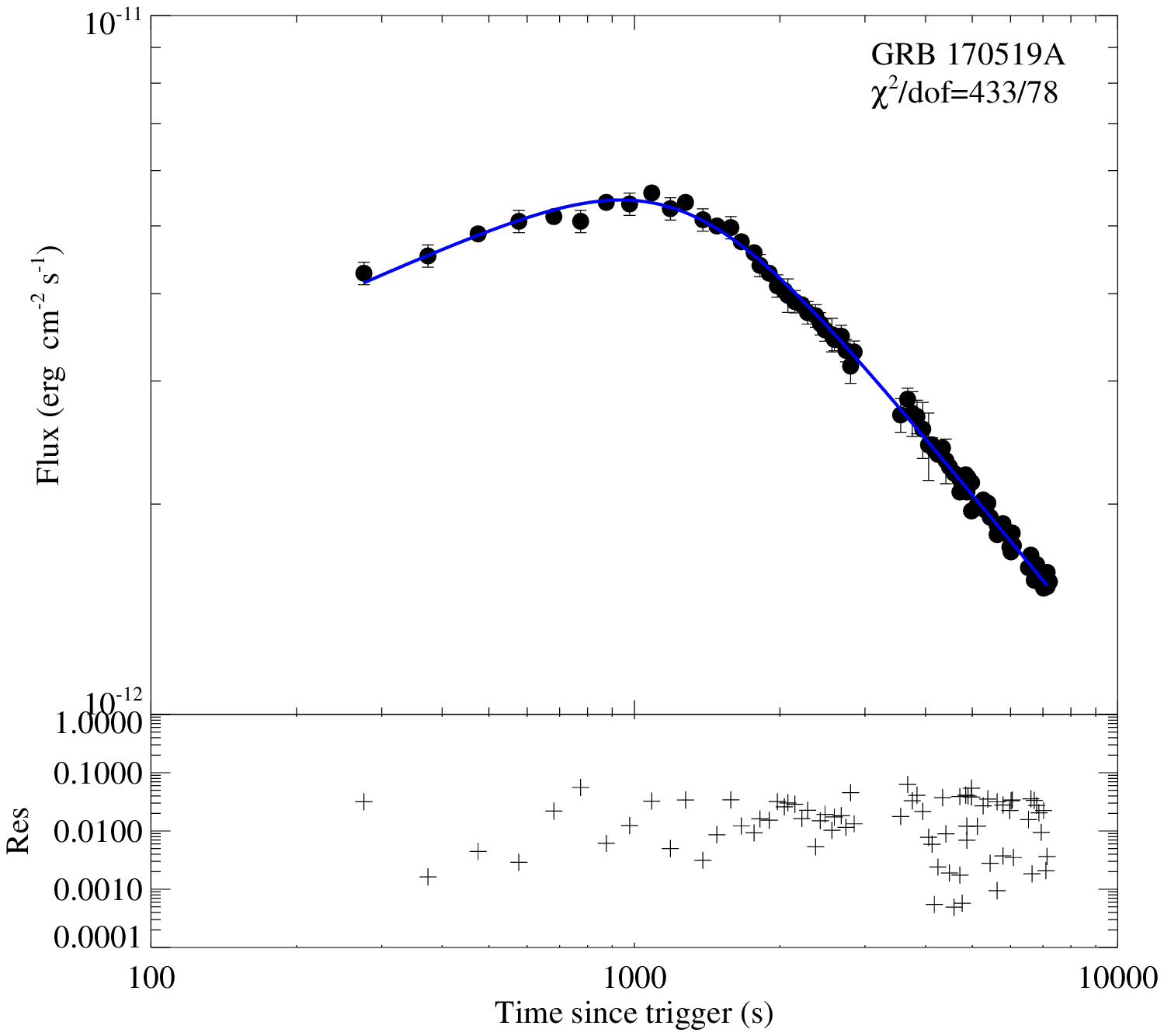}
\center{Fig. 1--- Continued}
\end{figure*}

\begin{figure*}
\includegraphics[angle=0,scale=0.32]{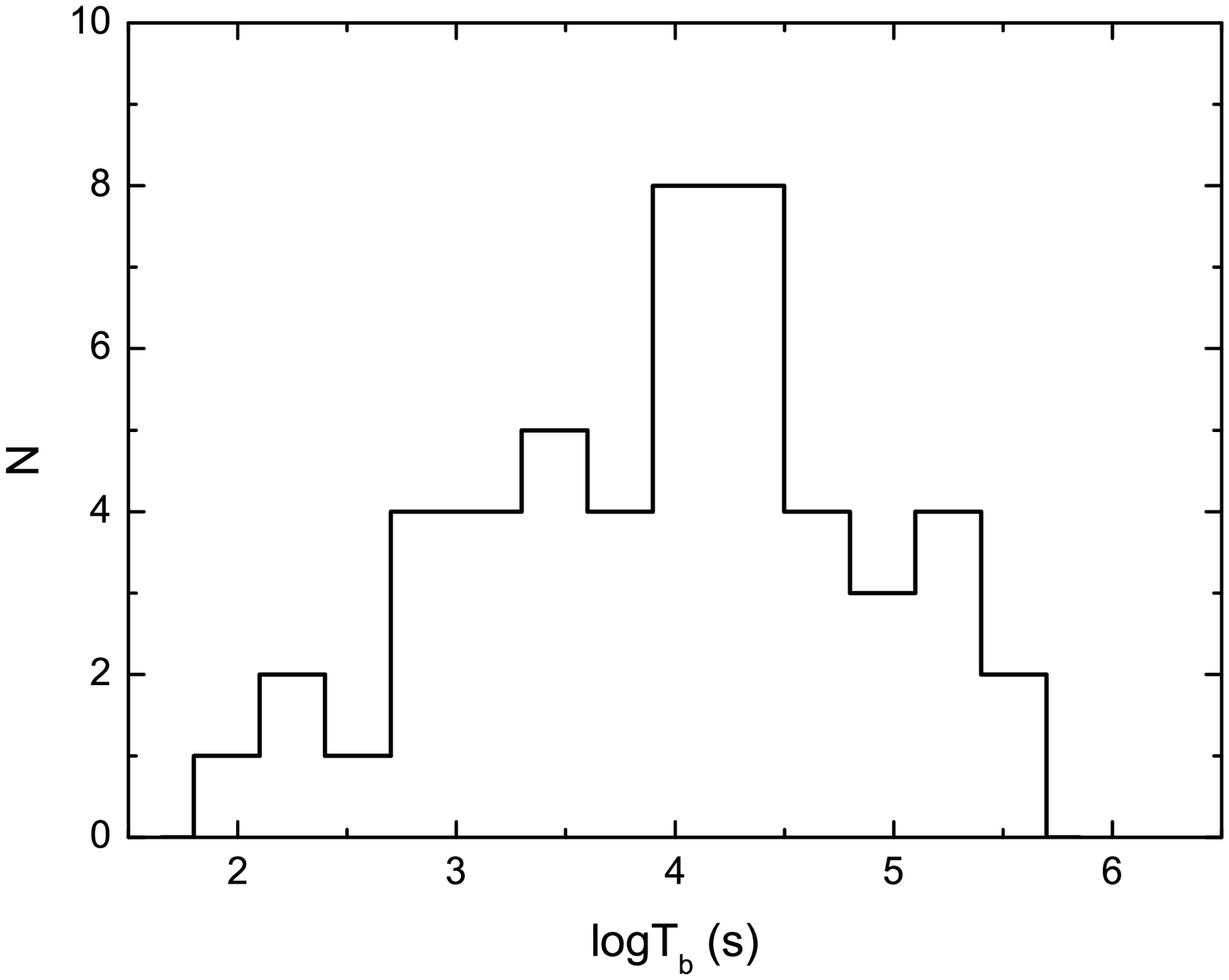}
\includegraphics[angle=0,scale=0.32]{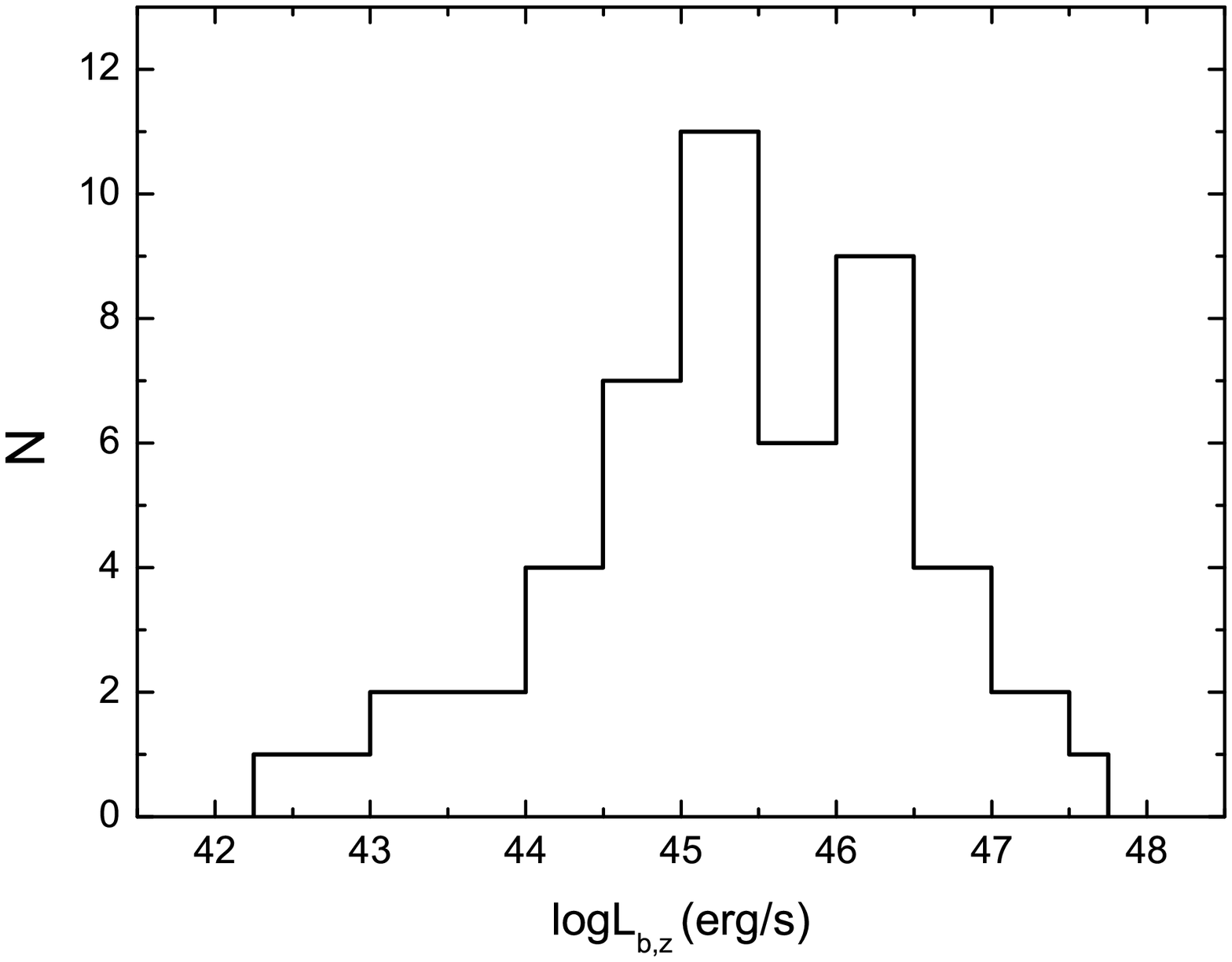}
\includegraphics[angle=0,scale=0.32]{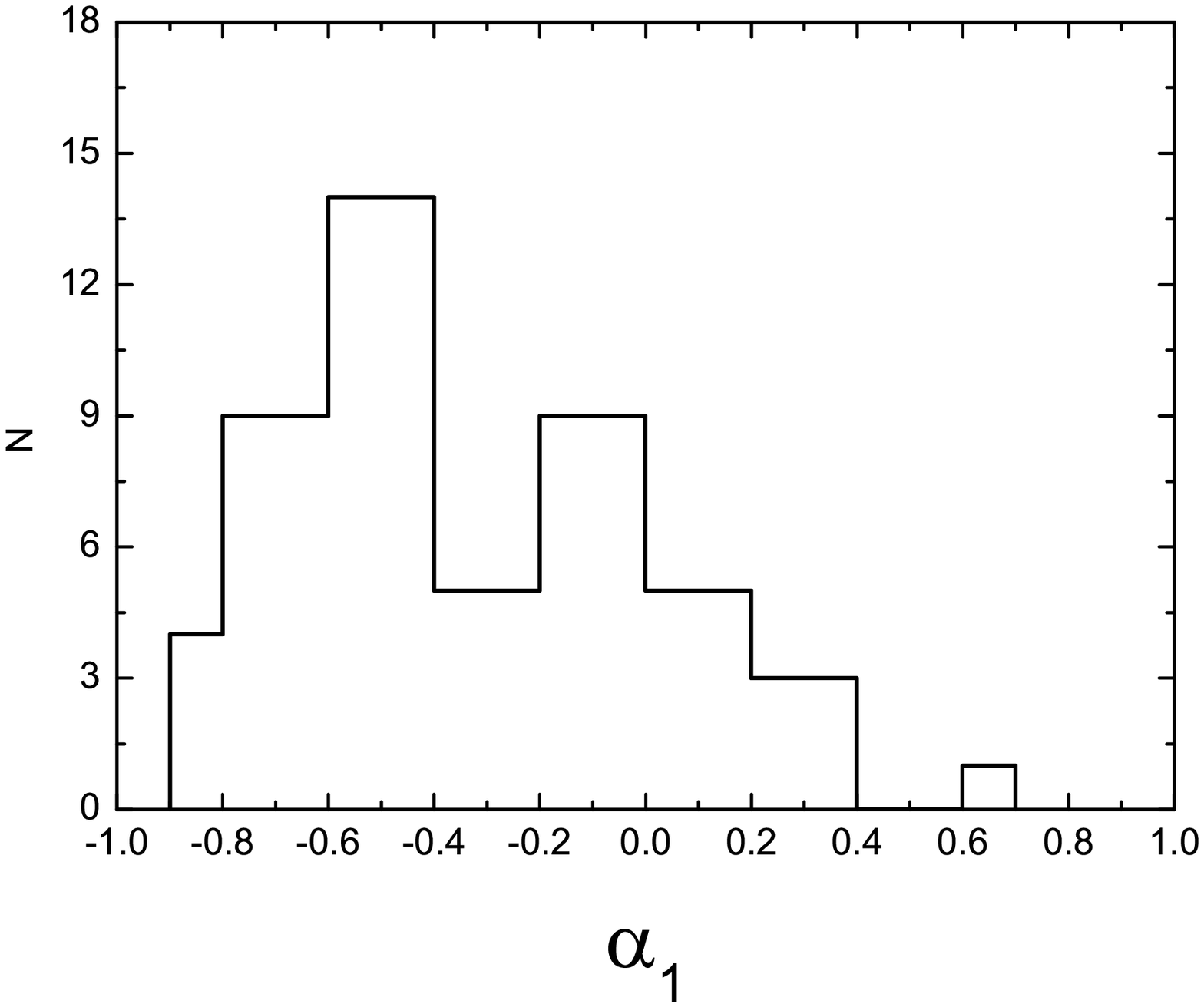}
\includegraphics[angle=0,scale=0.32]{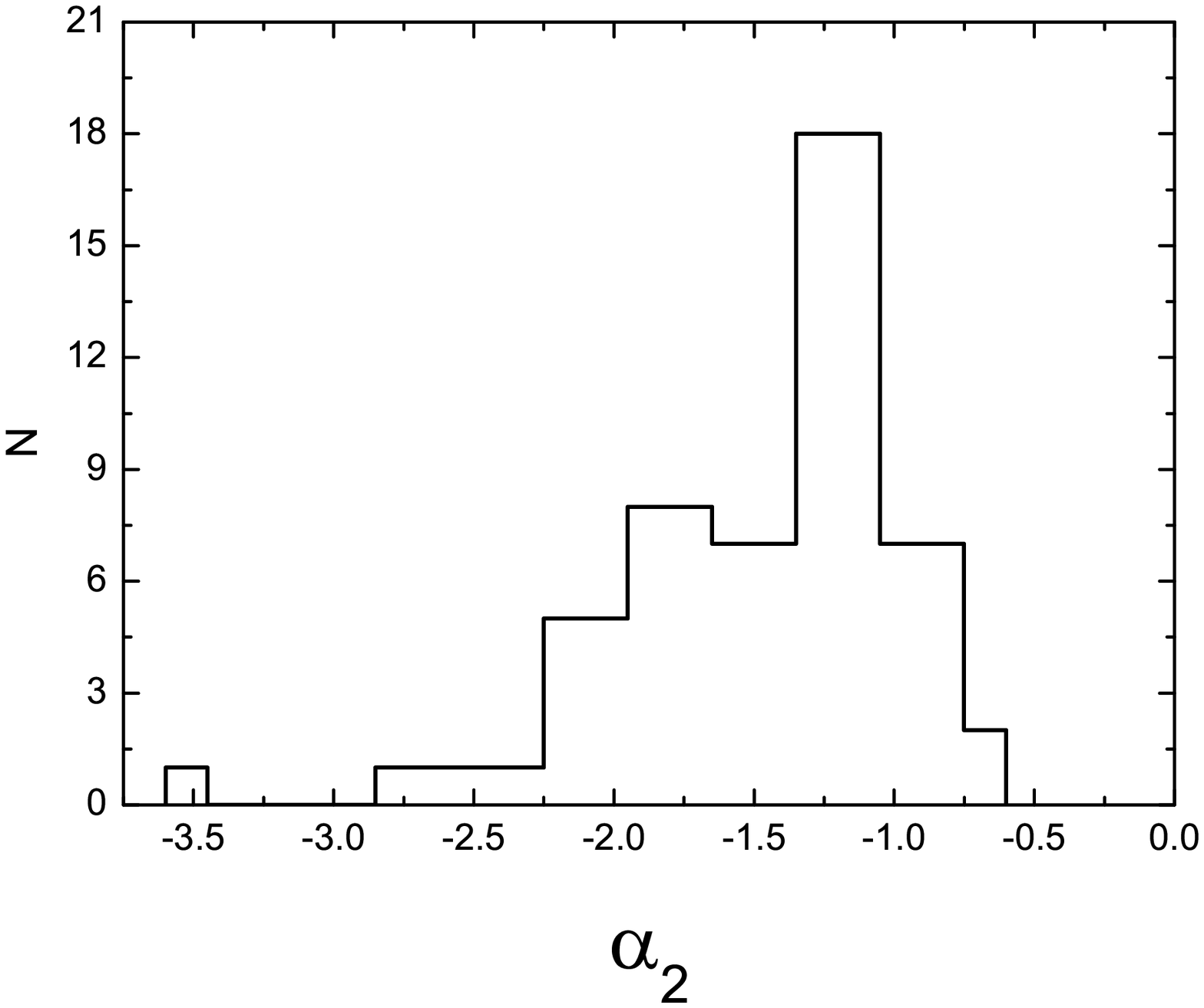}
\caption{The distributions of the parameters for optical plateaus, including the break time $T_{\rm b}$, the optical luminosity at the break $L_{\rm b,z}$, the slopes before and after the break ($\alpha_{\rm 1}$ and $\alpha_{\rm 2}$).}
\end{figure*}

\begin{figure*}
\includegraphics[angle=0,scale=0.32]{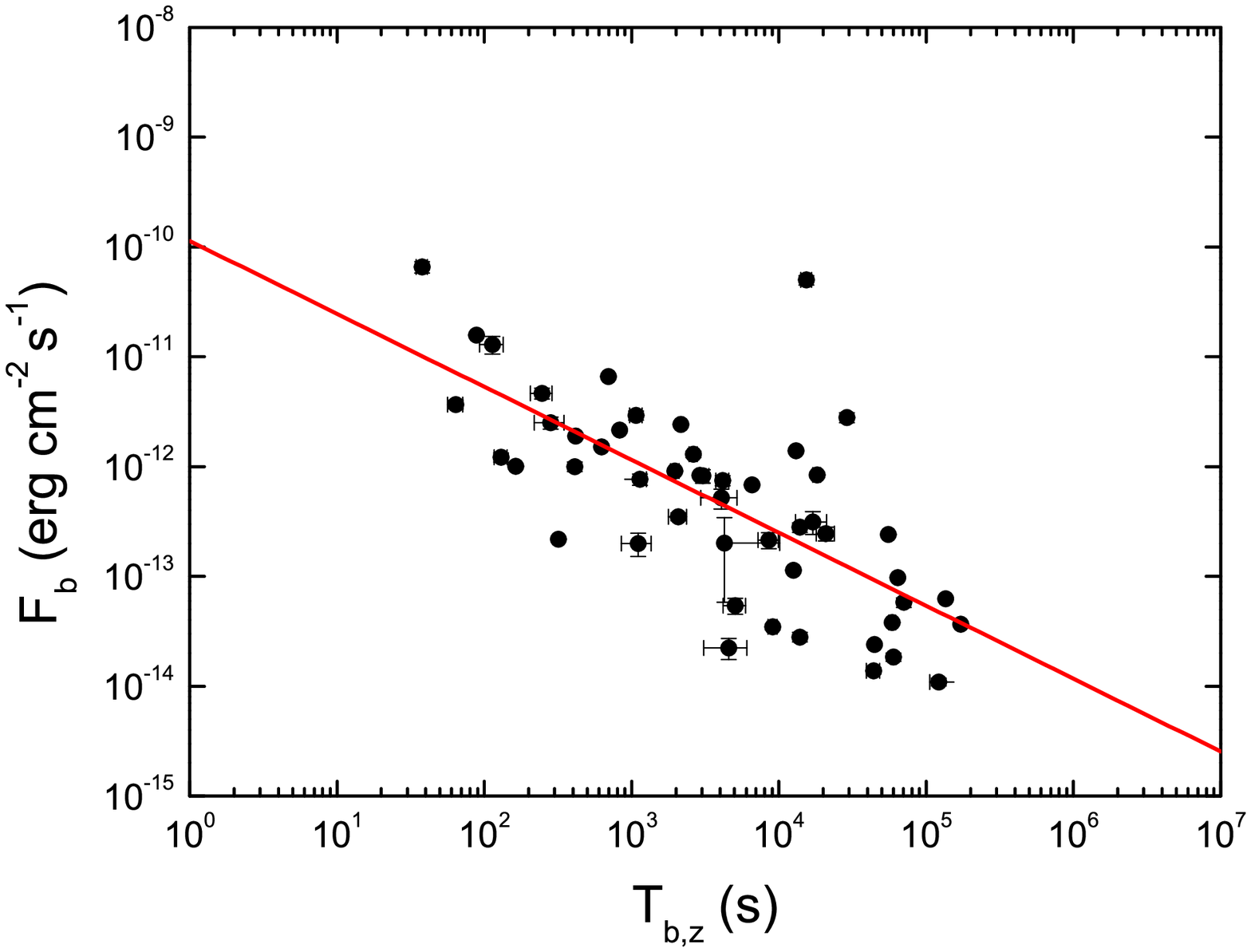}
\includegraphics[angle=0,scale=0.32]{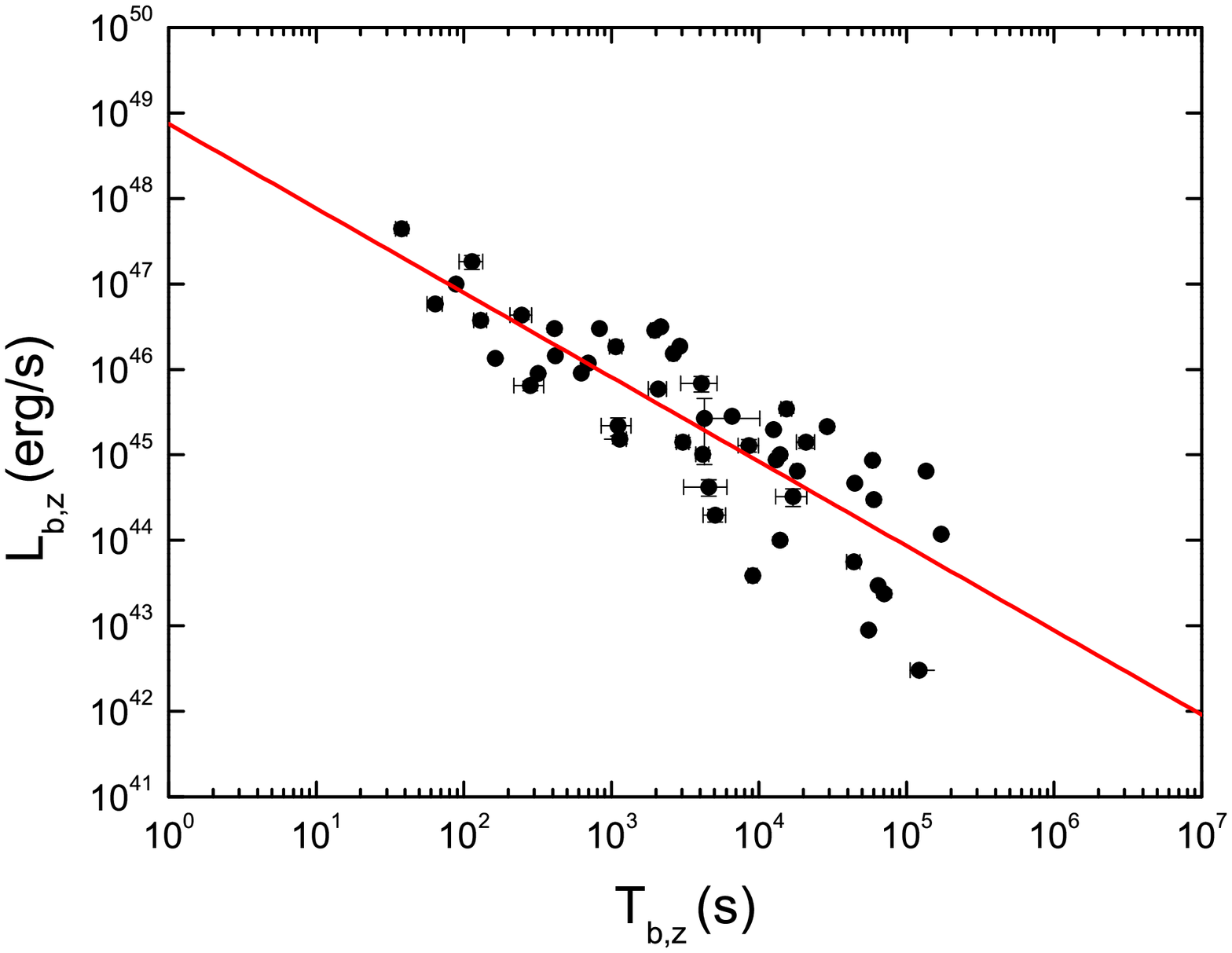}
\caption{Two correlations about $F_{\rm b}$-$T_{\rm b,z}$ and $L_{\rm b,z}$-$T_{\rm b,z}$ for optical plateaus.}
\end{figure*}

\begin{figure*}
\includegraphics[angle=0,scale=0.32]{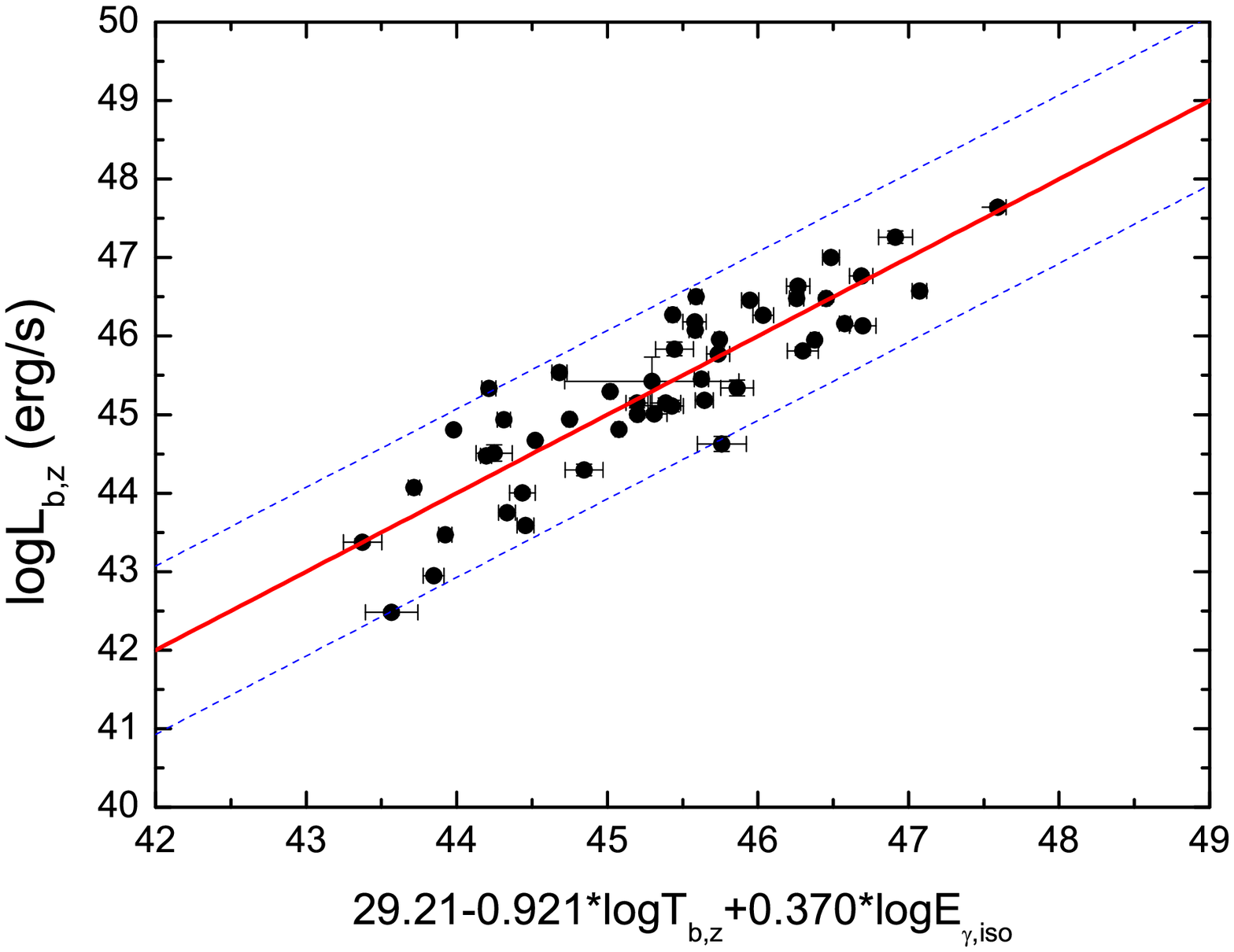}
\includegraphics[angle=0,scale=0.32]{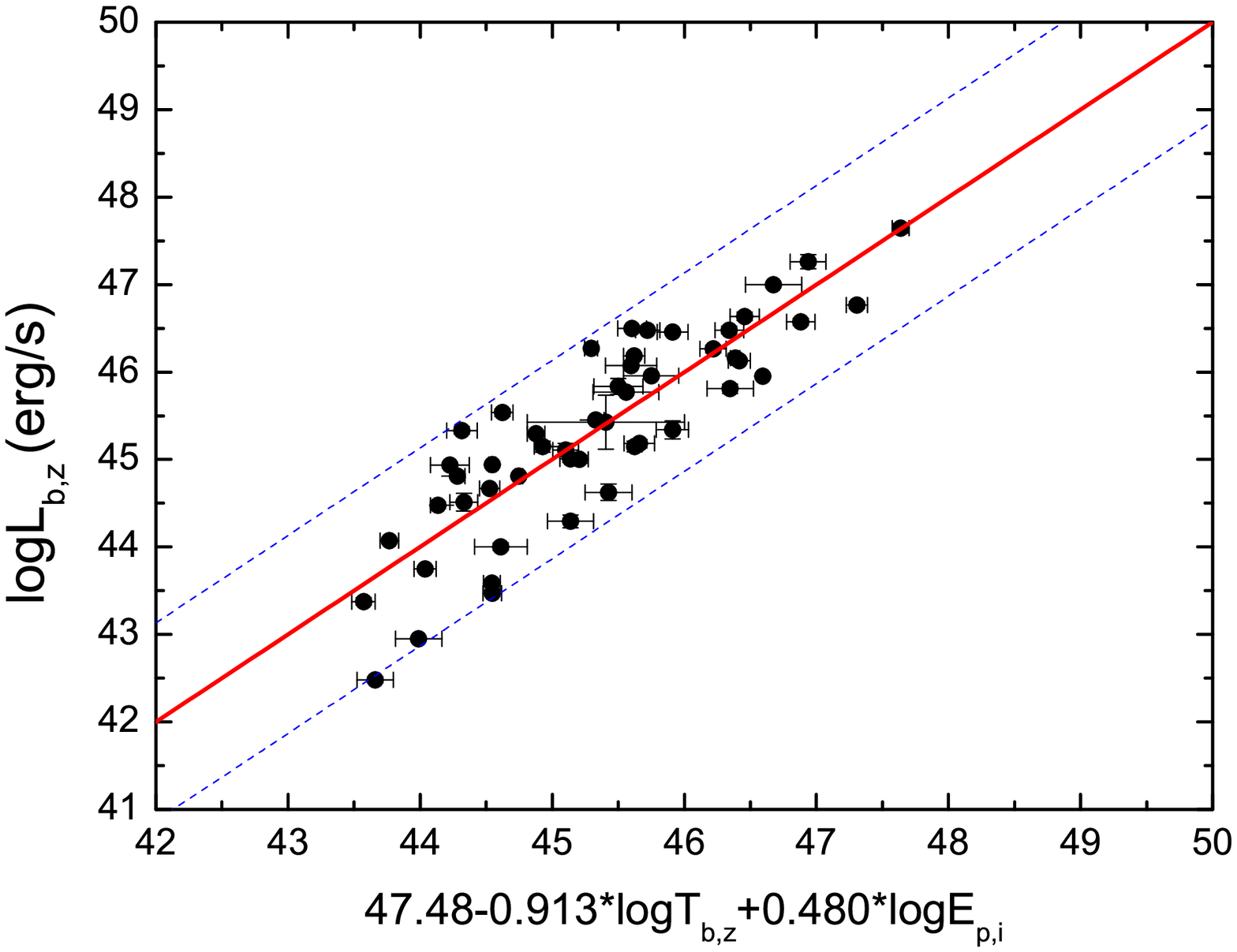}
\caption{The best-fit for two three-parameter correlations.}
\end{figure*}

\begin{figure*}
\includegraphics[angle=0,scale=0.32]{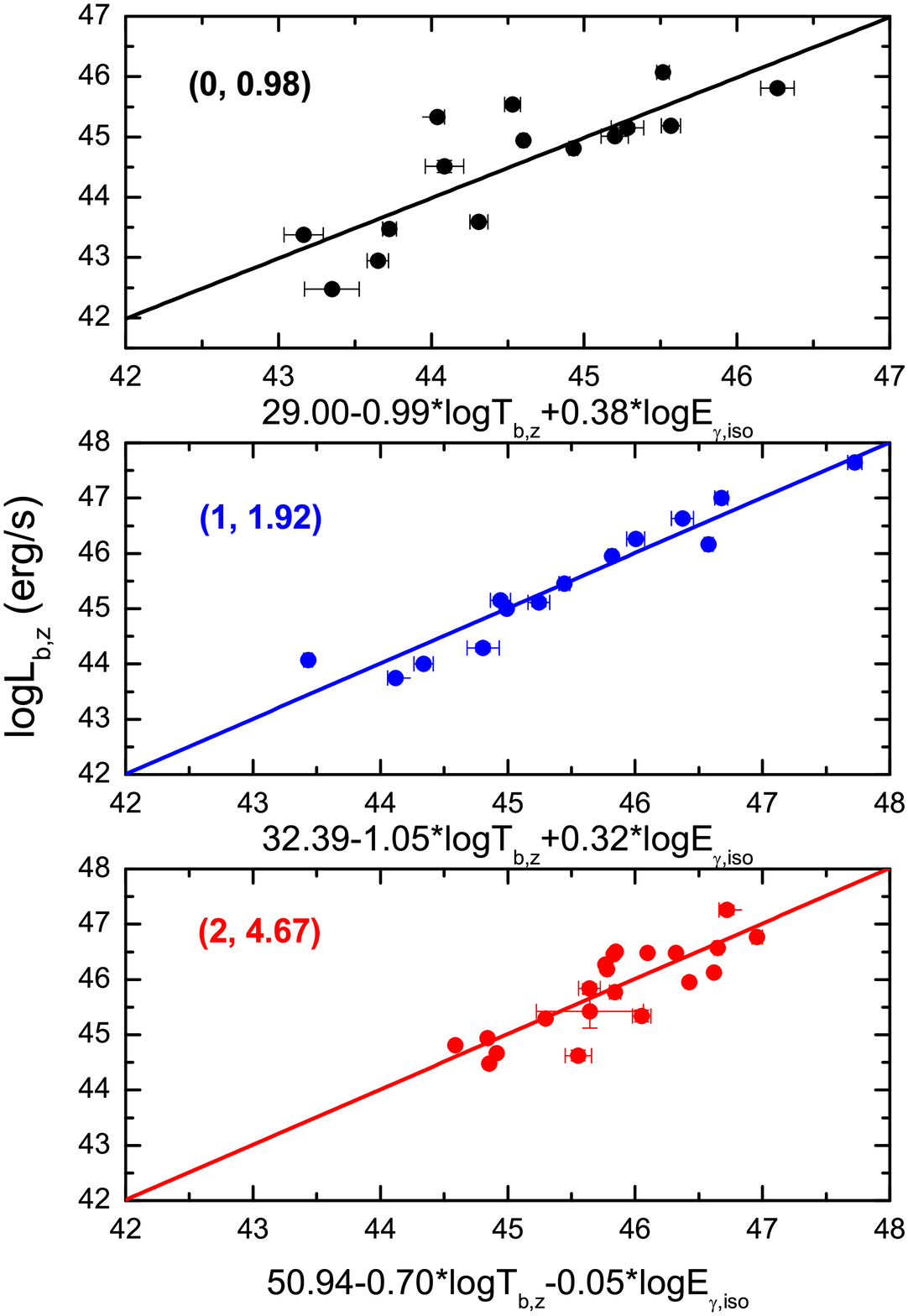}
\includegraphics[angle=0,scale=0.32]{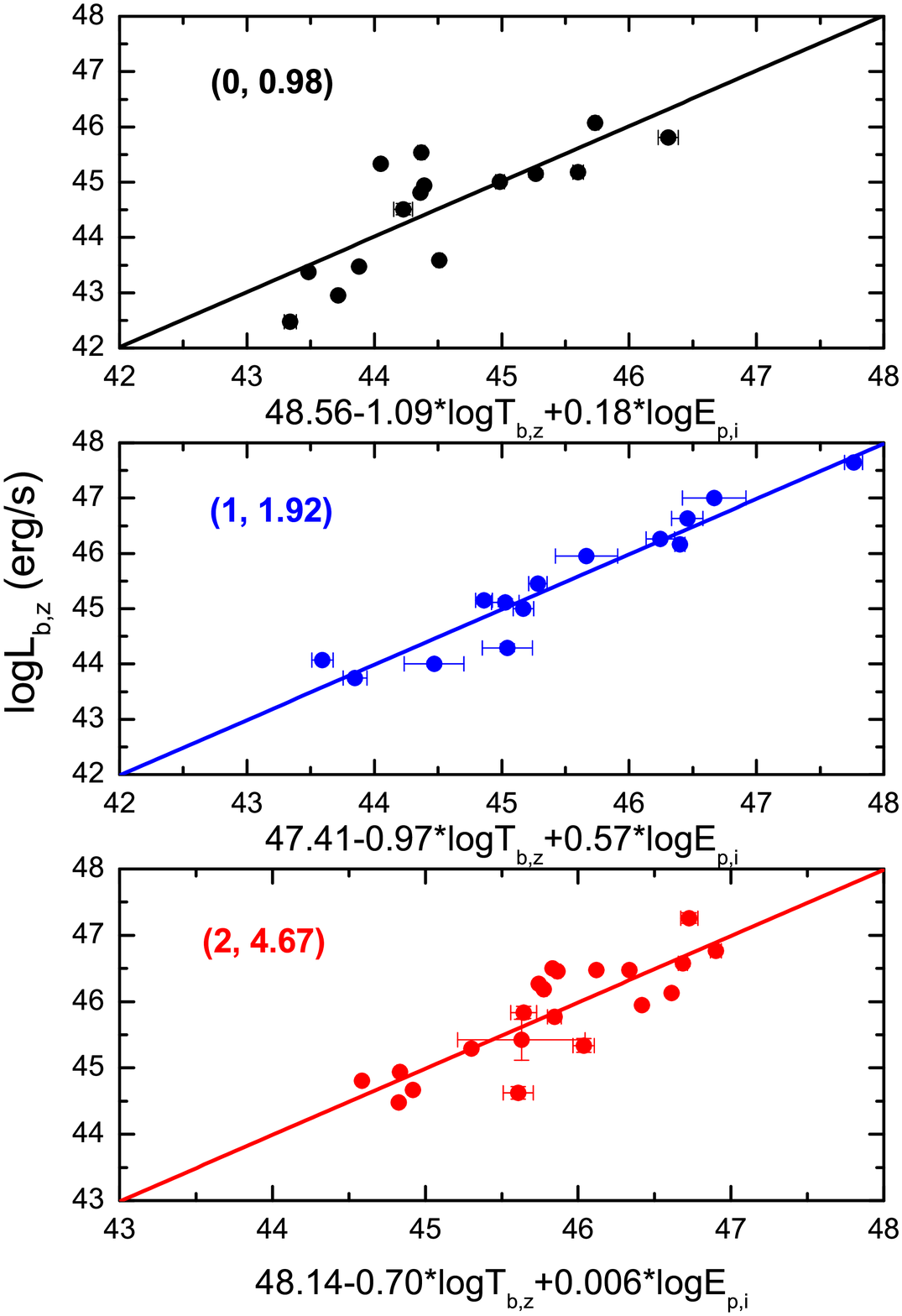}
\caption{The two three-parameter correlations divided into three redshift bins (0, 0.98), (1, 1.92) and (2, 467), respectively. The respective fitted lines are in the same colors.}
\end{figure*}

\begin{figure*}
\includegraphics[angle=0,scale=0.35]{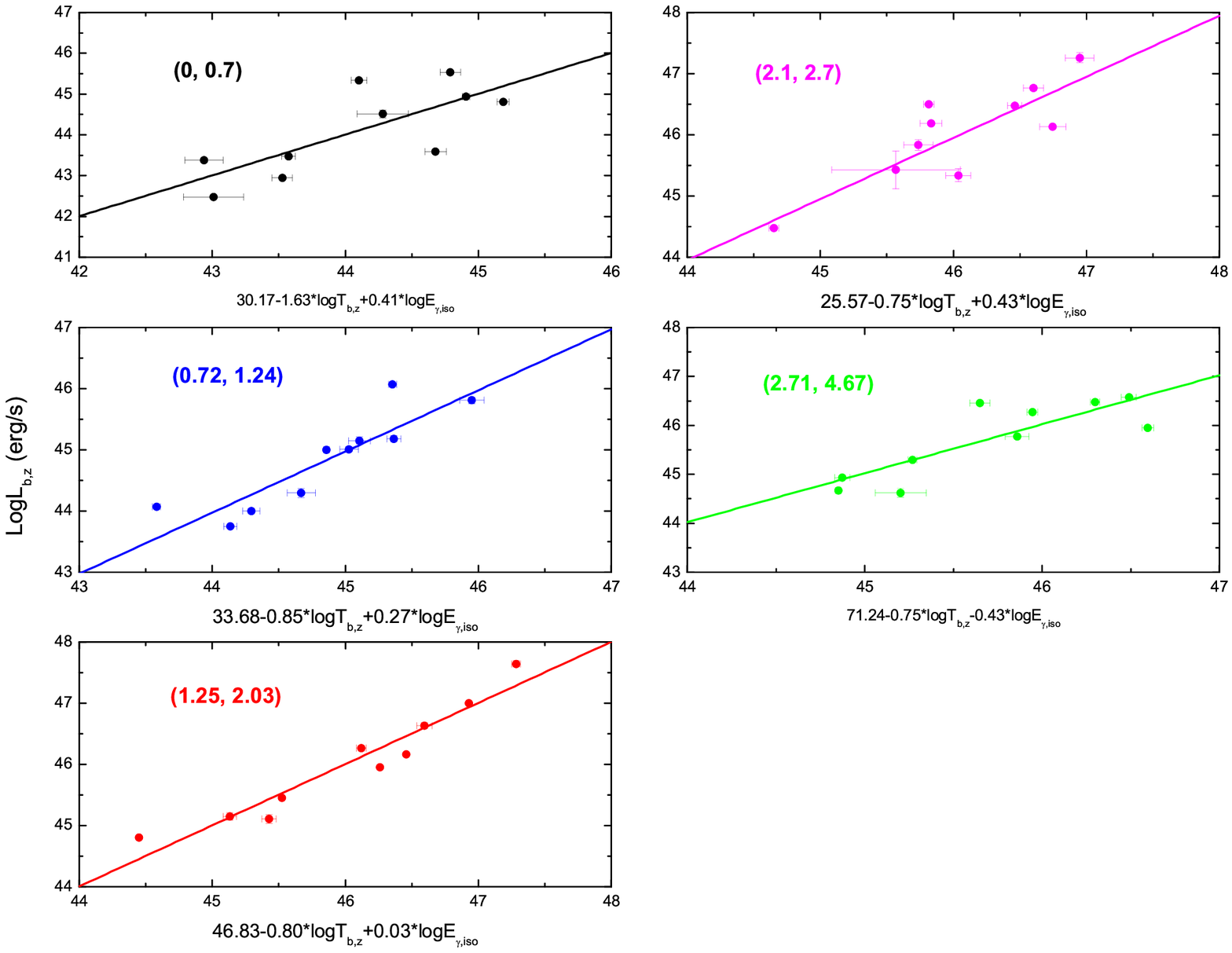}
\includegraphics[angle=0,scale=0.35]{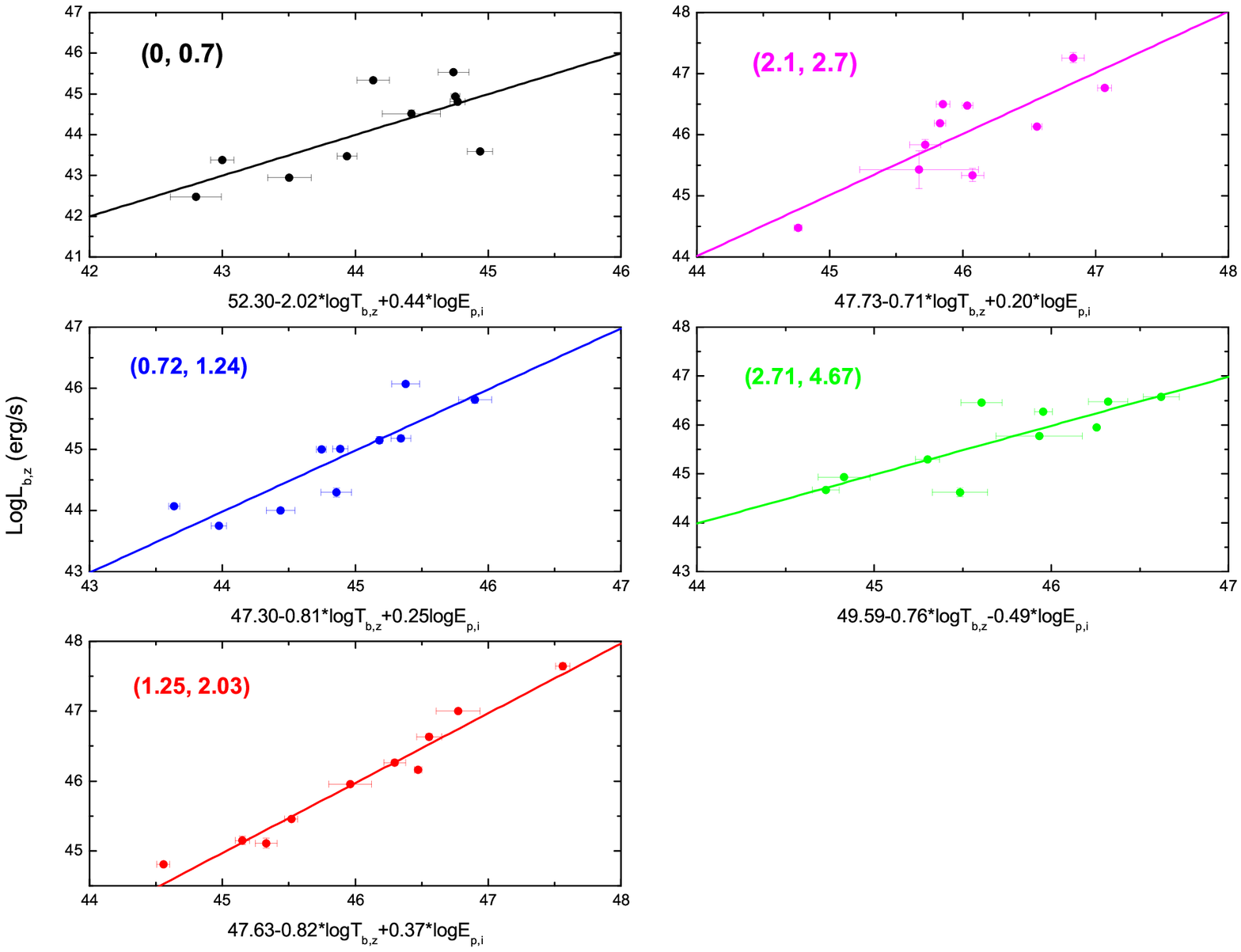}
\caption{The two three-parameter correlations divided into five redshift bins  (0, 0.70), (0.72, 1.24), (1.25, 2.03), (2.1, 2.7) and (2.71, 4,67), respectively. The respective fitted lines are in the same colors.}
\end{figure*}

\begin{figure*}
\includegraphics[angle=0,scale=0.32]{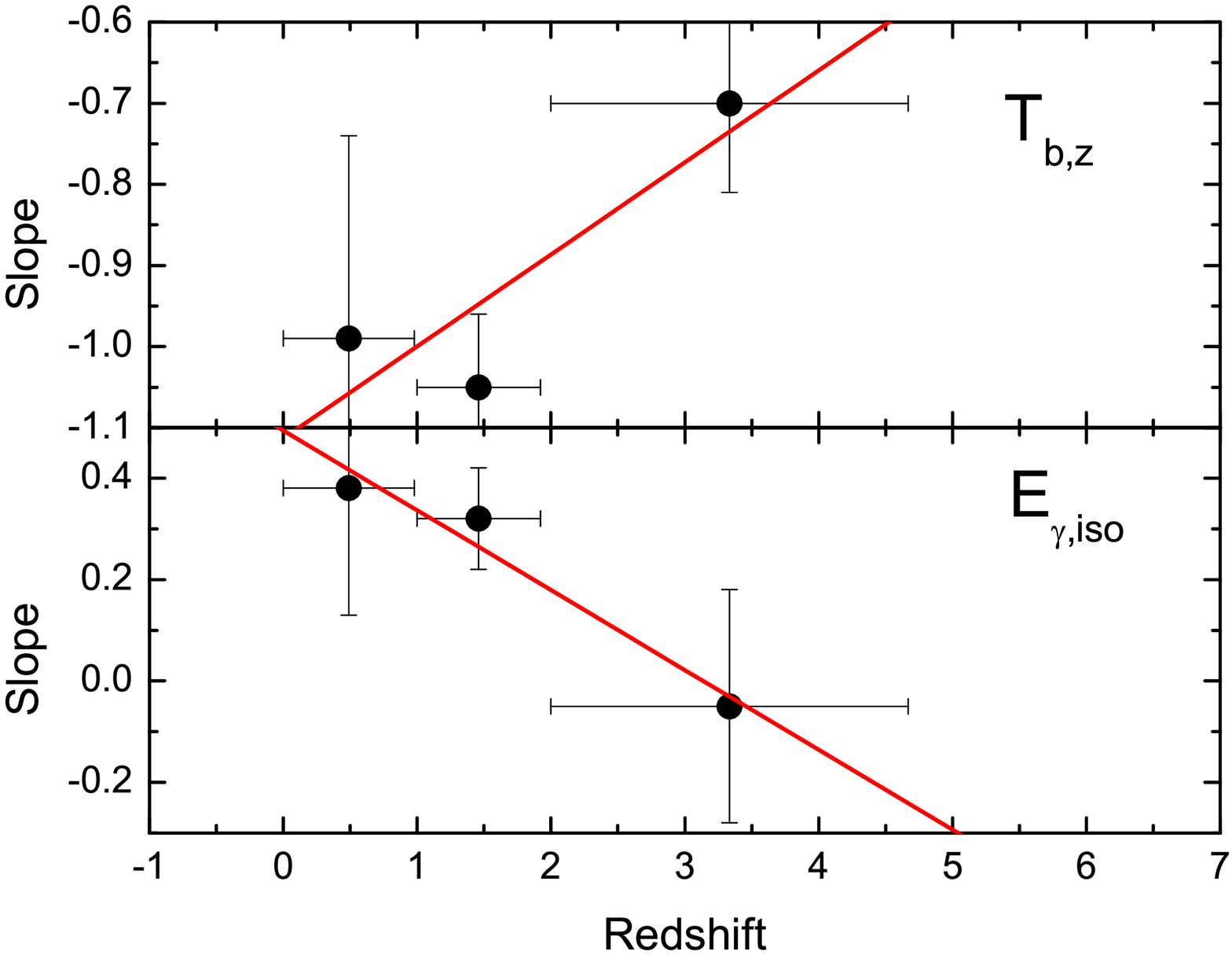}
\includegraphics[angle=0,scale=0.32]{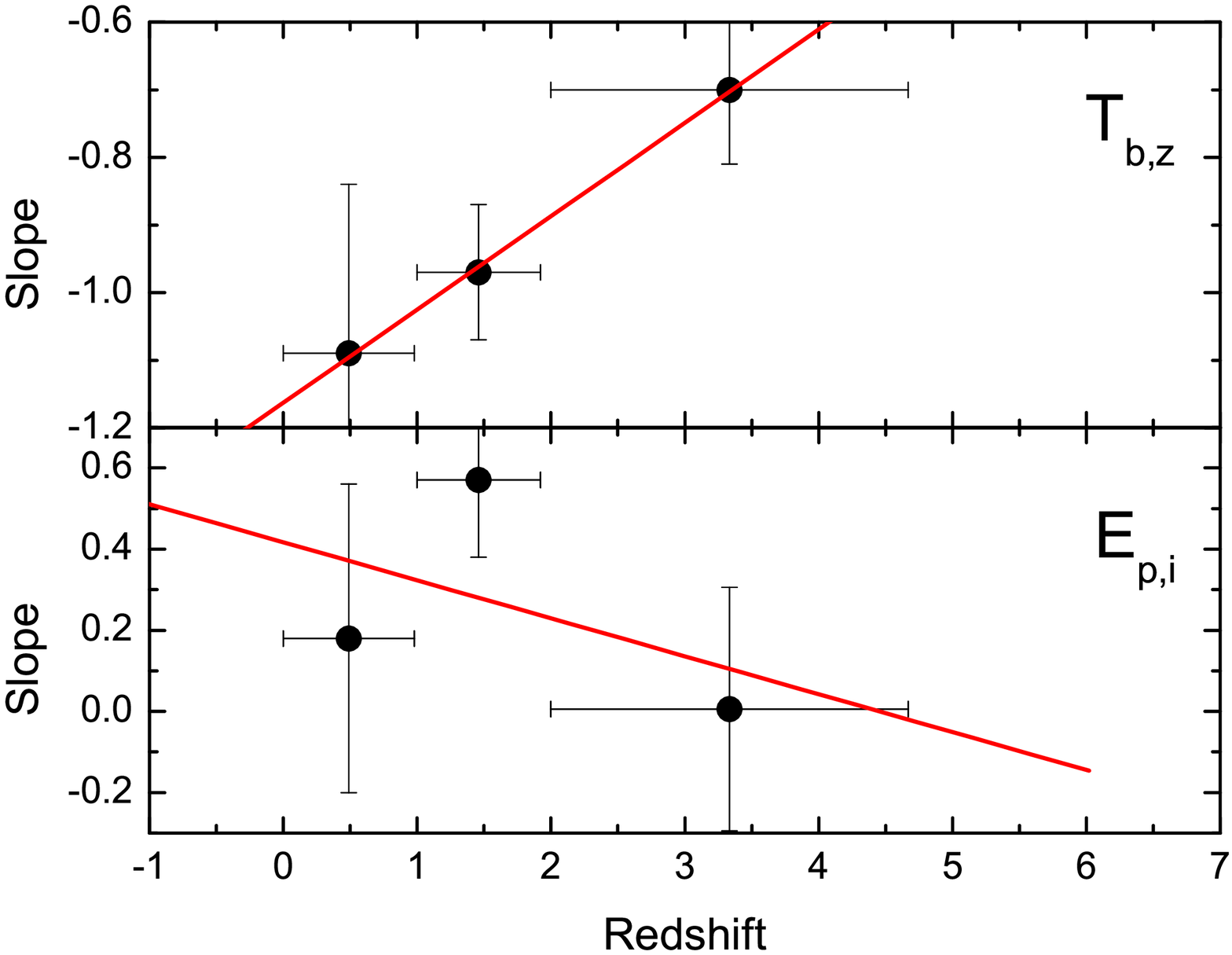}
\includegraphics[angle=0,scale=0.32]{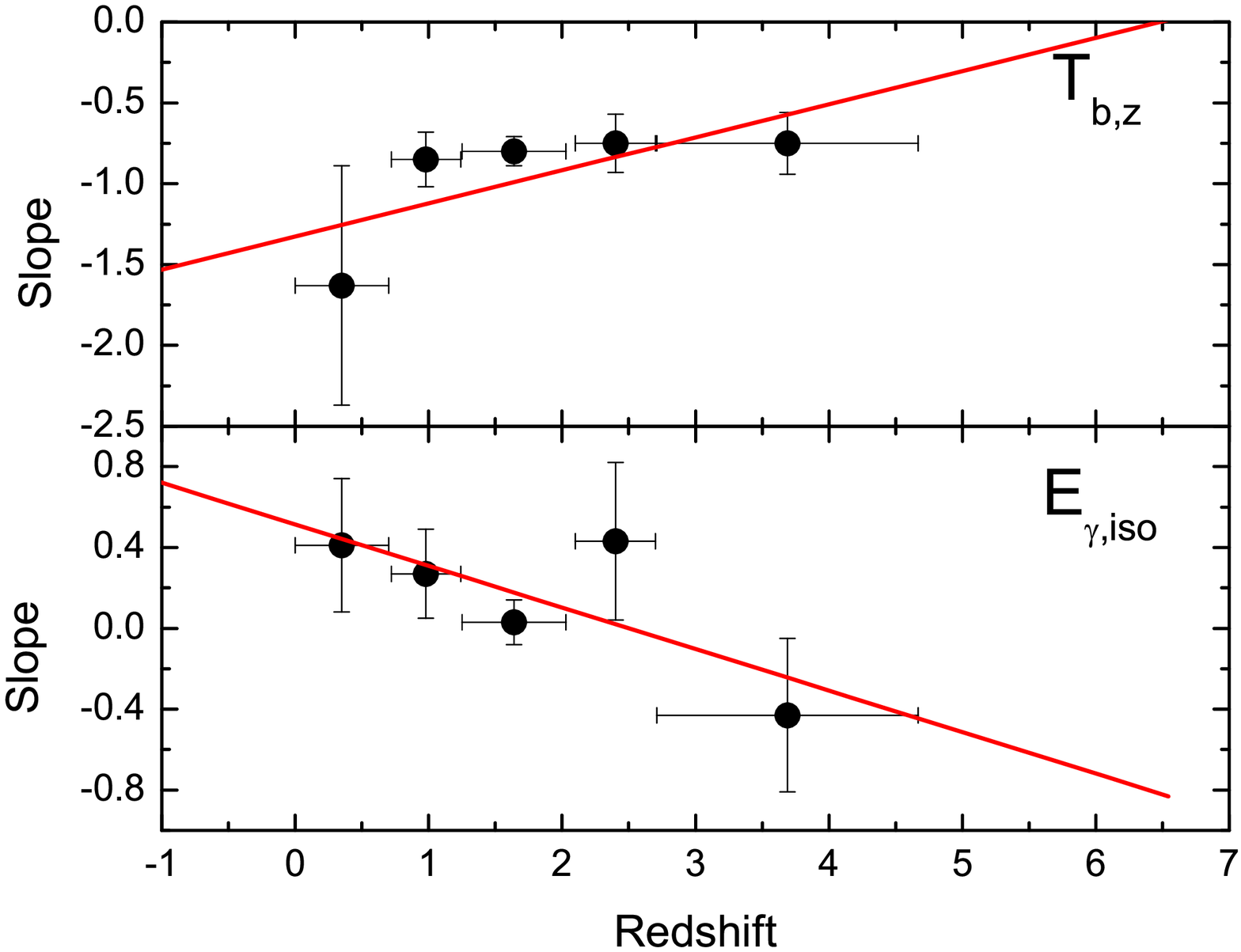}
\includegraphics[angle=0,scale=0.32]{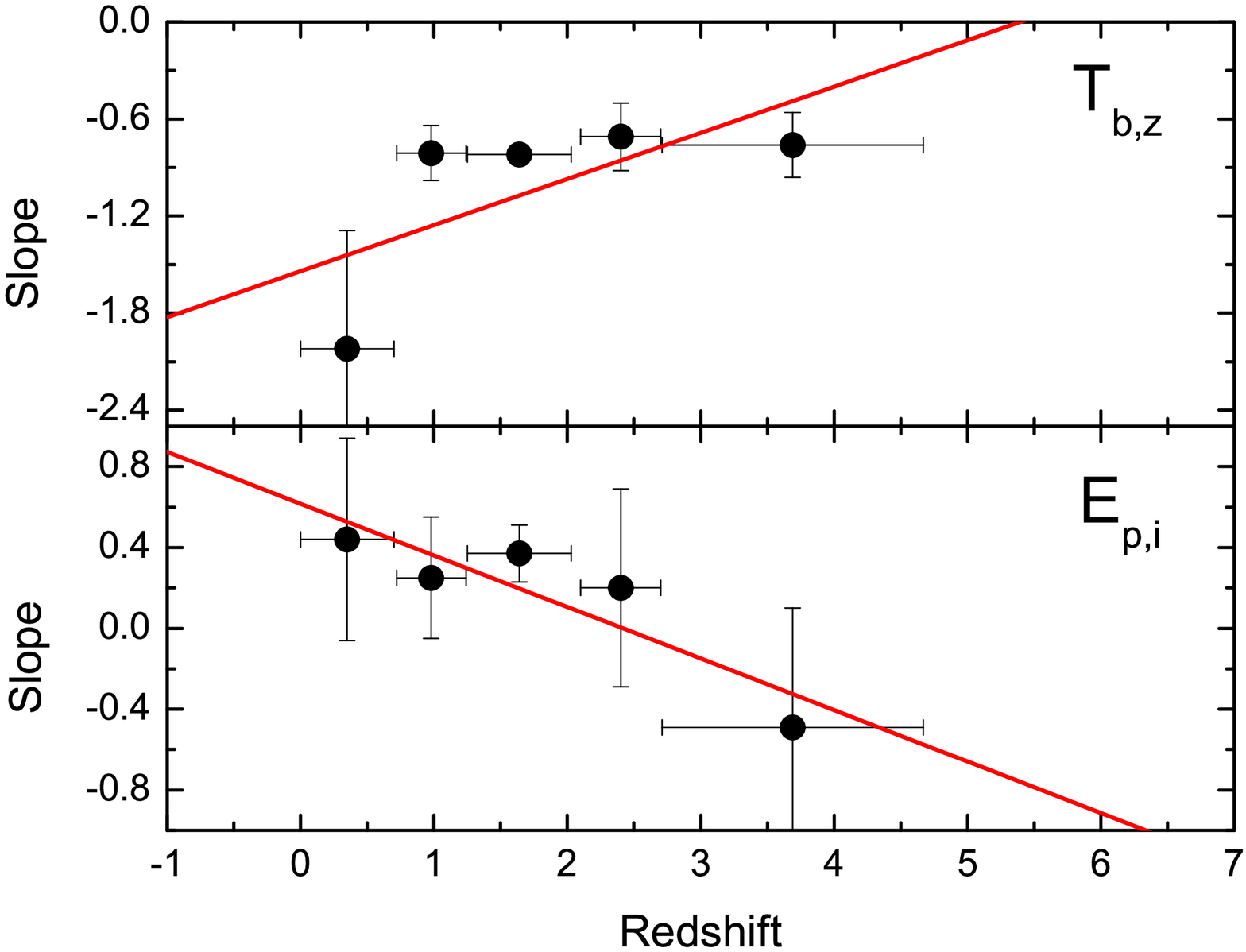}
\caption{The slope of the correlation in different redshift bins vary with the redshift for the two three-parameter correlations. The error bar is expressed as the redshift bin, and the black point is the mean value of the corresponding redshift bin. Above:
3 redshift bins; Bottom: 5 redshift bins.}
\end{figure*}

\begin{figure*}
\includegraphics[angle=0,scale=0.32]{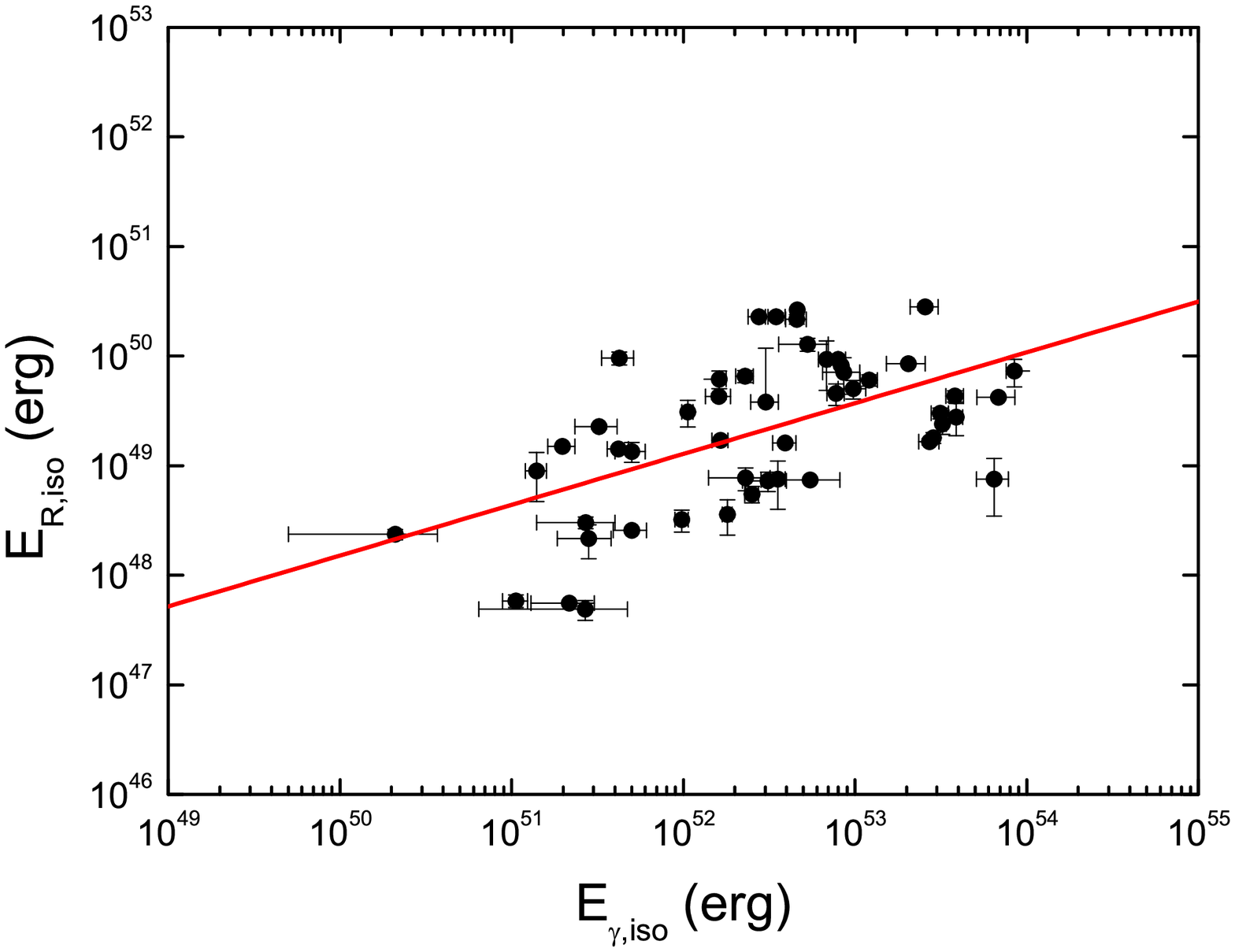}
\includegraphics[angle=0,scale=0.32]{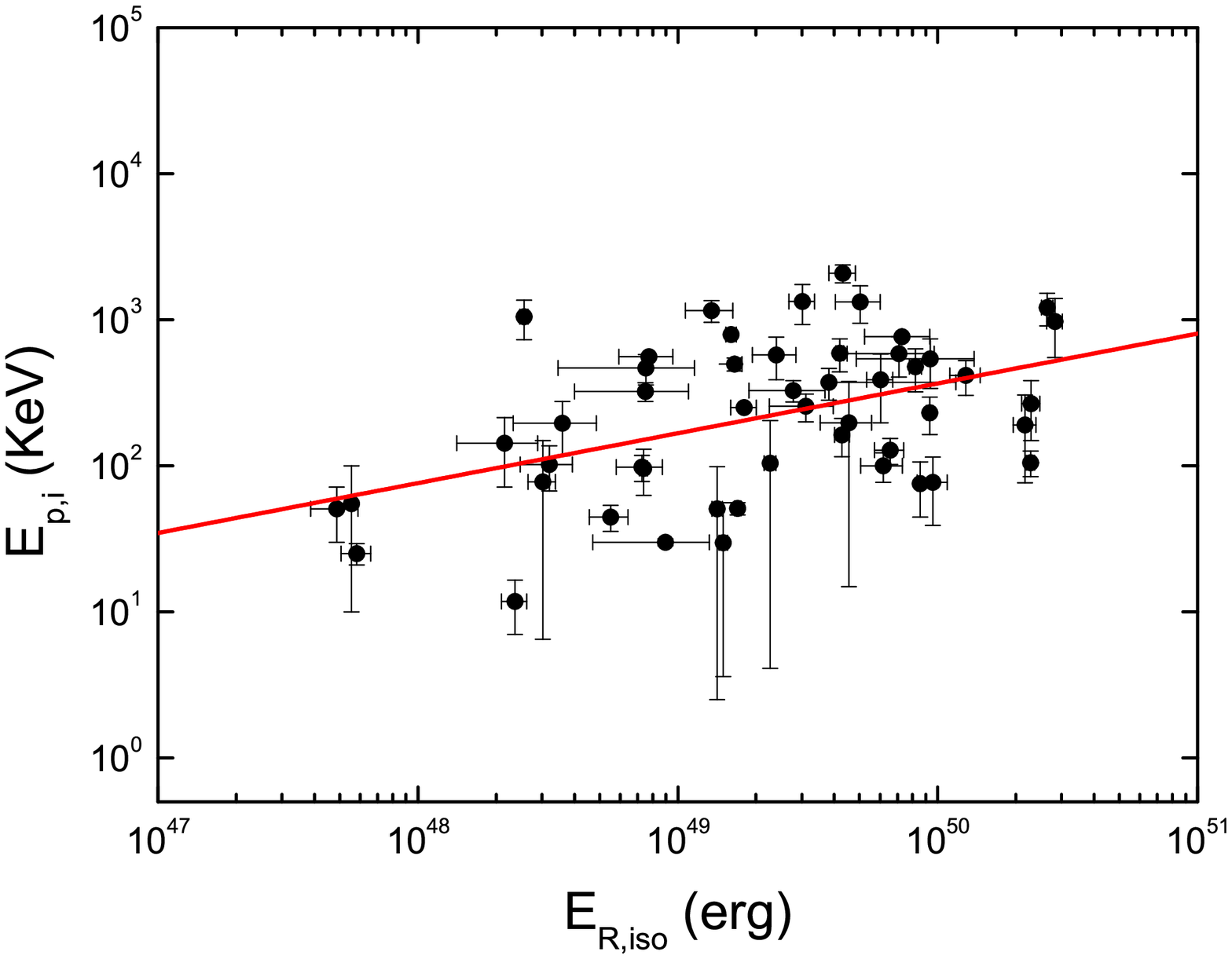}
\caption{The correlations between $E_{\rm R,iso}$-$E_{\rm \gamma,iso}$ (left pannel) and $E_{\rm p,i}$-$E_{\rm R,iso}$ (right pannel) for optical plateaus, where the optical energy $E_{\rm R,iso}=L_{\rm b,z}\times T_{\rm b}$.}
\end{figure*}


\clearpage
\begin{deluxetable}{ccccccccccccccccccccccccc}
\tabletypesize{\scriptsize}
\tablecaption{The fitting results for our optical sample. }
\tablewidth{0pt}
\tabletypesize{\tiny}

\tablehead{ \colhead{GRB}
&\colhead{$z$}
&\colhead{$F_{b}\tablenotemark{a}$}
&\colhead{$T_{b}\tablenotemark{b} $}
&\colhead{$\alpha_{1} $}
&\colhead{$\alpha_{2}$}
&\colhead{$L_{b,z}\tablenotemark{c}$}
&\colhead{$E_{\gamma,iso}\tablenotemark{d}$}
&\colhead{$E_{p,i}\tablenotemark{e}$}
&\colhead{Refs.\tablenotemark{f}}}

\startdata
000301C	&	2.03	&	6.27	$\pm$	0.20	&	410.60	$\pm$	7.72	&	-0.57	$\pm$	0.03	 &	-2.72	$\pm$	0.06	&	6.42	 $\pm$	0.20	&	4.60			&	1213.0	$\pm$	 303.0	&	1, 2	&\\
010222	&	1.48	&	24.48	$\pm$	3.50	&	51.78	$\pm$	7.20	&	-0.68	$\pm$	0.04	 &	-1.33	$\pm$	0.03	&	14.04	 $\pm$	2.01	&	84.90	$\pm$	9.03	&	766.0	 $\pm$	20.0	&	3, 4	&\\
020813	&	1.25	&	68.51	$\pm$	1.10	&	14.77	$\pm$	0.28	&	-0.20	$\pm$	0.01	 &	-1.04	$\pm$	0.01	&	28.47	 $\pm$	 0.46	&	68.35	$\pm$	17.09	&	590.0	 $\pm$	151.0	&	3, 4	&\\
021004	&	2.34	&	241.42	$\pm$	9.31	&	7.21	$\pm$	0.31	&	-0.30	$\pm$	0.01	 &	-1.03	$\pm$	0.00	&	316.60	 $\pm$	12.20	&	3.47	$\pm$	0.46	&	266.0	 $\pm$	117.0	&	3, 4	&\\
030328	&	1.52	&	21.39	$\pm$	3.59	&	21.58	$\pm$	3.40	&	-0.63	$\pm$	0.06	 &	-1.22	$\pm$	0.04	&	12.90	 $\pm$	2.16	&	38.86	$\pm$	3.62	&	328.0	 $\pm$	55.0	&	3, 4	&\\
030329	&	0.17	&	5000.0	$\pm$	500.0	&	18.00	$\pm$	1.50	&	-0.51	$\pm$	0.03	 &	-1.31	$\pm$	0.03	&	34.32	 $\pm$	 3.43	&	1.62	$\pm$	0.16	&	100.0	 $\pm$	23.0	&	5, 4	&\\
030429	&	2.65	&	1.84	$\pm$	0.16	&	218.49	$\pm$	8.85	&	-0.86	$\pm$	0.03	 &	-3.53	$\pm$	0.03	&	3.00	 $\pm$	0.26	&	2.29	$\pm$	0.27	&	128.0	 $\pm$	26.0	&	3, 4	&\\
030723	&	0.40	&	5.80	$\pm$	0.58	&	99.00	$\pm$	1.09	&	-0.03	$\pm$	0.05	 &	-1.81	$\pm$	0.02	&	0.24	 $\pm$	 0.02	&	0.02	$\pm$	0.02	&	11.8	 $\pm$	4.8	&	6, 6	&\\
040924	&	0.86	&	76.69	$\pm$	8.85	&	2.11	$\pm$	0.24	&	-0.42	$\pm$	0.16	 &	-1.28	$\pm$	0.02	&	15.20	 $\pm$	1.75	&	0.98	$\pm$	0.09	&	102.0	 $\pm$	35.0	&	3, 4	&\\
041006	&	0.72	&	74.62	$\pm$	7.60	&	7.11	$\pm$	0.70	&	-0.26	$\pm$	0.04	 &	-1.26	$\pm$	0.01	&	10.22	 $\pm$	1.04	&	3.11	$\pm$	0.89	&	98.0	 $\pm$	20.0	&	3, 4	&\\
050319	&	3.24	&	3.80	$\pm$	0.17	&	249.82	$\pm$	13.72	&	-0.54	$\pm$	0.01	 &	-1.95$\pm$	0.04		&	8.67	 $\pm$	 0.39	 &	4.57	$\pm$	0.63	&	190.8	 $\pm$	114.5	&	2, 2	&\\
050401	&	2.90	&	2.22	$\pm$	0.49	&	17.84	$\pm$	5.76	&	-0.50(fixed)	&	 -0.89$\pm$	0.04		&	4.22	$\pm$	 0.92	 &	64.70	$\pm$	13.60	&	467.0	$\pm$	 110.0	&	7, 4	&\\
050408	&	1.24	&	1.38	$\pm$	0.09	&	98.14	$\pm$	10.24	&	-0.59	$\pm$	0.04	 &	-1.20	$\pm$	0.02	&	0.56	 $\pm$	 0.04	&	2.51	$\pm$	0.23	&	44.6	 $\pm$	8.9	&	8, 8	&\\
050416A	&	0.65	&	3.46	$\pm$	0.20	&	14.98	$\pm$	1.09	&	-0.38	$\pm$	0.01	 &	-1.30	$\pm$	0.23	&	0.39	 $\pm$	 0.02	&	0.11	$\pm$	0.02	&	25.1	 $\pm$	4.2	&	3, 4	&\\
050730	&	3.97	&	91.07	$\pm$	0.05	&	9.80	$\pm$	0.68	&	-0.38	$\pm$	0.02	 &	-1.50	$\pm$	0.01	&	288.14	 $\pm$	 0.15	&	25.70	$\pm$	4.73	&	777.9	 $\pm$	345.3	&	9, 2	&\\
050801	&	1.56	&	1579.94	$\pm$	38.29	&	0.23	$\pm$	0.01	&	-0.04	$\pm$	0.02	 &	-1.20	$\pm$	0.01	&	1000.22	 $\pm$	24.24	&	0.32	$\pm$	0.09	&	104.1	 $\pm$	100.0	&	2, 2	&\\
050922C	&	2.20	&	129.63	$\pm$	8.93	&	8.40	$\pm$	0.55	&	-0.66	$\pm$	0.02	 &	-1.42	$\pm$	0.02	&	152.92	 $\pm$	10.53	&	5.30	$\pm$	1.70	&	415.0	 $\pm$	111.0	&	4, 4	&\\
051109A	&	2.35	&	51.87	$\pm$	10.65	&	13.62	$\pm$	3.73	&	-0.65	$\pm$	0.01	 &	-1.30	$\pm$	0.07	&	68.58	 $\pm$	14.08	&	6.85	$\pm$	0.73	&	539.0	 $\pm$	200.0	&	3, 4	&\\
051111	&	1.55	&	293.55	$\pm$	28.41	&	2.74	$\pm$	0.27	&	-0.79	$\pm$	0.01	 &	-1.77	$\pm$	0.16	&	183.62	 $\pm$	17.77	&	9.77	$\pm$	1.80	&	1328.0	 $\pm$	379.9	&	2, 2	&\\
060210	&	3.91	&	121.66	$\pm$	11.22	&	0.64	$\pm$	0.06	&	0.15	$\pm$	0.07	 &	-1.17	$\pm$	0.04	&	375.88	 $\pm$	34.65	&	32.23	$\pm$	1.84	&	574.0	 $\pm$	187.0	&	3, 3	&\\
060526	&	3.21	&	83.07	$\pm$	1.37	&	12.20	$\pm$	0.24	&	-0.54	$\pm$	0.08	 &	-1.13	$\pm$	0.00	&	186.68	 $\pm$	3.07	&	2.75	$\pm$	0.37	&	105.0	 $\pm$	21.0	&	3, 4	&\\
060614	&	0.13	&	24.15	$\pm$	0.26	&	62.43	$\pm$	0.62	&	0.07	$\pm$	0.02	 &	-1.95	$\pm$	0.02	&	0.09	 $\pm$	9.56E-4	&	0.22	$\pm$	0.09	&	55.0	 $\pm$	45.0	&	3, 4	&\\
060708	&	1.92	&	465.48	$\pm$	51.05	&	0.72	$\pm$	0.12	&	-0.06	$\pm$	0.05	 &	-0.87	$\pm$	0.02	&	431.45	 $\pm$	47.32	&	1.06	$\pm$	0.08	&	255.4	 $\pm$	55.3	&	5, 10	&\\
060714	&	2.71	&	34.94	$\pm$	2.88	&	7.69	$\pm$	1.09	&	-0.18	$\pm$	0.02	 &	-1.12	$\pm$	0.03	&	59.23	 $\pm$	4.87	&	7.76	$\pm$	0.89	&	196.7	 $\pm$	181.8	&	2, 2	&\\
060729	&	0.54	&	280.0	$\pm$	28.0	&	44.70	$\pm$	1.45	&	-0.11	$\pm$	0.02	 &	-1.24	$\pm$	0.04	&	21.46	 $\pm$	 2.15	&	0.42	$\pm$	0.09	&	77.0	 $\pm$	38.0	&	3, 3	&\\
061021	&	0.35	&	9.76	$\pm$	0.18	&	86.18	$\pm$	1.61	&	-0.66	$\pm$	0.06	 &	-2.09	$\pm$	0.34	&	0.30	 $\pm$	 0.01	&	0.50	$\pm$	0.11	&	1046.0	 $\pm$	319.0	&	5, 10	&\\
061126	&	1.16	&	28.00	$\pm$	2.80	&	30.00	$\pm$	0.40	&	-0.45	$\pm$	0.03	 &	-1.77	$\pm$	0.03	&	10.05	 $\pm$	 1.00	&	31.42	$\pm$	3.59	&	1337.0	 $\pm$	410.0	&	3, 4	&\\
070110	&	2.35	&	20.08	$\pm$	14.26	&	14.29	$\pm$	19.65	&	-0.16	$\pm$	0.27	 &	-0.98	$\pm$	0.78	&	26.66	 $\pm$	18.93	&	3.02	$\pm$	0.56	&	372.1	 $\pm$	90.5	&	2, 2	&\\
070208	&	1.17	&	5.42	$\pm$	0.92	&	10.95	$\pm$	1.91	&	-0.44	$\pm$	0.04	 &	-2.05	$\pm$	0.32	&	1.96	 $\pm$	 0.33	&	0.28	$\pm$	0.10	&	142.9	 $\pm$	71.4	&	2, 2	&\\
070411	&	2.95	&	2.39	$\pm$	0.08	&	175.95	$\pm$	3.75	&	-0.87	$\pm$	0.15	 &	-1.86	$\pm$	0.02	&	4.67	 $\pm$	0.16	&	8.31	$\pm$	0.45	&	475.5	 $\pm$	154.2	&	5, 2	&\\
070518	&	1.16	&	2.80	$\pm$	0.28	&	30.00	$\pm$	0.60	&	-0.45	$\pm$	0.03	 &	-1.85	$\pm$	0.34	&	1.01	 $\pm$	 0.10	&	0.27	$\pm$	0.13	&	77.8	 $\pm$	71.3	&	11, 7	&\\
071003	&	1.60	&	6589	$\pm$	780	&	0.1	$\pm$	0.01	&	-0.84	$\pm$	0.07	&	 -1.60	$\pm$	0.01	&	 4393	 $\pm$	 520	&	38.30	$\pm$	4.50	&	2077.0	$\pm$	 286.0	&	5, 3	&\\
080310	&	2.43	&	214.76	$\pm$	2.08	&	2.84	$\pm$	0.04	&	-0.12	$\pm$	0.01	 &	-1.25	$\pm$	0.01	&	301.08	 $\pm$	2.92	&	20.42	$\pm$	5.17	&	75.4	 $\pm$	30.8	&	9, 2	&\\
080330	&	1.51	&	151.32	$\pm$	3.32	&	1.57	$\pm$	0.04	&	0.10	$\pm$	0.02	 &	-1.14	$\pm$	0.01	&	90.13	 $\pm$	1.98	&	41.00	$\pm$	6.00	&	50.7	 $\pm$	48.2	&	2, 2	&\\
080413A	&	2.43	&	1292	$\pm$	238	&	0.39	$\pm$	0.07	&	-0.64	$\pm$	0.03	 &	-1.82	$\pm$	0.29	&	1819.83	 $\pm$	335.34	&	8.59	$\pm$	2.10	&	584.0	 $\pm$	180.0	&	3, 3	&\\
080413B	&	1.10	&	3.66	$\pm$	0.14	&	361.63	$\pm$	9.85	&	-0.50	$\pm$	0.01	 &	-2.42	$\pm$	0.03	&	1.18	 $\pm$	0.05	&	1.61	$\pm$	0.27	&	163.0	 $\pm$	47.5	&	3, 3	&\\
081029	&	3.85	&	100.0	$\pm$	10.0	&	2.00	$\pm$	0.03	&	-0.50	$\pm$	0.05	 &	-1.08	$\pm$	0.14	&	301.21	 $\pm$	 30.12	&	12.10	$\pm$	1.40	&	324.8	 $\pm$	63.0	&	5, 12	&\\
081109A	&	0.98	&	250.19	$\pm$	31.14	&	0.56	$\pm$	0.13	&	0.19	$\pm$	0.18	 &	-0.94	$\pm$	0.03	&	64.37	 $\pm$	8.01	&	1.81	$\pm$	0.12	&	195.9	 $\pm$	79.1	&	5, 7	&\\
090426	&	2.61	&	368.72	$\pm$	33.67	&	0.23	$\pm$	0.03	&	-0.27	$\pm$	0.07	 &	-1.23	$\pm$	0.04	&	585.50	 $\pm$	53.46	&	0.50	$\pm$	0.10	&	1154.9	 $\pm$	194.8	&	13, 14	&\\
090618	&	0.54	&	84.13	$\pm$	4.46	&	27.94	$\pm$	1.69	&	-0.66	$\pm$	0.01	 &	-1.49	$\pm$	0.06	&	6.45	 $\pm$	0.34	&	28.59	$\pm$	0.52	&	250.4	 $\pm$	4.6	&	3, 3	&\\
091029	&	2.75	&	11.35	$\pm$	0.12	&	47.18	$\pm$	0.84	&	-0.44	$\pm$	0.05	 &	-1.55(fixed)		&	19.74	 $\pm$	 0.20	&	7.97	$\pm$	0.82	&	230.0	$\pm$	 66.0	&	3, 3	&\\
091127	&	0.49	&	138.98	$\pm$	1.75	&	19.52	$\pm$	0.31	&	-0.43	$\pm$	0.06	 &	-1.26(fixed)		&	8.71	 $\pm$	0.11	&	1.65	$\pm$	0.18	&	51.0	$\pm$	 5.0	&	3, 3	&\\
100219A	&	4.67	&	21.89	$\pm$	0.74	&	1.80	$\pm$	0.02	&	-0.74	$\pm$	0.09	 &	-1.68	$\pm$	0.34	&	89.14	 $\pm$	 3.00	&	3.93	$\pm$	0.61	&	793.3	 	&	5, 7	&\\
100418A	&	0.62	&	31.42	$\pm$	7.37	&	27.65	$\pm$	6.64	&	0.68	$\pm$	0.17	 &	-1.37	$\pm$	0.13	&	3.24	 $\pm$	0.76	&	0.14	$\pm$	0.02	&	30.0	 $\pm$	1.6	&	11, 15	&\\
100728B	&	2.11	&	19.97	$\pm$	4.73	&	3.44	$\pm$	0.79	&	0.13	$\pm$	0.44	 &	-2.18	$\pm$	0.70	&	21.84	 $\pm$	5.18	&	3.55	$\pm$	0.36	&	323.0	 $\pm$	47.0	&	3, 3	&\\
101225A	&	0.33	&	1.10	$\pm$	0.09	&	161.83	$\pm$	20.88	&	-0.12	$\pm$	0.01	 &	-0.72	$\pm$	0.02	&	0.03	 $\pm$	 0.01	&	0.27	$\pm$	0.20	&	50.5	 $\pm$	20.7	&	16, 16	&\\
120119A	&	1.73	&	189.36	$\pm$	6.05	&	1.14	$\pm$	0.04	&	0.27	$\pm$	0.04	 &	-1.37	$\pm$	0.02	&	144.88	 $\pm$	4.63	&	27.20	$\pm$	3.63	&	496.0	 $\pm$	50.0	&	3, 3	&\\
120729A	&	0.80	&	82.28	$\pm$	11.32	&	5.48	$\pm$	0.54	&	-0.91	$\pm$	0.03	 &	-2.09	$\pm$	0.04	&	14.12	 $\pm$	1.94	&	2.30	$\pm$	0.90	&	559.1		 &	17, 17	&\\
120815A	&	2.36	&	101.05	$\pm$	1.04	&	0.55	$\pm$	0.01	&	0.24	$\pm$	0.02	 &	-0.64	$\pm$	0.03	&	134.80	 $\pm$	1.39	&	5.50	$\pm$	2.66	&	96.0	 $\pm$	33.5	&	9, 15	&\\
170519A	&	0.82	&	658.57	$\pm$	8.23	&	1.27	$\pm$	0.03	&	0.30	$\pm$	0.02	 &	-0.83	$\pm$	0.01	&	118.24	 $\pm$	1.48	&	0.20	$\pm$	0.04	&	29.6	 $\pm$	26.1	&	18, 18	&\\

\enddata
\tablenotetext{a}{In units of $10^{-14}$ erg cm$^{-2}$ s$^{-1}$.}
\tablenotetext{b}{In units of kilo seconds.}
\tablenotetext{c}{In units of $10^{44}$ erg/s.}
\tablenotetext{d}{In units of $10^{52}$ erg.}
\tablenotetext{e}{In units of KeV.}
\tablenotetext{f}{References for $E_{\rm \gamma,iso}$ and $E_{\rm p,i}$.}
\tablerefs{(1) Jensen et al. 2001; (2) Kann et al. 2010; (3) Demianski et al. 2017; (4) Amati et al. 2008; (5) Ruffini et al. 2016; (6) Butler et al. 2005; (7) Li et al. 2012; (8) Wei et al. 2013; (9) Beskin et al. 2015; (10) Yu et al. 2015; (11) Wang et al. 2015; (12) Cummings et al. 2008; (13) Antonelli et al. 2009; (14) Zaninoni et al. 2016; (15) Zitouni et al. 2014; (16) Thone et al. 2011; (17) Cano et al. 2014;
(18) Krimm et al. 2017}

\end{deluxetable}

\clearpage
\begin{deluxetable}{ccccccccccccccccccccccccc}
\tabletypesize{\scriptsize} \tablecaption{Results of the linear
regression analysis for optical plateaus. $R$ is the Spearman
correlation coefficient, $P$ is the chance probability, and $\delta$
is the correlation dispersion.} \tablewidth{0pt}

\tablehead{ \colhead{Correlations}& \colhead{Expressions}&
\colhead{$R$}& \colhead{$P$}& \colhead{$\delta$} }

\startdata
\hline
$F_{\rm b}(T_{\rm b,z})$ & $\log F_{\rm b}=(-9.94\pm 0.36)+(-0.66\pm0.10)\times \log T_{\rm b,z}$ & -0.71 & $<10^{-4}$ & 0.65 \\
$L_{\rm b,z}(T_{\rm b,z})$ & $\log L_{\rm b,z}=(48.87\pm 0.34)+(-0.99\pm0.09)\times \log T_{\rm b,z}$ & -0.84 & $<10^{-4}$ & 0.62\\
\hline
$L_{\rm b,z}(T_{\rm b,z},E_{\rm \gamma,iso})$ & $\log L_{\rm b,z}=(29.22\pm 5.04)+(-0.92\pm0.08)\times \log T_{\rm b,z}$ & 0.89 & $<10^{-4}$ & 0.54\\& $+(0.37\pm0.09)\times \log E_{\rm \gamma,iso})$ &  &  & \\
$L_{\rm b,z}(T_{\rm b,z},E_{\rm p,i})$ & $\log L_{\rm b,z}=(47.48\pm 0.56)+(-0.91\pm0.09)\times \log T_{\rm b,z}$ & 0.87 & $<10^{-4}$ & 0.57\\
 & $+(0.48\pm0.16)\times \log E_{\rm p,i})$ &  &  & \\
 \hline
$E_{\rm R,iso}(E_{\rm \gamma,iso})$ & $\log E_{\rm R,iso}=(24.99\pm 5.29)+(0.46\pm0.10)\times \log E_{\rm \gamma,iso}$ & 0.55 & $<10^{-4}$ & 0.59\\$E_{\rm p,i}(E_{\rm R,iso})$ & $\log E_{\rm p,i}=(-14.51\pm 4.79)+(0.34\pm0.09)\times \log E_{\rm R,iso}$ & 0.45 & $9.62\times10^{-4}$ & 0.48\\
\enddata
\end{deluxetable}

\clearpage
\begin{deluxetable}{ccccccccccccccccccccccccc}
\tabletypesize{\scriptsize} \tablecaption{Results of the linear
regression analysis of optical plateaus for three redshift bins and five bins, respectively.} \tablewidth{0pt}

\tablehead{ \colhead{Redshift Bins}& \colhead{Expressions}&
\colhead{$R$}& \colhead{$P$}& \colhead{$\delta$}& \colhead{N} }

\startdata
\hline
0-0.98 & $\log L_{\rm b,z}=(29.00\pm 13.52)+(-0.99\pm0.25)\times \log T_{\rm b,z}$ & 0.82 & $1.88\times 10^{-4}$ & 0.66 & 15\\ & $+(0.38\pm0.25)\times \log E_{\rm \gamma,iso})$ &  &  & \\
1-1.92 & $\log L_{\rm b,z}=(32.39\pm 5.43)+(-1.05\pm0.09)\times \log T_{\rm b,z}$ & 0.96 & $<10^{-4}$ & 0.34 & 14\\ & $+(0.32\pm0.10)\times \log E_{\rm \gamma,iso})$ &  &  & \\
2-4.67 & $\log L_{\rm b,z}=(50.94\pm 12.14)+(-0.70\pm0.11)\times \log T_{\rm b,z}$ & 0.83 & $<10^{-4}$ & 0.45& 21\\ & $+(-0.05\pm0.23)\times \log E_{\rm \gamma,iso})$ &  &  & \\
\hline
0-0.98 & $\log L_{\rm b,z}=(48.56\pm 1.35)+(-1.09\pm0.25)\times \log T_{\rm b,z}$ & 0.79 & $5.13\times 10^{-4}$ & 0.71& 15\\ & $+(0.18\pm0.38)\times \log E_{\rm p,i})$ &  &  & \\
1-1.92 & $\log L_{\rm b,z}=(47.41\pm 0.65)+(-0.97\pm0.10)\times \log T_{\rm b,z}$ & 0.96 & $< 10^{-4}$ & 0.35& 14\\ & $+(0.57\pm0.19)\times \log E_{\rm p,i})$ &  &  & \\
2-4.67 & $\log L_{\rm b,z}=(48.14\pm 0.89)+(-0.70\pm0.11)\times \log T_{\rm b,z}$ & 0.83 & $< 10^{-4}$ & 0.46& 21\\ & $+(0.006\pm0.30)\times \log E_{\rm p,i})$ &  &  & \\
\hline
\hline
0-0.70 & $\log L_{\rm b,z}=(30.17\pm 18.36)+(-1.63\pm0.74)\times \log T_{\rm b,z}$ & 0.76 & $1.1\times 10^{-2}$ & 0.73&10\\ & $+(0.41\pm0.33)\times \log E_{\rm \gamma,iso})$ &  &  & \\
0.72-1.24 & $\log L_{\rm b,z}=(33.68\pm 11.32)+(-0.85\pm0.17)\times \log T_{\rm b,z}$ & 0.88 & $7.0\times10^{-4}$ & 0.39&10\\ & $+(0.27\pm0.22)\times \log E_{\rm \gamma,iso})$ &  &  & \\
1.25-2.03 & $\log L_{\rm b,z}=(46.83\pm 5.79)+(-0.80\pm0.09)\times \log T_{\rm b,z}$ & 0.96 & $<10^{-4}$ & 0.27&10\\ & $+(0.03\pm0.11)\times \log E_{\rm \gamma,iso})$ &  &  & \\
2.1-2.7 & $\log L_{\rm b,z}=(25.57\pm 20.25)+(-0.75\pm0.18)\times \log T_{\rm b,z}$ & 0.85 & $2.0\times10^{-3}$ & 0.45&10\\ & $+(0.43\pm0.39)\times \log E_{\rm \gamma,iso})$ &  &  & \\
2.71-4.67 & $\log L_{\rm b,z}=(71.24\pm 20.17)+(-0.75\pm0.19)\times \log T_{\rm b,z}$ & 0.84 & $2.6\times10^{-3}$ & 0.45&10\\ & $+(-0.43\pm0.38)\times \log E_{\rm \gamma,iso})$ &  &  & \\
\hline
0-0.70 & $\log L_{\rm b,z}=(52.30\pm 3.30)+(-2.02\pm0.73)\times \log T_{\rm b,z}$ & 0.73 & $1.6\times 10^{-2}$ & 0.77&10\\ & $+(0.44\pm0.50)\times \log E_{\rm p,i})$ &  &  & \\
0.72-1.24 & $\log L_{\rm b,z}=(47.30\pm 0.89)+(-0.81\pm0.17)\times \log T_{\rm b,z}$ & 0.87 & $1.1\times 10^{-3}$ & 0.41&10\\ & $+(0.25\pm0.30)\times \log E_{\rm p,i})$ &  &  & \\
1.25-2.03 & $\log L_{\rm b,z}=(47.63\pm 0.37)+(-0.82\pm0.06)\times \log T_{\rm b,z}$ & 0.98 & $< 10^{-4}$ & 0.19&10\\ & $+(0.37\pm0.14)\times \log E_{\rm p,i})$ &  &  & \\
2.1-2.7 & $\log L_{\rm b,z}=(47.73\pm 1.53)+(-0.71\pm0.21)\times \log T_{\rm b,z}$ & 0.82 & $3.7\times10^{-3}$ & 0.49&10\\ & $+(0.20\pm0.49)\times \log E_{\rm p,i})$ &  &  & \\
2.71-4.67 & $\log L_{\rm b,z}=(49.59\pm 1.91)+(-0.76\pm0.20)\times \log T_{\rm b,z}$ & 0.82 & $3.6\times 10^{-3}$ & 0.46&10\\ & $+(-0.49\pm0.59)\times \log E_{\rm p,i})$ &  &  & \\
\enddata
\end{deluxetable}

\end{document}